\documentclass{jcd}
\usepackage{amsmath,amssymb}

\usepackage{algorithmic}
\usepackage{mathrsfs}
\usepackage{mathtools}

\usepackage{pdflscape}
\usepackage{afterpage}
\usepackage{caption}
\usepackage{cases}
  \usepackage{paralist}
  \usepackage{graphics} 
  \usepackage{epsfig} 
\usepackage{graphicx}  \usepackage{epstopdf}
 \usepackage[colorlinks=true]{hyperref}
\hypersetup{urlcolor=blue, citecolor=red, pdfstartview=FitH, bookmarks=false, bookmarksopen=false}

\usepackage[T1]{fontenc}

\def\currentvolume{X}
 \def\currentissue{X}
  \def\currentyear{200X}
   \def\currentmonth{XX}
    \def\ppages{X--XX}
     \def\DOI{10.3934/xx.xxxxxxx}

\theoremstyle{definition}

\newcommand{\cref}[1]{Conjecture~\ref{#1}}

\newcommand{\be}{\begin{equation}}
\newcommand{\ee}{\end{equation}}

\newcommand{\udot}{\dot{u}}
\newcommand{\vdot}{\dot{v}}
\newcommand{\zdot}{\dot{z}}
\newcommand{\uddot}{\ddot{u}}
\newcommand{\R}{\mathbb{R}}

\newcommand{\cP}{\mathcal{P}}
\renewcommand{\leq}{\leqslant}
\renewcommand{\geq}{\geqslant}
\renewcommand{\phi}{\varphi}

\title[Bistability, Bifurcations and Chaos in Mackey-Glass] 
      {Bistability, Bifurcations and Chaos\\ in the Mackey-Glass Equation}

\author[V. Duruisseaux  and A. R. Humphries]{}



 \email{vduruiss@ucsd.edu}
 \email{tony.humphries@mcgill.ca}

\begin{document}
\maketitle

\centerline{\scshape Valentin Duruisseaux}
\medskip
{\footnotesize
 \centerline{Department of Mathematics}
 \centerline{University of California, San Diego}
 \centerline{La Jolla, California 92093, USA}
} 

\medskip

\centerline{\scshape Antony R. Humphries}
\medskip
{\footnotesize
 \centerline{Departments Mathematics \& Statistics, and, Physiology}
 \centerline{McGill University}
 \centerline{Montreal, Quebec H3A 0B9, Canada}
} 

\bigskip


\begin{abstract}
Numerical bifurcation analysis, and in particular two-parameter continuation, is used in consort with numerical simulation to reveal complicated dynamics in the Mackey-Glass equation for moderate values of the delay close to the onset of chaos. In particular a cusp bifurcation of periodic orbits and resulting branches of folds of periodic orbits effectively partition the parameter space into regions where different behaviours are seen. The cusp bifurcation leads directly to bistability between periodic orbits, and subsequently to bistability between a periodic orbit and a chaotic attractor.
This leads to two different mechanisms by which the chaotic attractor is destroyed in a global bifurcation with a periodic orbit in either an interior crisis or a boundary crisis.
In another part of parameter space a sequence of subcritical period-doublings is found to give rise to bistability between a periodic orbit and a chaotic attractor. Torus bifurcations, and a codimension-two fold-flip bifurcation are also identified, and Lyapunov exponent computations are used to determine chaotic regions and attractor dimension.
\end{abstract}

\section{Introduction}
\label{sec:intro}

The Mackey-Glass equation \cite{Glass_Mackey_1979,Mackey1977}
\begin{equation}
\label{eq:MG}
\udot(t) = -\gamma u(t)+\beta f(u(t-\tau)), \quad f(x) = \frac{\theta^n x}{\theta^n+x^n}
\end{equation}
was originally introduced as a simple nonlinear delayed feedback model for circulating blood cell concentrations in hematopoiesis. It is a scalar delay-differential equation (DDE) with a constant delay $\tau>0$.
As DDEs have become the subject of intense research in recent decades, this
equation has become one of the archetypal examples of a constant-delay scalar DDE which can exhibit chaotic dynamics.

The Mackey-Glass equation has been studied extensively since its introduction in 1977, but much of that work consists of theoretical results for idealised equations, or numerical simulation of the full equation \eqref{eq:MG}, often with very large delay.
In the current work we will use numerical bifurcation analysis and continuation to identify interesting dynamics
for moderate values of the delay $\tau$ near to the onset of chaotic dynamics.

Unlike numerical simulation which can only reveal stable invariant objects, numerical bifurcation analysis can identify unstable periodic orbits, and also bifurcations leading to more complicated invariant objects. Our main tool will be the \textsc{Matlab}~\cite{Matlab} package DDE-BIFTOOL \cite{DDEBiftool15}.
We will show that numerical bifurcation analysis is particularly powerful when it is applied in consort with
numerical simulation to identify interesting parameter regimes and study the dynamics and bifurcations that arise.

Farmer \cite{Farmer82} was the first to systematically study chaos in the Mackey-Glass equation,
using numerical simulation.  For large delays, he computed the Lyapunov exponents and Lyapunov dimension, and also power spectra. 
He also observed that the attractor dimension increases linearly with the delay.
This led to further study of high-dimensional chaos in the
Mackey-Glass equation \cite{Lepri94,Longtin1998}, as well as related equations such as the Ikeda DDE \cite{Ikeda_1987}.
Mensour and Longtin \cite{Longtin1998} study power spectra in detail for the Mackey-Glass and Ikeda DDEs and hence demonstrate the connections between the power spectra, Lyapunov dimension, and the linearisation about the steady state which lead to the dimension increasing linearly with the delay $\tau$. More recently,
routes to chaos are studied using numerical simulation in \cite{Junges_2012}.
R\"ost and Wu \cite{RostWu2007} study DDEs with unimodal feedback, including the Mackey-Glass equation, and find bounds on the global attractor, which limit where chaos can occur.
There is still no rigorous non-numerical proof of chaos in the Mackey-Glass equation, but the search for such a result has been influential; see \cite{walther2020impact} for a review of the
impact of the Mackey-Glass equation on mathematics.
Very recent work includes \cite{Krisztin_StableOrbitsMG_JDE2021} where stable periodic orbits are shown to exist for large $n$ in the Mackey-Glass equation.
Many other authors study modified Mackey-Glass equations with additional delay terms, nonconstant delays, stochasticity or coupled equations.

The Mackey-Glass equation \eqref{eq:MG} is a very simplified model of blood cell production, and there
has been much work in recent decades to develop more sophisticated models of hematopoiesis,
as reviewed by Pujo-Menjouet \cite{PujoMenjouet2016}.
The resulting models cover most aspects of hematopoiesis encompassing erythropoiesis, granulopoiesis and thrombopoiesis, using a range of mathematical models including DDEs with discrete delays \cite{BMM95},
DDEs with distributed delays \cite{Craig_2016}, and large systems of ordinary differential equations (ODEs) \cite{SS20}.
While these models are mostly very well behaved with parameters in normal physiological ranges, large perturbations of the parameters can lead to very complicated dynamics. One example of this is the Burns-Tannock cell-cycle model, which was originally formulated as a scalar constant delay DDE in
\cite{Glass_Mackey_1979}, and which was recently shown to exhibit very complicated dynamics \cite{DeSouza2019}.
Mackey \cite{mcm2020chaos} reviews the work showing how
periodic hematological disorders can be regarded mathematically as dynamical diseases.

The rest of this work is organised as follows. In Section~\ref{sec:mg} we briefly review fundamental aspects of the Mackey-Glass equation, including its formulation as an infinite-dimensional dynamical system, and the existence and stability of steady states.
In Section~\ref{sec:nummethods} we describe the numerical methods that we use to study the non-constant recurrent dynamics in the Mackey-Glass equation, specifically DDE-BIFTOOL in Section~\ref{sec:ddebiftool}, and \verb+dde23+ in Section~\ref{sec:dde23}. Lyapunov dimension computations are described in Section~\ref{sec:LyapDim}.
In Section~\ref{sec:dynamics} we apply these methods to study the bifurcations and dynamics in the Mackey-Glass equation, highlighting several different routes that can lead to chaotic dynamics and bistability.  First, in Section~\ref{sec:contn} we demonstrate one-parameter continuation while revisiting the period-doubling cascade first explored by Glass and Mackey \cite{Glass_Mackey_1979}. Then two-parameter continuation in Section~\ref{sec:twoparcont} leads to the main bifurcation diagram shown in Figure~\ref{fig:2paramcont}. In subsequent sections we explore the dynamics revealed by the bifurcation diagram. In Section~\ref{sec:bistab}
we highlight
a cusp bifurcation leading to bistability between two periodic orbits or between a periodic orbit and a chaotic attractor. We investigate two
global bifurcations through which the chaotic attractor is destroyed when it collides with a periodic orbit
in a boundary crisis (in Section~\ref{sec:cusp}) or an interior crisis (in Section~\ref{sec:snoca}).
In Section~\ref{sec:subcritpd} we study bistability between a chaotic attractor and a periodic orbit arising from a sequence of subcritical period doubling bifurcations. In Section~\ref{sec:foldflip} we show that the Mackey-Glass equation presents fold-flip bifurcations and determine the type of one of these bifurcations.
In Section~\ref{sec:dimension} we look at how the attractor dimension depends on the parameters in the model.
Finally in Section~\ref{sec:conclusions} we present our conclusions.

\section{Mackey-Glass Equation}
\label{sec:mg}

Here we review the basic theory of the Mackey-Glass equation as an infinite-dimensional dynamical system.
The theory of constant discrete delay DDEs as infinite-dimensional dynamical systems is well developed (see \cite{Hale_1993,Smith_2010}). Since the Mackey-Glass equation \eqref{eq:MG} models a population of blood cells, we restrict attention to non-negative solutions.
Thus, let $C_+=C([-\tau,0],\R_+)$ where $\R_+$ denotes the non-negative reals. We equip
$C_+$ with the norm
$\|\phi\|=\max_{\theta\in[-\tau,0]}\phi(\theta)$, and note that $C_+$ is a cone within the Banach space
$C([-\tau,0],\R)$.
For $t\geq0$ let the function segment $u_t\in C_+$ be given by
$$u_t(\theta)=u(t+\theta), \quad \theta\in[-\tau,0].$$
Then posing \eqref{eq:MG} as an initial value problem (IVP) to solve for $t\geq0$
requires an initial function $u_0\in C_+$.
For any such $u_0$ it is easily seen that solutions remain non-negative and can be continued while they remain bounded. Letting
$v(t)=\max_{s\in[-\tau,t]}u(t)$ we find that
$$
\vdot(t) \leq \max\big\{0,\theta^n\beta[v(t)]^{1-n}-\gamma v(t)\big\}
$$
which ensures that
$$\limsup_{t\to\infty}u(t)\leq\max\left\{\|u_0\|,\theta\left(\frac{\beta}{\gamma}\right)^\frac1n\right\}$$
and the Mackey-Glass equation \eqref{eq:MG} defines a dynamical system on $C_+$. The DDE \eqref{eq:MG} may be rewritten as a retarded functional differential equation
$\dot{u}(t)=G(u_t)$,
where $G:C_+\to\R$ is defined by
$G(u_t) = -\gamma u_t(0)+\beta f(u_t(-\tau))$. The function $G$ is easily seen to be completely continuous,
and then it follows from standard results (see \cite{Hale_1988,Hale_1993,Smith_2010}) that the system has compact $\omega$-limit sets, and it can also be shown to have a global attractor in $C_+$ \cite{RostWu2007}.

The steady states of the Mackey-Glass equation\eqref{eq:MG} are found by solving $0=-\gamma u + f(u)$.
With $f$ defined as in \eqref{eq:MG}, this leads to one steady state $\xi_0=0$, and
a second steady state
\be \label{eq:MGxi1}
\xi_1=\theta\left(\frac{\beta}{\gamma}-1\right)^{\frac1n}
\ee
when $\beta>\gamma$. The steady state $\xi_0$ is globally asymptotically stable when it is the unique steady state, hence we consider only the case $\beta>\gamma$ from now on, so that $\xi_1>0$ exists.
Conditions for global asymptotic stability of $\xi_1$ (among strictly positive solutions) can be found in \cite{GTB98}.

Local stability of the steady states $\xi$ is determined by linearising
the DDE~\eqref{eq:MG} around $\xi$, with $z(t)\coloneqq u(t)-\xi$ to get
\begin{equation}\label{eq:MGlin}
\zdot(t)=az(t)+bz(t-\tau),
\end{equation}
with
$$a=-\gamma, \qquad b=\beta f'(\xi).$$
Seeking a nontrivial solution $z(t)=z_{0}\mathnormal{e}^{\lambda t}$ of equation~\eqref{eq:MGlin},
leads to the characteristic equation
\begin{equation}\label{eq:hayes}
p(\lambda)\coloneqq  \lambda - a  - b\mathnormal{e}^{-\lambda\tau} = 0,
\end{equation}
first studied by Hayes~\cite{Hayes_1950}.
Applying the \textit{Principle of Linearised Stability}~\cite{Smith_2010}
the stability analysis of the steady states for the Mackey-Glass equation \eqref{eq:MG} is reduced to the stability analysis of the steady state of the linearised equation~\eqref{eq:hayes}. Stability analysis of equation~\eqref{eq:hayes} is a standard example in DDEs, and can be found in~\cite{Breda_2015,Hale_1993,Insperger_2011,Smith_2010}.

For the Mackey-Glass equation \eqref{eq:MG} with $\beta>\gamma$ the characteristic equation \eqref{eq:hayes} always has a positive root at $\xi_0$, so this steady state is unstable. The steady state $\xi_1$ is asymptotically stable for all delays $\tau\geq0$ if $|f'(\xi_1)|<\gamma/\beta$. This is often called the delay-independent stability region for \eqref{eq:MGlin}. Using \eqref{eq:MGxi1} and differentiating the expression for $f$ in \eqref{eq:MG} we find that
\be \label{eq:fdxi1}
f'(\xi_1)=\left(\frac\gamma\beta\right)^2\Bigl(\frac\beta\gamma-n\Bigl(\frac\beta\gamma-1\Bigr)\Bigr),
\ee
and the delay-independent stability region for $\xi_1$
is given by
\be \label{eq:ndelind}
n<\frac{2\beta}{\beta-\gamma}.
\ee

If $|f'(\xi_1)|>\gamma/\beta$, equivalently if $n>2\beta/(\beta-\gamma)$ there is a sequence of Hopf bifurcations as the curves
\be \label{eq:C2n}
C_{2n}=\bigl\{(-\gamma,\beta f'(\xi))=(\phi\cot(\phi)/\tau,-\phi\csc(\phi)/\tau), \;
\phi\in(2n\pi,(2n+1)\pi)\bigr\},
\ee
for $n=0,1,2,\ldots$
are crossed in parameter space, with $\xi_1$ losing stability at $C_0$.

When $n>2\beta/(\beta-\gamma)$, equation \eqref{eq:fdxi1} implies that $f'(\xi_1)<0$ and then,
as shown for example in \cite{JWei_2007}, this sequence of Hopf bifurcations occurs
at
\be \label{eq:tauk}
\tau_k=\frac{1}{\sqrt{\beta^2(f'(\xi_1))^2-\gamma^2}}
\left[\arccos\Bigl(\frac{\gamma}{\beta f'(\xi_1)}\Bigr)+2k\pi\right],
\quad k=0,1,\ldots
\ee
with $\xi_1$ asymptotically stable for $\tau<\tau_0$ and unstable for $\tau>\tau_0$. Moreover, as $\tau$ is increased, a pair of complex conjugate characteristic values pass from the left to the right half complex plane at each $\tau_k$, and the dimension of the unstable manifold of $\xi_1$ increases by two.

In the rest of this work we will numerically study the periodic orbits emanating from the first Hopf bifurcation, and the further bifurcations from these orbits.

\section{Numerical Methods}
\label{sec:nummethods}

Our numerical computations are performed in \textsc{Matlab}~\cite{Matlab} using the DDE-BIFTOOL \cite{DDEBiftool15} numerical bifurcation analysis suite of routines, as well as numerical simulation of solutions using \verb+dde23+, the built-in \textsc{Matlab}~IVP solver for DDEs. We often use both approaches in combination on the same problem to reveal details that could not be obtained by applying only one of these approaches. We
describe essential details of our DDE-BIFTOOL and \verb+dde23+ implementations below.

\subsection{DDE-BIFTOOL}
\label{sec:ddebiftool}

DDE-BIFTOOL was originally developed by Engelborghs and collaborators \cite{Engelborghs_2002}, but its maintenance and development are now overseen by Jan Sieber and it is downloadable from SourceForge, with the manual published on the arXiv \cite{DDEBiftool15}. It is in active development with several versions available at the time of writing
(beta version; v3.1.1, alpha version; v3.2a, and a daily snapshot). We mainly used a daily snapshot from March 2021 for our computations, as the newer versions of the code have additional capabilities not available in older versions.

DDE-BIFTOOL
solves directly for steady states and finds periodic orbits by posing a periodic boundary value problem (BVP). This is done by rescaling time by the unknown period $T$, and imposing periodic boundary conditions on the resulting differential equation on $t\in[0,1]$. A phase condition is required for local uniqueness of the solution, and the resulting BVP is discretized using a collocation method. Exactly the same strategy has long been used to find periodic orbits in ODE dynamical systems. Equations with delays (or even with advances) can be handled exactly the same way, with the delays just creating additional entries, potentially far away from the diagonal, in the resulting numerical algebra problem.

To get started all that is needed is a reasonable first approximation to an orbit. Nearly all DDE-BIFTOOL users get this by starting from a Hopf bifurcation. However, it is also possible to use an apparently periodic orbit found through numerical simulation as an initial approximation (this is particularly useful when computing phase-locked orbits on tori \cite{Calleja_2017}).
Once an orbit is found, pseudo-arclength continuation is used to continue it as a parameter is varied. In this way unstable solutions can be computed just as easily as stable ones. DDE-BIFTOOL
has routines to efficiently compute the linear stability of steady states and periodic orbits, returning the important characteristic values and Floquet multipliers respectively. In this way it is possible to detect changes in stability and hence bifurcations. For one-parameter continuation these tasks are all automated within DDE-BIFTOOL, as is branch switching to enable computation of new branches of solutions, such as the period-doubled solutions emanating form a period-doubling bifurcation of a periodic orbit. Normal form calculations can be performed for local bifurcations of steady states
using the \verb+ddebiftool_nmfm+ extension to DDE-BIFTOOL.
DDE-BIFTOOL can also solve problems  state-dependent as well as constant discrete delays, but not distributed delays.

DDE-BIFTOOL comes equipped with a manual \cite{DDEBiftool15}, but it is best used as a reference document, once the user already has some idea of how to use the code. The package contains extensive demos, which provide a very good window into its capabilities and how to use the code. We also strongly recommend the tutorials of
Maikel Bosschaert \cite{ddebiftooltutorials}.

DDE-BIFTOOL can compute steady states, periodic orbits and heteroclinic and homoclinic connections.
While it cannot be used to directly compute more exotic invariant sets such as tori and chaotic attractors, it can still be very useful for finding such objects. It can locate torus bifurcations, and then for a stable torus it is possible to find the torus by simulation, as we did in \cite{Calleja_2017,Hum-DeM-Mag-Uph-12}. Since we know that the Mackey-Glass equation has a global attractor, it can be revealing to perform a numerical simulation for any parameter values for which the invariant objects identified by DDE-BIFTOOL are unstable. In this way we are able to find chaotic dynamics; see Section~\ref{sec:dde23} for more details.

DDE-BIFTOOL has long been able to perform two-parameter continuation of bifurcation points for steady states, and the recent \verb+ddebiftool_extra_psol+ extension adds the functionality for two-parameter continuation of fold, torus and period-doubling bifurcations of periodic orbits. This feature is essential to the current work, and used to calculate the bifurcations in Figure~\ref{fig:2paramcont}. Although it is easy to perform two-parameter continuation of the bifurcations, this feature is not as well developed as other aspects of DDE-BIFTOOL, and in particular the computation of stability and codimension-two bifurcations is not automated. Nevertheless, with some work this information can be recovered.

DDE-BIFTOOL performs two-parameter continuation of bifurcations by solving an augmented determining system which essentially consists of the equations defining the steady state or periodic orbit plus the additional conditions required for the bifurcation of interest. These systems are defined in
Section~10.1 of \cite{DDEBiftool15} for bifurcations of steady states only (see Section 10.3.2 of \cite{Kuznetsov_2004} for bifurcations of periodic orbits), but the details are not essential here. The key point is that since the equations defining the periodic orbit are embedded within the defining system, the solution of the defining system contains the periodic orbit embedded within it. The DDE-BIFTOOL command
\texttt{pfuncs.get\_comp(x,'solution')} can be used to extract periodic solution(s) from a point or branch of points \verb+x+. If an entire branch from a two-parameter continuation is extracted then the usual stability calculations can be applied. Since we are performing a two-parameter continuation of a bifurcation, there will always be multiple Floquet multipliers on the unit circle, with two at $1$ for a fold bifurcation, two at $\pm1$ for a period-doubling bifurcation, and a complex conjugate pair and one at $1$ for a torus bifurcation. If all the other Floquet multipliers are less than one in magnitude then the branch is on a stability boundary. If additional Floquet multipliers cross the unit circle a codimension-two bifurcation occurs.

\subsection{\textsc{Matlab} \texttt{dde23} solver}
\label{sec:dde23}

We make extensive use of the \textsc{Matlab}~DDE IVP solver \verb+dde23+ to simulate \eqref{eq:MG} with parameter values and initial functions $u_0\in C_+$ for which DDE-BIFTOOL computations have indicated the
possibility of interesting dynamics. The \verb+dde23+ solver is an adaptive time-step solver, based on an embedded pair of continuous Runge-Kutta methods (see \cite{BZ2003} or \cite{BGMZ2009}), with the time-step controlled by an estimate of the local error.  The acceptable local error is controlled by a combination of relative and absolute error with default tolerances of $\text{RelTol}=10^{-3}$ and $\text{AbsTol}=10^{-6}$.
While these default tolerances allow for fast solutions we prefer tighter tolerances, and when performing a single simulation of \eqref{eq:MG}, such as in Figures~\ref{fig:contnps}(e)-(f), \ref{fig:contn20r2p}(a), \ref{fig:cusp}(d) and \ref{fig:bistabpoca}(a)-(d), we set $\text{RelTol}=10^{-6}$ and $\text{AbsTol}=10^{-8}$.

In each of
Figures~\ref{fig:contnps}(e)-(f), \ref{fig:contn20r2p}(a), \ref{fig:cusp}(d) and \ref{fig:bistabpoca}(a)-(d)
DDE-BIFTOOL had detected an unstable periodic orbit, and we were interested in finding an $\omega$-limit set at the boundary of its unstable manifold. Consequently, unlike many users, we did not apply \verb+dde23+ with a constant function, but rather we used the unstable periodic orbit detected by DDE-BIFTOOL as the initial function to solve \eqref{eq:MG}. This is relatively easy to do using the DDE-BIFTOOL function \verb+psol_eva+ to evaluate the initial function, taking care to rescale time by the period since DDE-BIFTOOL rescales time so that the periodic orbit has period $1$, and wrapping multiple copies of the periodic orbit if the delay $\tau$ is larger than the period $T$.

\verb+dde23+ is also used to compute the orbit diagrams shown in Figures~\ref{fig:ODtau2}, \ref{fig:ODtau2details}, \ref{fig:subcritpd}(c)-(d). For each of these plots, equation~\eqref{eq:MG} is solved hundreds or thousands of times with just a small change in the value of the parameter $n$ between each simulation (except Figure~\ref{fig:subcritpd} for which $\tau$ is varied). Each problem was solved over $1500$ time units, with the transient behaviour over the first $1200$ time units discarded.
An \texttt{event} function
was used to detect local maxima over the last part of the solution, and these are plotted for each value of $n$ to obtain the orbit diagram.

For each orbit computation we use the solution from the adjacent parameter value as initial function for the
next computation. A little care is needed in doing this, since recent versions of
\textsc{Matlab} will accept a \verb+dde23+ solution structure as an initial function and then extend the solution; but this is not appropriate for these computations because the initial function and solution extension are computed with slightly different parameters. Instead we define a new solution structure for each simulation and use the
\textsc{Matlab} function \verb+deval+ to evaluate the initial function from the previous solution.
For the large orbit diagram in Figure~\ref{fig:ODtau2} which required 8000 \verb+dde23+ simulations we used
$\text{RelTol}=2\times10^{-5}$ and $\text{AbsTol}=10^{-7}$ to reduce the computation time, but for the other orbit diagrams we used the same tolerances as for the other simulations reported above.

By using an initial function which is the solution for a nearby parameter value, we are usually starting very close to the attractor for the new parameter value, resulting in much cleaner orbit diagrams than would be achieved using a constant history function.
The constant initial function
\be \label{eq:if2}
u(t)=2, \quad\text{for}\;t\in[-\tau,0]
\ee
was used for the orbit plot in
Figure~\ref{fig:subcritpd}(d), this time solving over $3000$ time units, and discarding the transient dynamics over the first $2700$ time units.
Despite this much longer transient, comparing the periodic window close to $\tau=2.74$ in
Figure~\ref{fig:subcritpd}(c) and (d) it is clear that the shorter transient combined with using the solution from the adjacent parameter value results in a cleaner diagram.

\subsection{Lyapunov Exponents and Dimension}
\label{sec:LyapDim}

Kaplan and Yorke~\cite{KaplanYorke1979}
defined an attractor dimension, now known as the Lyapunov dimension, to be
\begin{equation} \label{eq:lyapdim}
d=k+\frac{1}{|\lambda_{k+1}|}\sum_{j=1}^k\lambda_j
\end{equation}
where the Lyapunov exponents are ordered so $\lambda_1\geq \lambda_2\geq\ldots$, and $k$ is the largest integer so that the sum of the first $k$ exponents is non-negative, thus necessarily $\lambda_{k+1}<0$, and $d\in[k,k+1)$. One generally accepted indication of chaos is the presence of a positive Lyapunov exponent.
There is always a zero Lyapunov exponent (corresponding to translation along the trajectory) which is not calculated exactly in numerical algorithms, so
detecting chaos by testing whether the largest Lyapunov exponent is positive is not a numerically stable process. Instead, we perform the equivalent test of checking the sum of the two largest Lyapunov exponents. If this is positive the solution is chaotic, and in that case it follows from \eqref{eq:lyapdim} that
the Lyapunov dimension will be larger than two.

We will investigate the existence of chaotic solutions for~\eqref{eq:MG} by
numerically computing the Lyapunov exponents using the method
of Breda and Van Vleck~\cite{Breda_2014}. We do this in two ways, either by computing just a few Lyapunov exponents (as in Figure~\ref{fig:2paramcont}) to determine whether the dynamics is chaotic or not, or by computing many Lyapunov exponents in order to determine the Lyapunov dimension.

For Figure~\ref{fig:2paramcont}, the Mackey-Glass equation was first solved using \texttt{dde23} and our usual tight tolerances over 5000 time units. Then the method of Breda and Van Vleck \cite{Breda_2014} was used to estimate the first 8 Lyapunov exponents. To obtain the chaotic region, the computation was repeated
for 78000 different pairs of values for the parameters $n$ and $\tau$
on an equally spaced mesh $(n,\tau) \in [7,20] \times [1,4]$ with steps of 0.05 in $n$ and $0.01$ in $\tau$.

This task was very expensive computationally, so we used the Calcul Qu\'ebec supercomputer Guillimin
to accelerate the computations. More precisely, we partitioned the $(n,\tau)$ grid between the different computing threads of the supercomputer, so that each thread solved the Mackey-Glass equation using \texttt{dde23}  for its own set of $(n,\tau)$ pairs. To facilitate parallel computations, we solved \eqref{eq:MG} with the constant initial  function \eqref{eq:if2}.

In the region of parameter space near $(n,\tau)=(13.4,1.44)$ there is bistability between a periodic orbit and a chaotic attractor,
and the constant function \eqref{eq:if2} was found not to be in the basin of attraction of the chaotic attractor.
In that case a different strategy was needed, and
we started with a parameter pair $(n,\tau)$ leading to chaotic dynamics
and sequentially computed solutions using \verb+dde23+ along constant $\tau$ lines.
For each computation the value of $n$ was increased by $0.01$, with the previous solution
used as the initial function for the next computation.
An initial function on an attractor will be in the basin of attraction of the perturbed attractor for a nearby parameter set, provided the steps in parameter space are small with respect to the size of the basin of attraction. This works well in practice as can be seen in Figure~\ref{fig:2paramcont} in Section~\ref{sec:twoparcont}.

For the Lyapunov Dimension computations shown in Section~\ref{sec:LyapDim} we needed to compute more Lyapunov exponents (25 for Figure~\ref{fig:dimensionA}, 15 for Figure~\ref{fig:dimensionB}, and between 80 and 120 exponents for Figure~\ref{fig:dimensionC}). We also did not have access to the supercomputer so the computations for Figure~\ref{fig:dimensionA} were performed on a coarser mesh of $26300$ points in parameter space, with the initial condition \eqref{eq:if2}, shorter time interval of $2000+500\tau$ time units, and slightly relaxed tolerances of
$\text{RelTol}=10^{-7}$ and $\text{AbsTol}=10^{-5}$. The computation took 18 days to complete on a Lenovo X270!
For Figure~\ref{fig:dimensionB} we used the same tolerances, on about 6000 points in parameter space, and a time interval of $1500$ time units, using the solution from adjacent parameter values as the initial function.
The computations for Figure~\ref{fig:dimensionC}
were less intensive, since only $\tau$ is varied, and were performed over $10000$ time units with our usual strict tolerances.

%
%
%

\section{Dynamics and Bifurcations}
\label{sec:dynamics}

Rescaling time and space it is possible to eliminate two parameters from \eqref{eq:MG}, consequently we will fix $\gamma=\theta=1$ throughout.  We also require $\beta>\gamma$ for interesting dynamics, and it is  convenient to set $\beta=2$, since then the non-zero steady state is always $\xi_1=1$, independent of the values of $n$ and $\tau$. Glass and Mackey already made the same parameter choices in \cite{Glass_Mackey_1979}.
Now the Mackey-Glass equation \eqref{eq:MG} becomes
\be \label{eq:MG12}
\udot(t) = - u(t)+ 2f(u(t-\tau)), \qquad f(\xi) = \frac{\xi}{1+\xi^n}.
\ee
From \eqref{eq:ndelind} the steady state $\xi_1=1$ is asymptotically stable if $n<4$ for arbitrary values of the delay $\tau$. However for $n>4$, the steady state undergoes a sequence of Hopf bifurcations when $\tau=\tau_k$, defined by \eqref{eq:tauk}. Hence we will consider
the remaining parameters $n>4$ and $\tau>0$ as continuation and bifurcation parameters, and study the ensuing dynamics. We note that some authors, including Farmer \cite{Farmer82}, prefer $\gamma=0.1$ and $\beta=0.2$; but the resulting system is equivalent to \eqref{eq:MG12} on rescaling $t$ and $\tau$ by a factor of $10$.

\subsection{One-Parameter Continuation and Period-Doubling Cascade}
\label{sec:contn}

We begin our numerical investigation by considering \eqref{eq:MG12} with $\tau=2$ fixed,
but for a sequence of increasing values of $n$. This scenario was originally considered by Glass and Mackey in \cite{Glass_Mackey_1979}, who numerically simulated the dynamical system to find asymptotically stable periodic orbits. In contrast we use DDE-BIFTOOL as explained in Section~\ref{sec:ddebiftool} to perform a one-parameter
continuation of the dynamical system. This
allows us to compute and continue periodic orbits, whether they are stable or unstable, as well as to detect bifurcations and to switch branches at bifurcation points.

\begin{figure}[t!]
   \includegraphics[width=\textwidth]{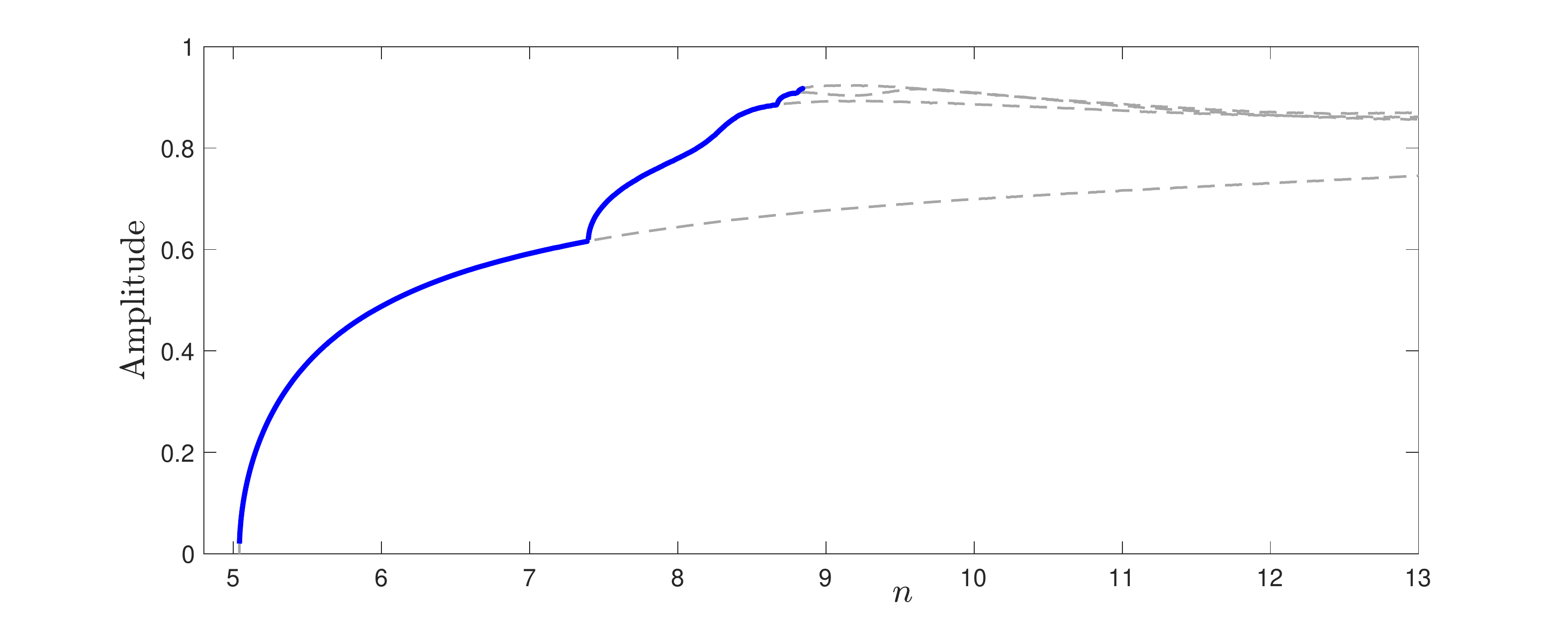}
  \caption{Bifurcation Diagram for the Mackey-Glass Equation \eqref{eq:MG12} with $\tau=2$ as $n$ is increased showing the 4 branches of periodic orbits computed using DDE-BIFTOOL, with solid/dashed lines indicating respectively stable/unstable orbits.}
  \label{fig:contnamp}
\end{figure}

\begin{figure}[t!]
\includegraphics[width=0.49\textwidth]{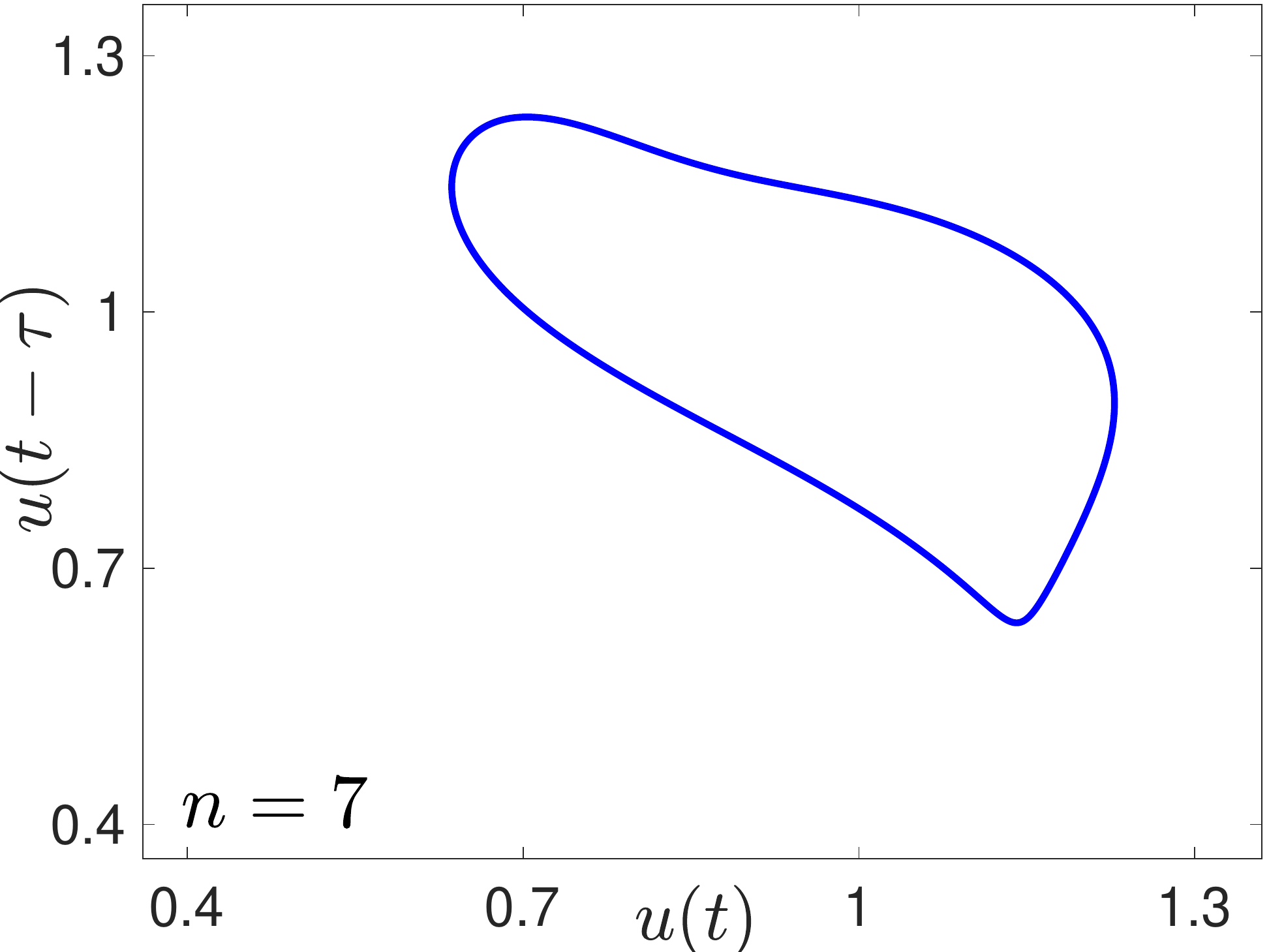}\hspace*{0.02\textwidth}\includegraphics[width=0.49\textwidth]{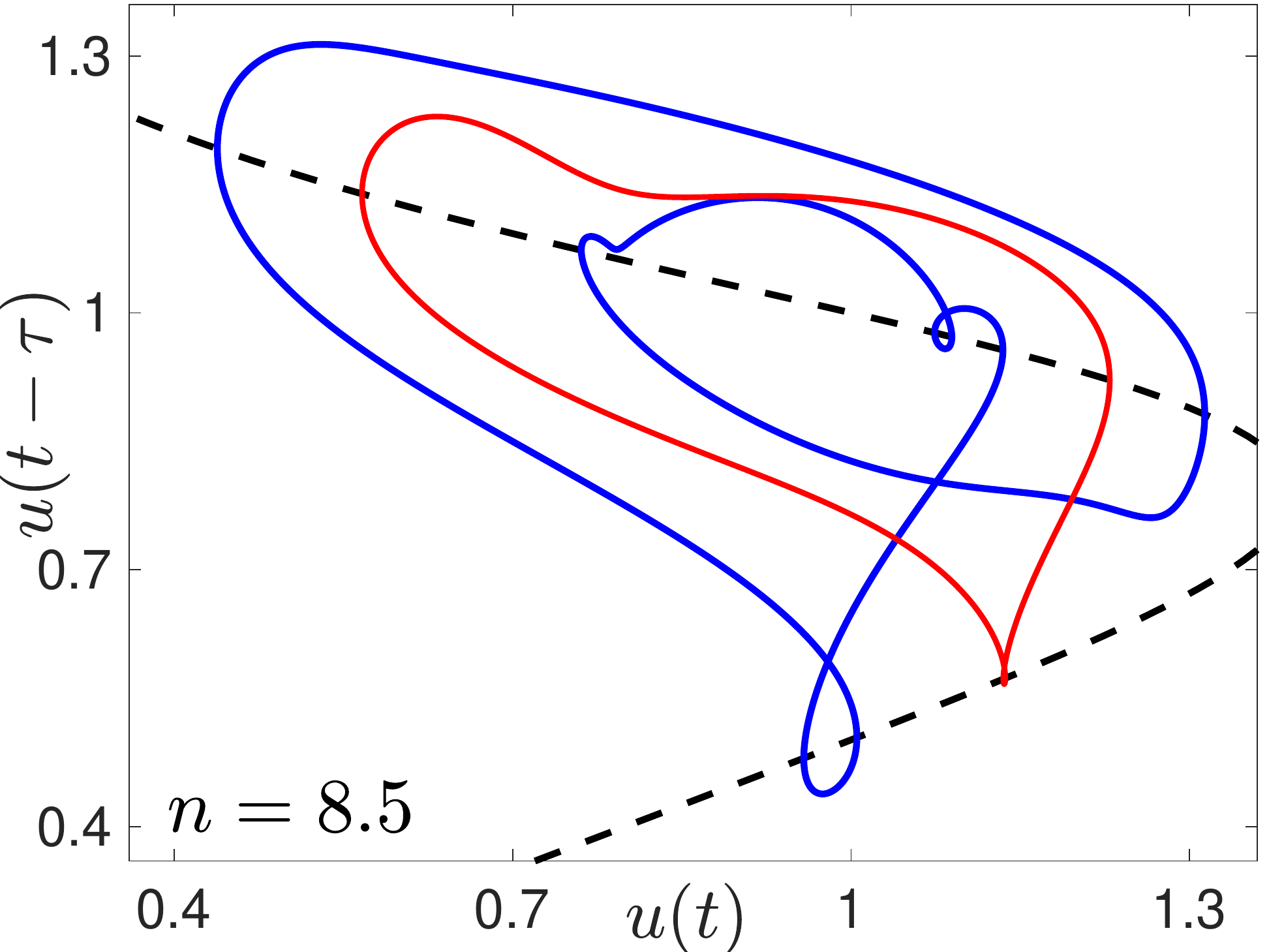}
\put(-335,32){(a)}
\put(-152,32){(b)}	

\vspace*{2ex}		\includegraphics[width=0.49\textwidth]{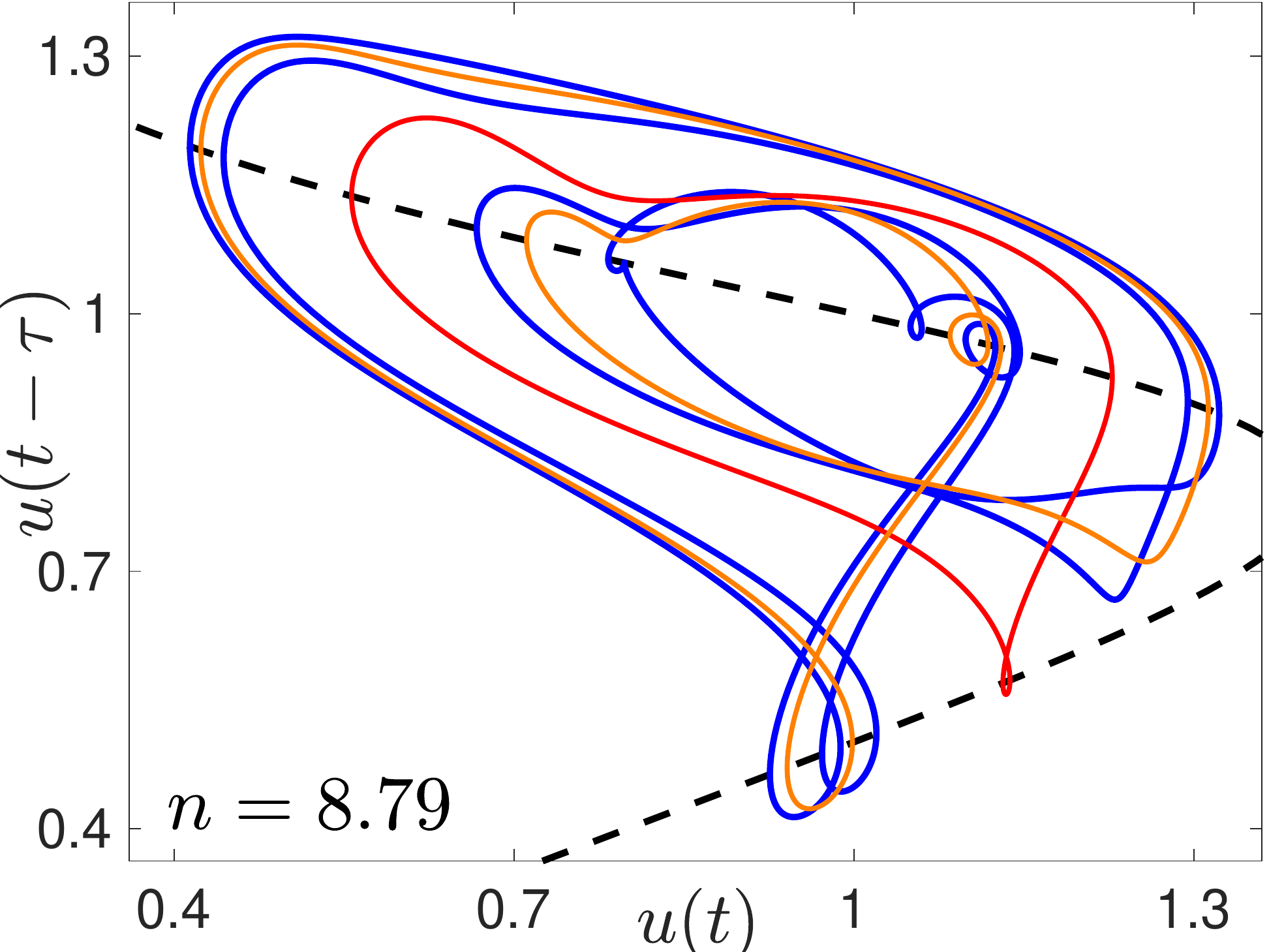}\hspace*{0.02\textwidth}\includegraphics[width=0.49\textwidth]{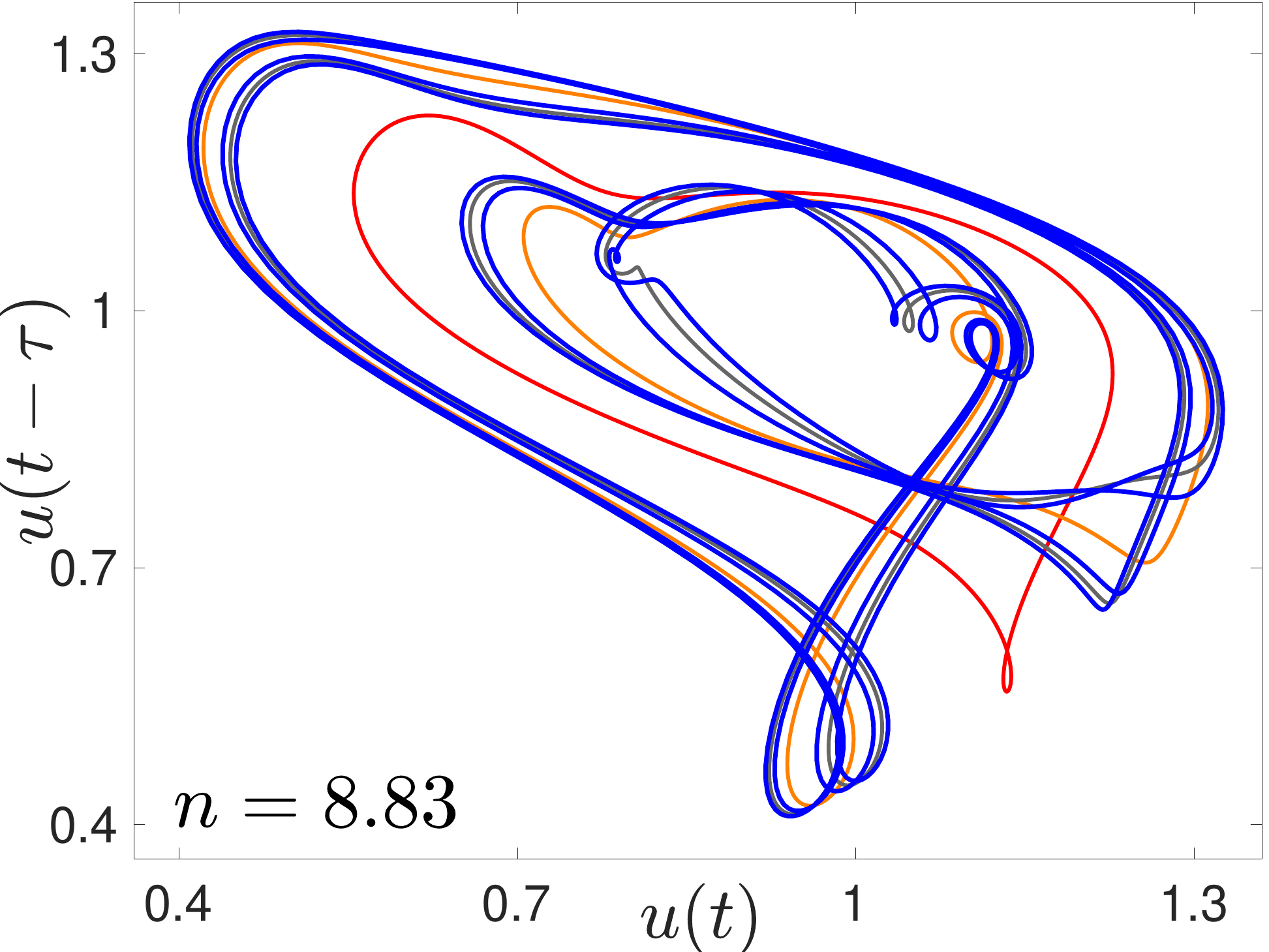}
\put(-335,32){(c)}
\put(-152,32){(d)}	

\vspace*{2ex}	
\includegraphics[width=0.49\textwidth]{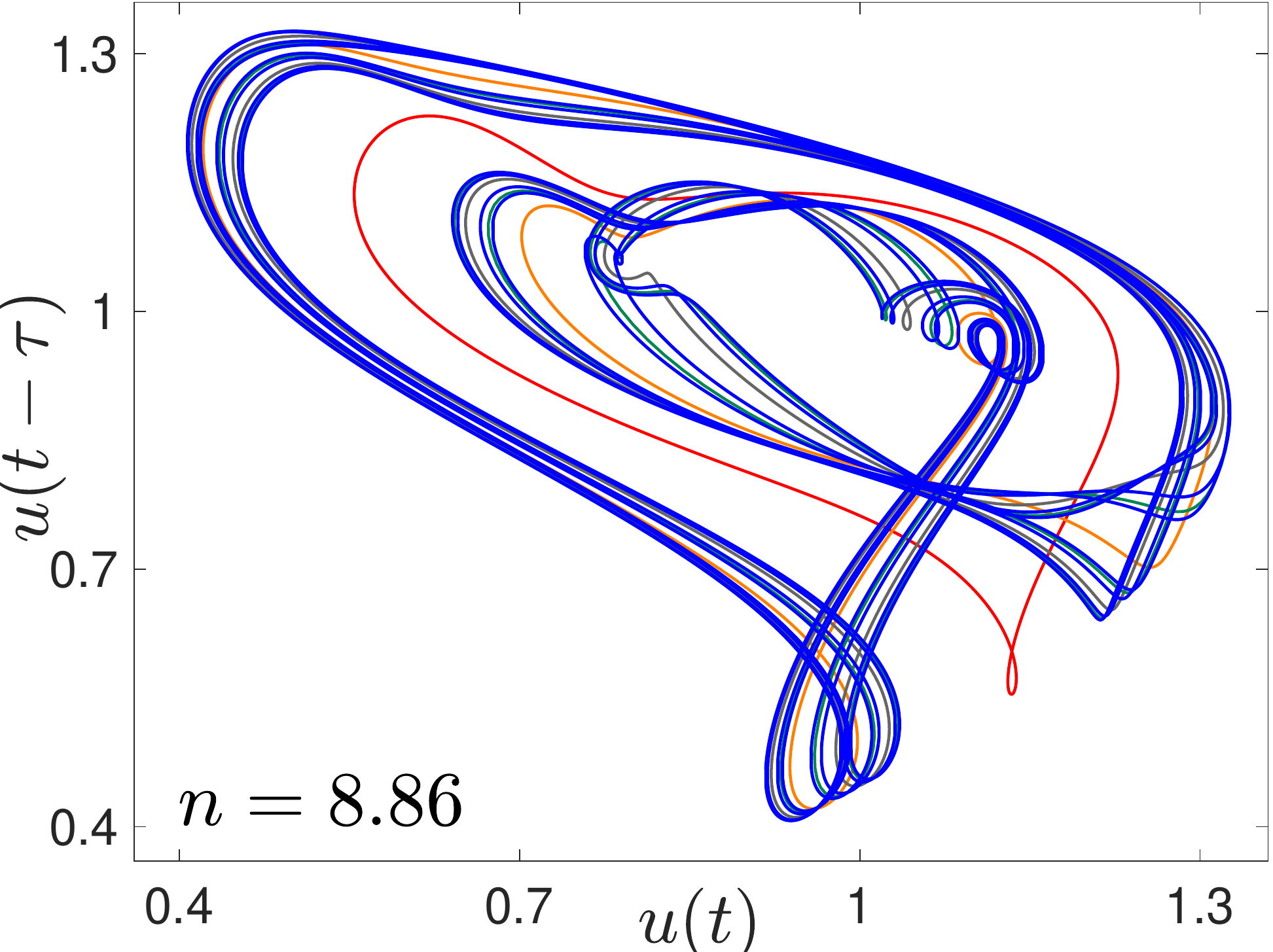}\hspace*{0.02\textwidth}\includegraphics[width=0.49\textwidth]{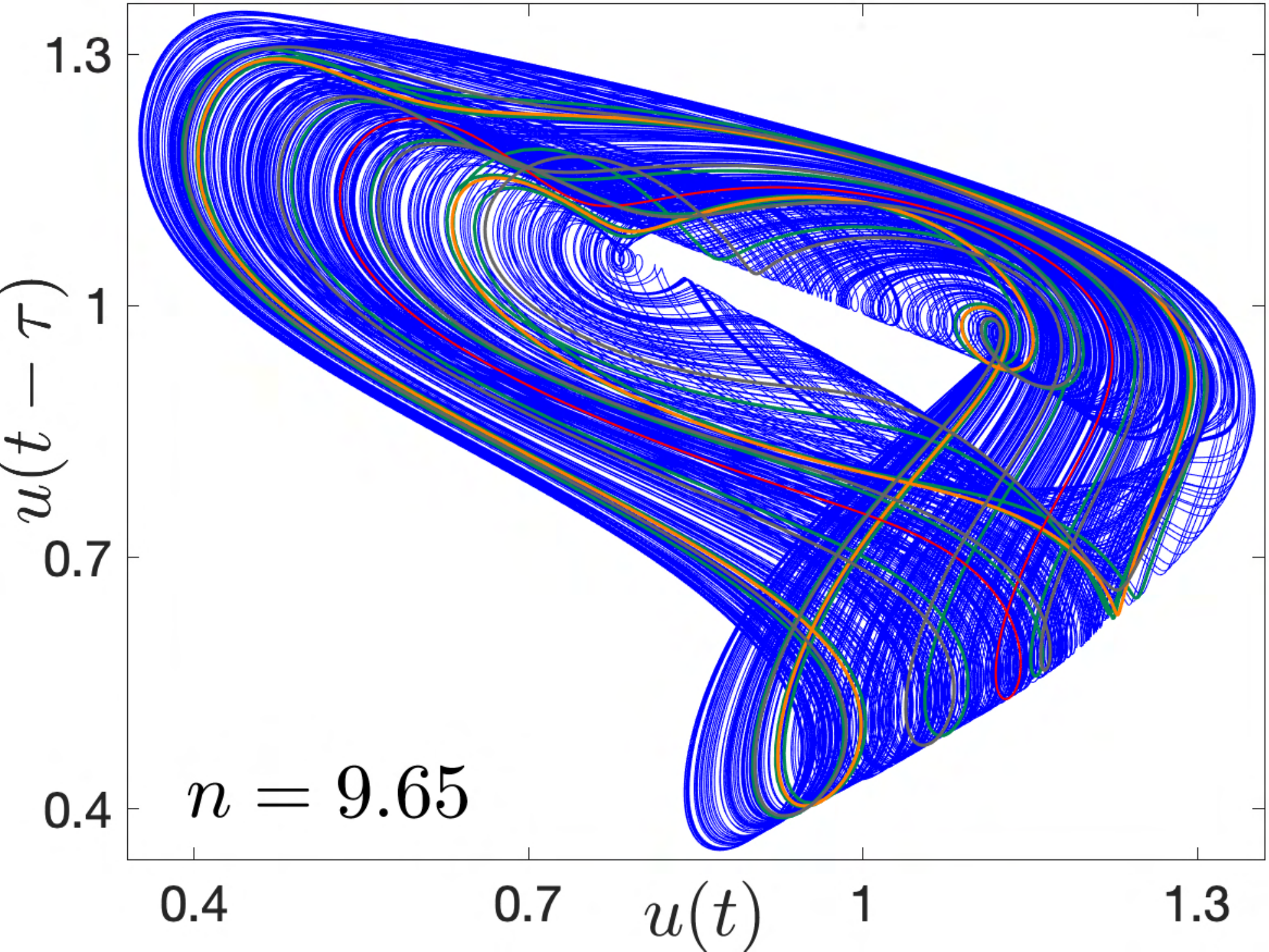}
\put(-335,32){(e)}
\put(-152,32){(f)}	

\caption{Phase portraits for the Mackey-Glass Equation \eqref{eq:MG12} with $\tau=2$. Stable orbits are coloured blue, while all other colours indicate
unstable orbits. For (a)-(d) all orbits computed using DDE-BIFTOOL. For (e) and (f) all DDE-BIFTOOL computed periodic orbits are unstable, and the stable object shown is the result of simulation using \texttt{dde23}. On panels (b) and (c) the $\dot{u}=0$ nullcline is also shown.}
  \label{fig:contnps}
\end{figure}

Figure~\ref{fig:contnamp} shows the branches of periodic orbits that DDE-BIFTOOL detects, and these orbits are illustrated in Figure~\ref{fig:contnps} for certain values of $n$. The steady state $\xi_1=1$ loses stability in a Hopf bifurcation at $n=5.0396$ (consistent with \eqref{eq:tauk} which gives $\tau_0=2$ when $n=5.0396$).
The periodic orbit created in the bifurcation is initially stable but loses stability in a period-doubling bifurcation at $n=7.3951$.
Further period-doubling bifurcations occur at $n=8.6731$, $8.8122$, $8.8516$.
At the first three period-doubling bifurcations we are able to switch to and continue the new stable period-doubled branch of periodic orbits.
After losing stability each periodic orbit continues to exist at least until $n=20$ as an unstable periodic orbit.

Figure~\ref{fig:contnps} shows phase portraits of the DDE-BIFTOOL computed orbits for different values of $n$. Since the phase space $C_+$ is infinite-dimensional, a finite-dimensional projection of phase space is required. We use the simple projection
\be \label{eq:P2}
\cP_2:u_t\mapsto(u_t(0),u_t(-\tau)),
\ee
or equivalently plot $u(t-\tau)$ against $u(t)$; which was first used by Glass and Mackey in \cite{Glass_Mackey_1979} and has since become the ubiquitous two-dimensional phase space projection for DDEs. The first 4 panels of Figure~\ref{fig:contnps} show the DDE-BIFTOOL computed phase portraits through the first three period-doubling bifurcations, with the stable orbit coloured blue, and the unstable orbits also shown.
The effect of the period-doubling is clearly seen in the phase portrait as each new period-doubled orbit is seen to wind twice around the previous orbit. The periodic orbit can also develop more complexity by gaining additional local maxima in its profile through inflection points in the orbit, without a bifurcation occurring. This happens when the phase portrait just touches the nullcline $\udot(t)=0$.
Panels (b) and (c) of Figure~\ref{fig:contnps} show this nullcline, which from \eqref{eq:MG12} is defined by $u(t)=2f(u(t-\tau)$. Since $\dot{u}(t)=0$ on the nullcline, orbits cross this curve vertically in the phase portrait, as can be seen in the figure. This phenomenon is already well known; see \cite{Junges_2012}
for a good explanation.

For $n=8.86$, just after the last period-doubling bifurcation that DDE-BIFTOOL detects,
all the DDE-BIFTOOL periodic orbits are unstable, so we use a \verb+dde23+ simulation, starting from an initial function corresponding to the largest period unstable orbit, and after discarding the transient dynamics
find the further period-doubled orbit shown in Figure~\ref{fig:contnps}(e). It is believed that there is then a period-doubling cascade to chaos. The final panel shows an apparently chaotic attractor
for $n=8.86$
again obtained by \verb+dde23+ simulation, with the unstable periodic orbits obtained from DDE-BIFTOOL embedded in the attractor. The Lyapunov dimension is computed to be 2.27, which numerically verifies the chaotic dynamics.

If we could follow the period-doubling cascade further in DDE-BIFTOOL it would be interesting
to see whether a
Feigenbaum constant emerges (see \cite{Feigenbaum78}). However, that is much more computationally challenging for periodic orbits of differential equations as opposed to periodic points of mappings. To maintain the same computational accuracy it is necessary to double the number of collocation mesh points at each period-doubling and numerical errors and computational capacity quickly overwhelm the computation.

\begin{figure} [t!]
	\centering
	\includegraphics[width=\textwidth]{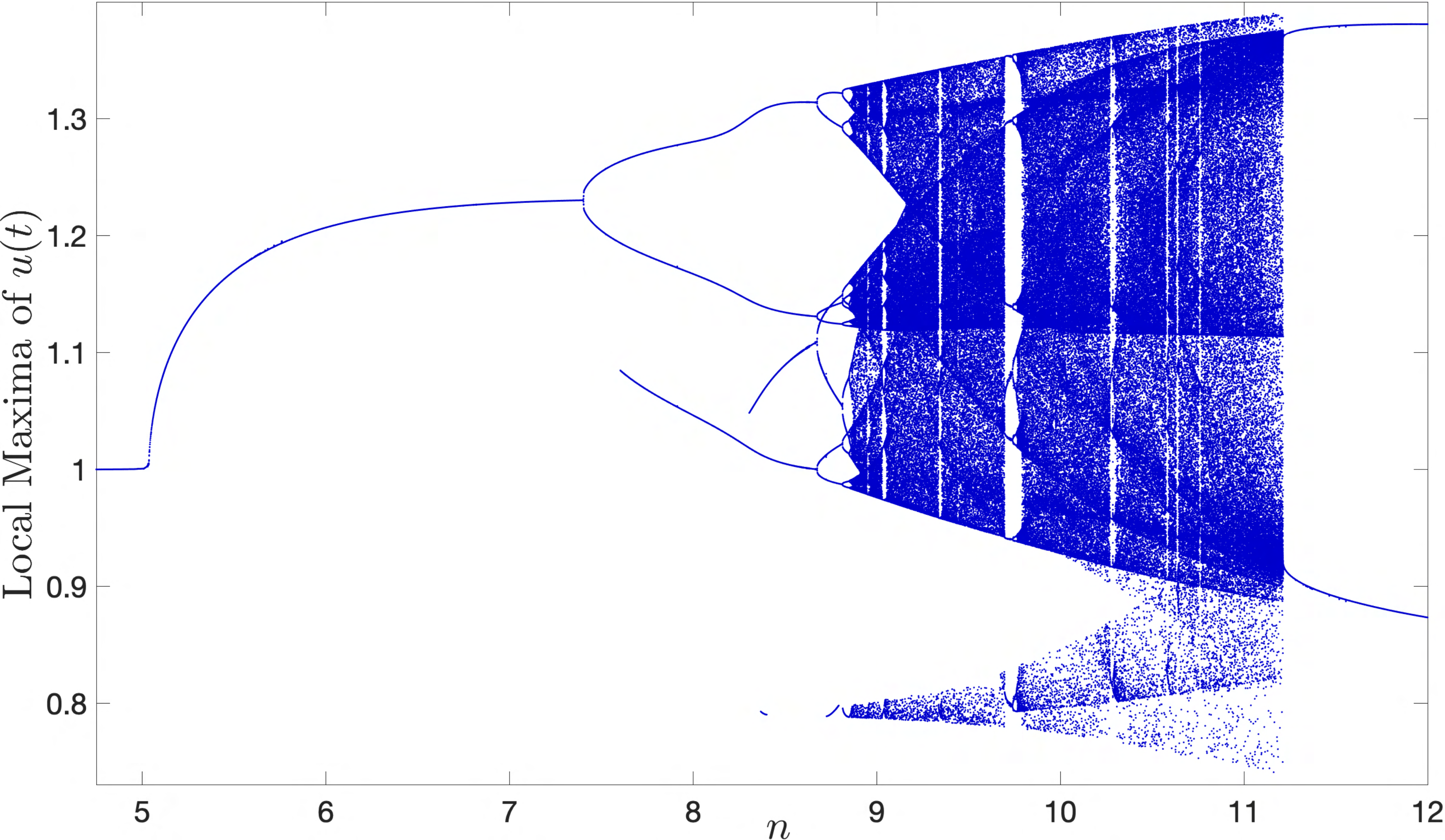}
	\caption{Orbit Diagram for the Mackey-Glass Equation \eqref{eq:MG12} with $\tau=2$, with $n$ varied between $4$ and $12$ in increments of $10^{-3}$ between each simulation. See Section~\ref{sec:dde23} for further details on the numerical computation.}
\label{fig:ODtau2}
\end{figure}

\begin{figure} [t!] \includegraphics[width=0.49\textwidth]{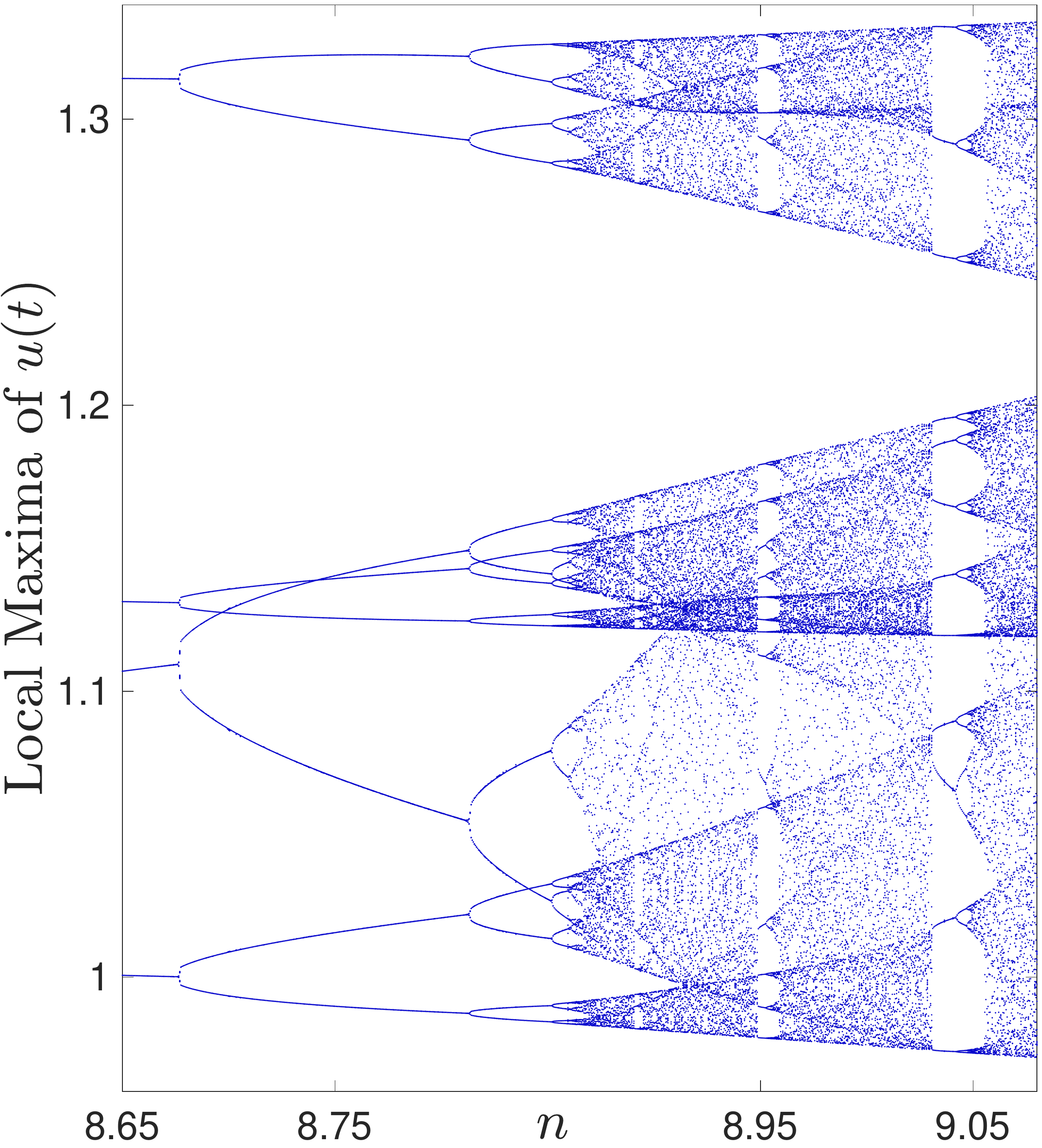}\hspace*{0.02\textwidth}\includegraphics[width=0.49\textwidth]{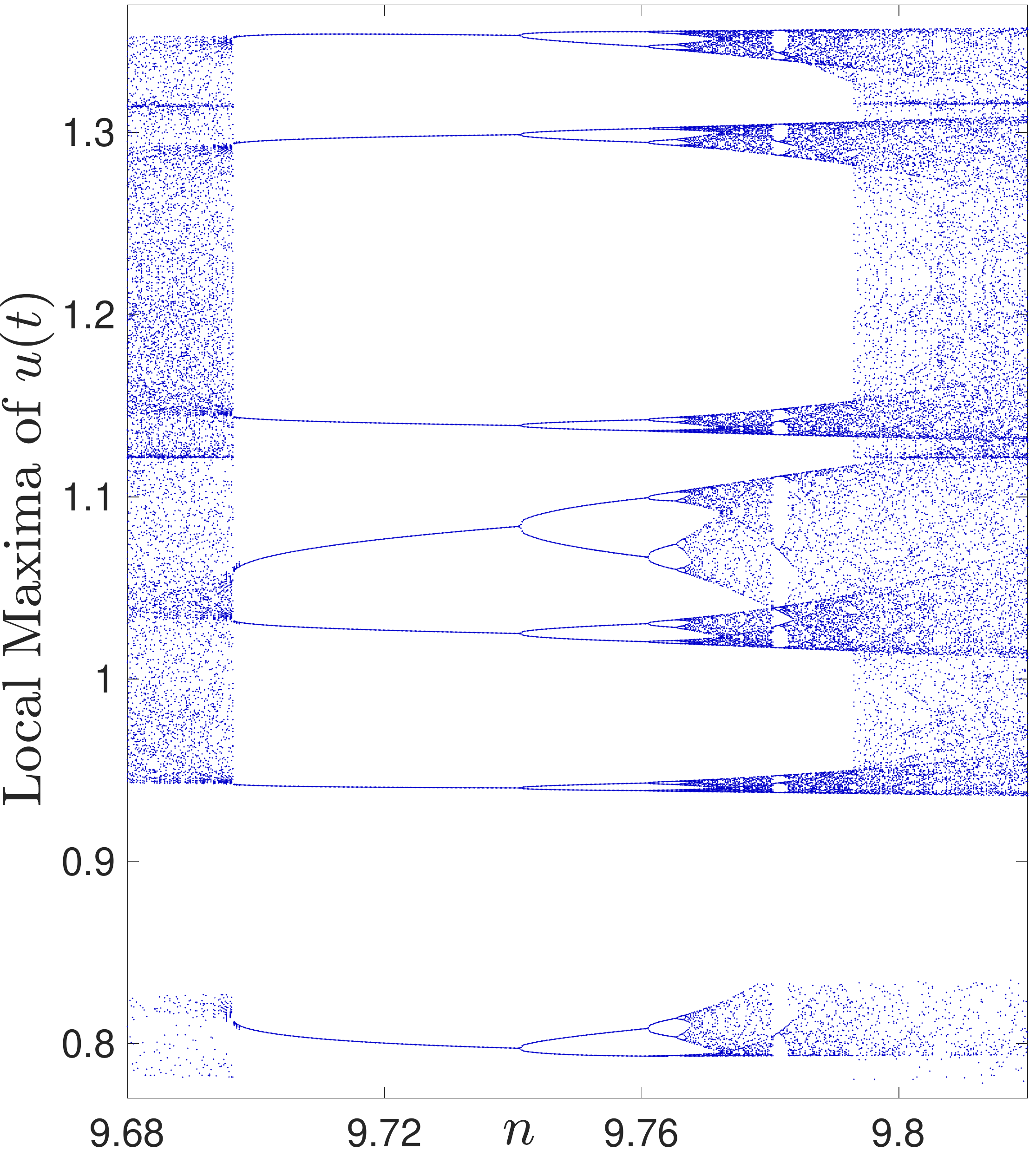}
\put(-100,145){(b)}
\put(-335,145){(a)}
\caption{Orbit Diagrams for the Mackey-Glass Equation \eqref{eq:MG12} with $\tau=2$.
(a) A close up of the dynamics during the period-doubling cascade for $n\approx8.85$,
with $n$ increased in increments of $5\times10^{-4}$ between each computation.
(b) The periodic window near $n\approx9.7$, obtained with increments of $n$ of $2\times10^{-4}$.}
\label{fig:ODtau2details}
\end{figure}

The period-doubling cascade can be investigated numerically using an orbit diagram. Orbit diagrams are usually presented for one-dimensional maps; see \cite{Strogatz} for some good examples revealing period-doubling cascades and chaotic dynamics. Orbit diagrams can also be produced for differential equations by using a Poincar\'e section to reduce them to a map, then projecting the Poincar\'e section into one dimension. For ODEs it is common to determine and plot the local maxima along a solution trajectory
(so the values of $u(t)$ when $\dot{u}(t)=0$ and $\ddot{u}(t)<0$) and this works just as well for DDEs.
We adopt this approach for the Mackey-Glass equation \eqref{eq:MG12}
(see Section~\ref{sec:dde23} for numerical details).
The result is shown in Figure~\ref{fig:ODtau2}.
A few branches on the graph appear to be born spontaneously, but this is not due to bifurcations, and occurs
when the periodic orbit develops additional local maxima through inflection points in the orbit when the solution touches the $\udot(t)=0$ nullcline.

The period-doubling cascade as $n$ approaches $8.85$ happens so fast that it is difficult to distinguish clearly in Figure~\ref{fig:ODtau2}, but is clearer in the zoomed view in
Figure~\ref{fig:ODtau2details}(a) with chaotic behaviour apparent for $n>8.87$.
Several windows of periodic behaviour are visible in Figures~\ref{fig:ODtau2}
and~\ref{fig:ODtau2details}(a). These are very reminiscent of the periodic windows seen in the orbit diagram for the one-dimensional logistic map (see for example \cite{Strogatz}).
An enlargement of the widest such window for $n\approx9.7$ is shown in
Figure~\ref{fig:ODtau2details}(b); Glass and Mackey showed two solutions from within this window in
\cite{Glass_Mackey_1979}.

With $\tau=2$ and for all $n$ sufficiently large ($n$ slightly larger than $11$ is sufficient) the chaotic dynamics vanish, and once again a stable periodic solution is obtained. This is illustrated for $n=20$ in
Figure~\ref{fig:contn20r2p}(a). All the DDE-BIFTOOL computed orbits seen in Figure~\ref{fig:contnamp} can be continued to $n=20$ and the resulting periodic orbits are shown in Figure~\ref{fig:contn20r2p}(a), however these orbits are all unstable. The stable periodic orbit seen in Figure~\ref{fig:contn20r2p}(a) is found by simulating with \verb+dde23+. This leads to several interesting questions. One of these is to consider the limiting behaviour as $n\to\infty$. This problem was recently tackled in \cite{Krisztin_StableOrbitsMG_JDE2021} where stable periodic orbits were constructed in this limiting case and proved to persist for large $n$. In the current work we ask two simpler questions. Where does the stable orbit seen in Figure~\ref{fig:contn20r2p}(a) originate, and what is the mechanism leading to chaos if instead of increasing $n$, we start from large $n$ and decrease $n$? The right side of Figure~\ref{fig:ODtau2} suggests the chaotic dynamics might be born spontaneously in an attractor crisis. To investigate these bifurcations and answer these questions properly will require two-parameter continuation of the Mackey-Glass equation \eqref{eq:MG12}.

\subsection{Two-Parameter Continuation}
\label{sec:twoparcont}

\begin{figure}[t!]	\includegraphics[width=0.49\textwidth]{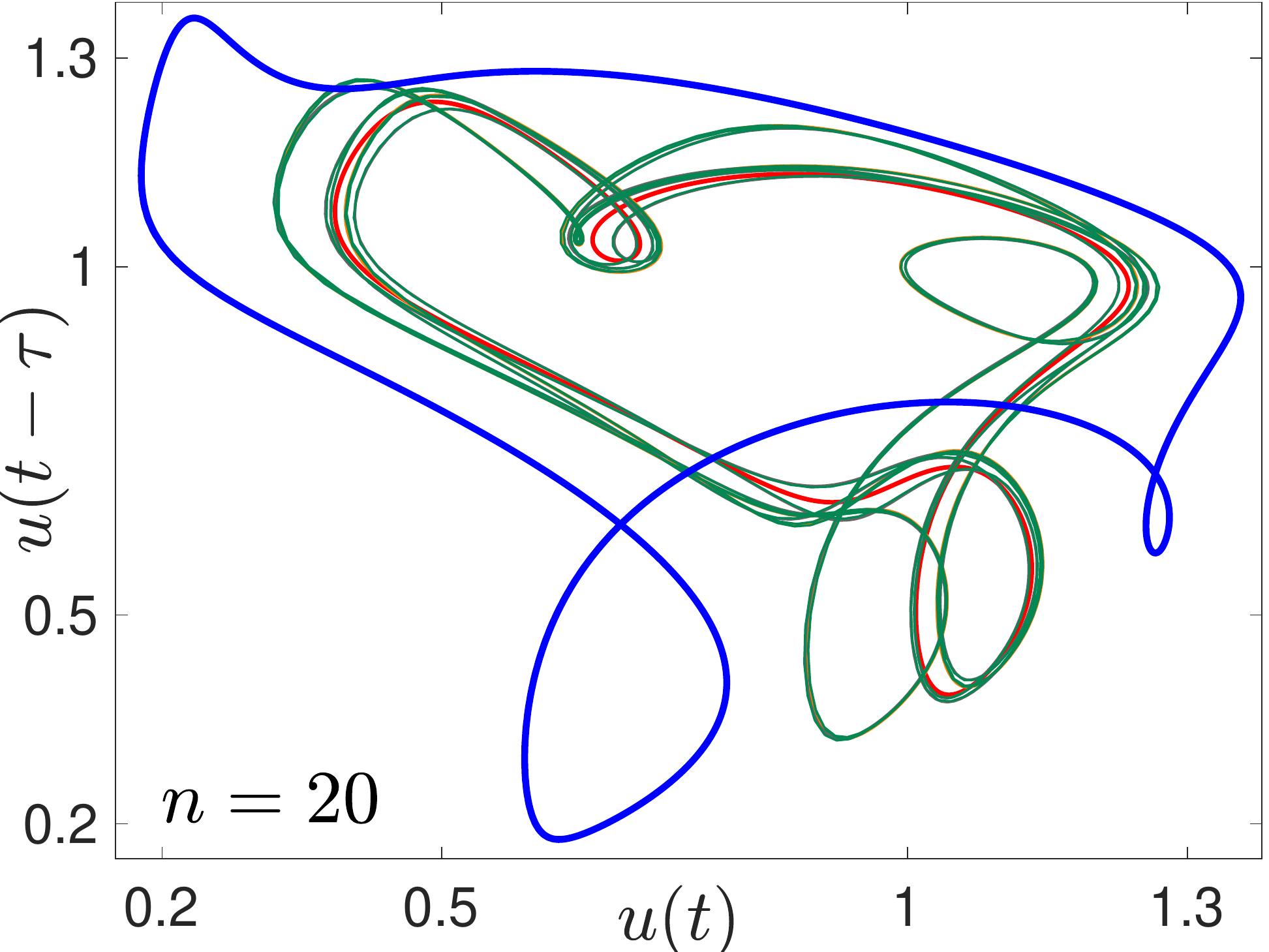}\hspace*{0.02\textwidth}\includegraphics[width=0.49\textwidth]{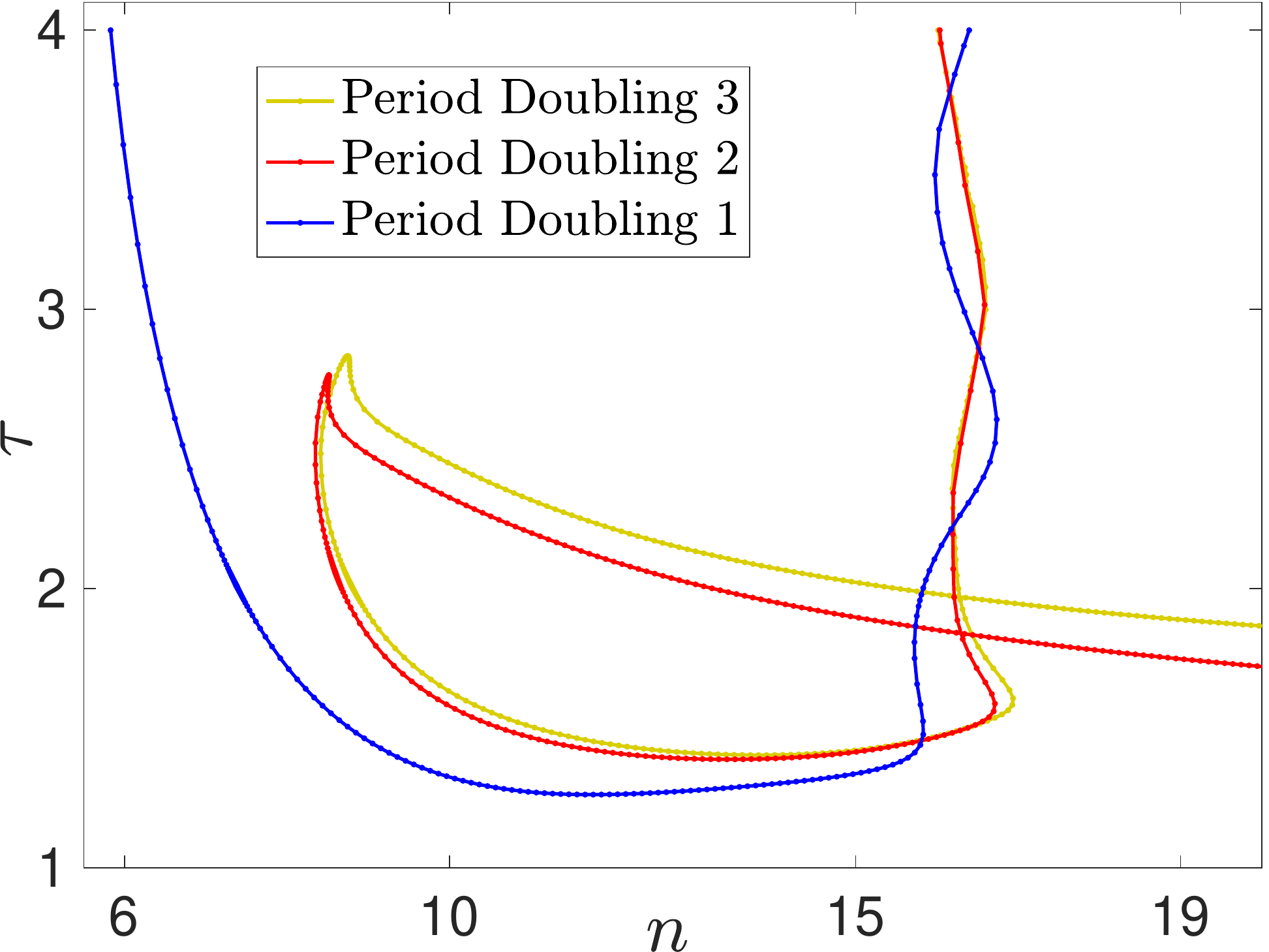}
\put(-338,35){(a)}
\put(-160,25){(b)}	
\caption{(a) Phase portrait of the Mackey-Glass Equation \eqref{eq:MG12} with $\tau=2$ and $n=20$.
The stable periodic orbit (shown in blue) was computed using \texttt{dde23}, while all the
DDE-BIFTOOL computed periodic orbits (all other colours) are unstable.
		(b) Two-parameter continuation of the first three period-doubling bifurcations in $n$ and $\tau$.}
	\label{fig:contn20r2p}
\end{figure}

DDE-BIFTOOL has the capability to perform two-parameter continuation of bifurcations of steady states and periodic orbits, and we use this to study the bifurcations of the Mackey-Glass equation \eqref{eq:MG12} more systematically. Figure~\ref{fig:contn20r2p}(b) shows a two-parameter continuation of the first three period-doubling bifurcations, constructed by starting from the period-doubling bifurcations illustrated in
Figure~\ref{fig:contnamp}. The Mackey-Glass demo in version 3.2a of DDE-BIFTOOL produces similar curves.

The left-hand side of Figure~\ref{fig:contn20r2p}(b) looks largely as expected; with $\tau=2$ fixed and increasing $n$ from zero, or with $n=12$ fixed and increasing $\tau$ from zero, we observe a sequence of three period-doubling bifurcations with the third one following very soon after the second.
As already observed in Section~\ref{sec:contn} for $\tau=2$ these are just the first three period-doublings on an apparent period-doubling cascade to chaos.

The right-hand side of Figure~\ref{fig:contn20r2p}(b) is troubling, as two of the branches snake towards large $\tau$ values with $n$ about $16$. Consequently if $n\geq18$ is fixed and $\tau$ is increased from zero, the first bifurcation branch encountered in the bifurcation diagram is the second period doubling, but obviously the second period doubling cannot occur before the first one. If a one-parameter continuation of the periodic orbit is performed from the Hopf bifurcation with $n=20$ fixed and $\tau$ increased up to $2$, the periodic orbit remains stable with no bifurcation, even as the Period Doubling 2 and 3 curves are crossed. At $(n,\tau)=(20,2)$ the stable periodic orbit shown in blue in Figure~\ref{fig:contn20r2p}(a) is obtained.

So it seems that two-parameter continuation of bifurcations can be very misleading if not performed carefully.
In computing a bifurcation diagram for the Mackey-Glass equation \eqref{eq:MG12} we are interested in understanding the bifurcations through which stable periodic orbits are destroyed or become unstable.
The problem essentially is that in a two-parameter continuation we should expect to find codimension-two bifurcations. If we follow in two parameters a bifurcation curve at which a periodic orbit loses stability and we step over such a codimension-two bifurcation point, then even though the bifurcation may continue to exist, it may no longer delineate a loss of stability. DDE-BIFTOOL does not automatically detect codimension-two bifurcations, so some work is required to resolve this issue.

\afterpage{%
\clearpage
\begin{landscape}
\begin{figure}[!htb]
\makebox[\textwidth][c]{\includegraphics[width=1.66\textwidth]{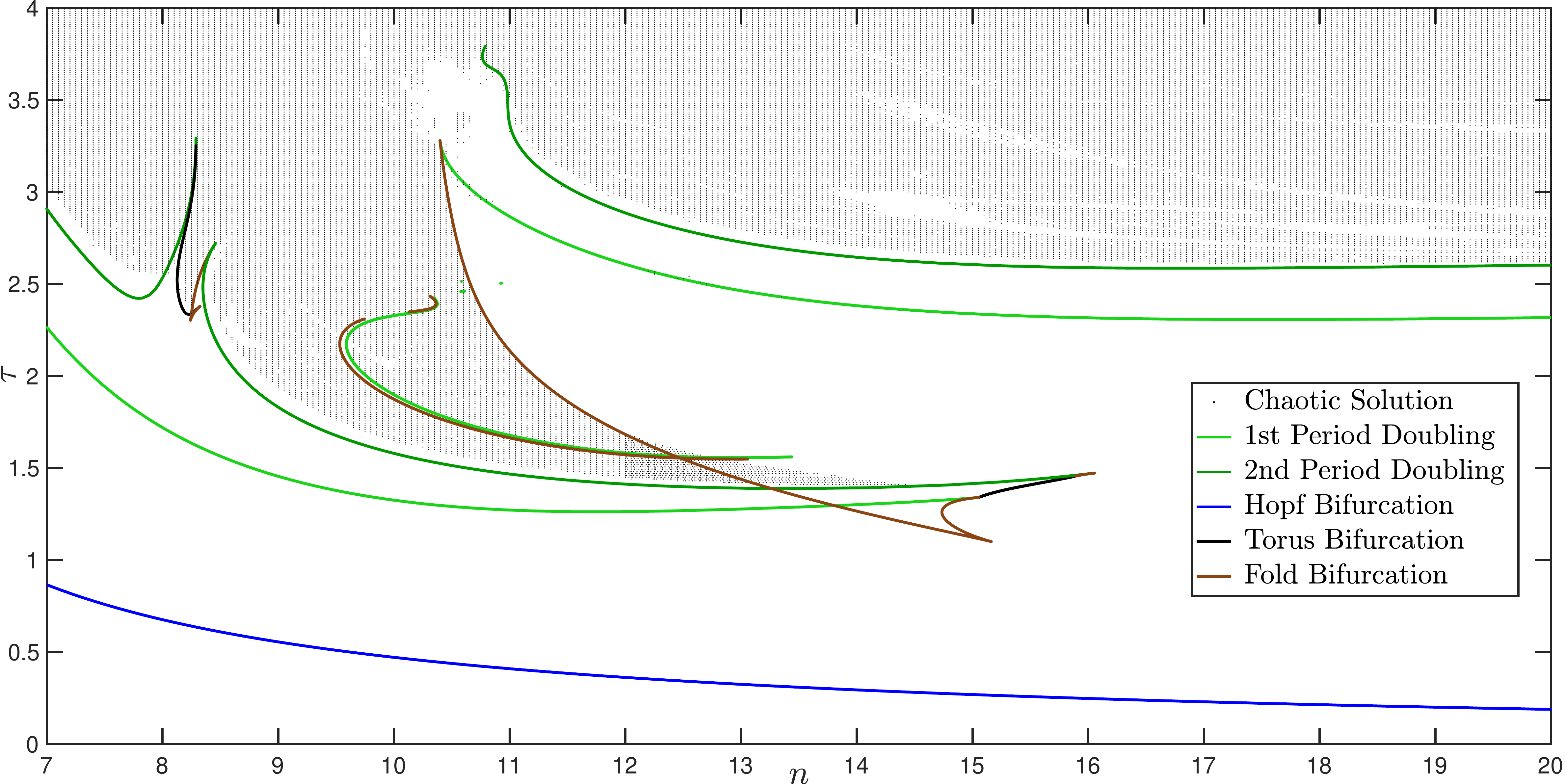}}
\put(-100,59){Cusp}
\put(-78,90){Fold-flip}
\put(-450,100){Bistability}
\put(-150,150){Bistability}
\put(-400,275){Chaos}
\put(-60,275){Chaos}
\put(-80,96){\vector(-2,1){24}}
\put(-90,66){\vector(-1,3){9}}
\put(-151,149){\vector(-1,-1){30}}
\put(-436,110){\vector(1,4){22}}
\put(-127,148){\vector(0,-1){43}}

\vspace*{-2.5mm}
\captionsetup{width=0.9\linewidth,font=small}
\caption{Bifurcation Diagram for the Mackey-Glass Equation \eqref{eq:MG12} showing the first Hopf bifurcation, and the bifurcations from the stable periodic orbit created at the Hopf bifurcation, as well as the bifurcations from the first (stable) period-doubled orbit. Also shown are regions where chaotic dynamics are detected. Several valleys of non-chaotic behaviour are visible in the chaotic regions, and the fold and period-doubling bifurcations that bound one of these valleys are shown. Also indicated are several interesting codimension-two bifurcations and regions of bistability of solutions.
}
\label{fig:2paramcont}
\end{figure}
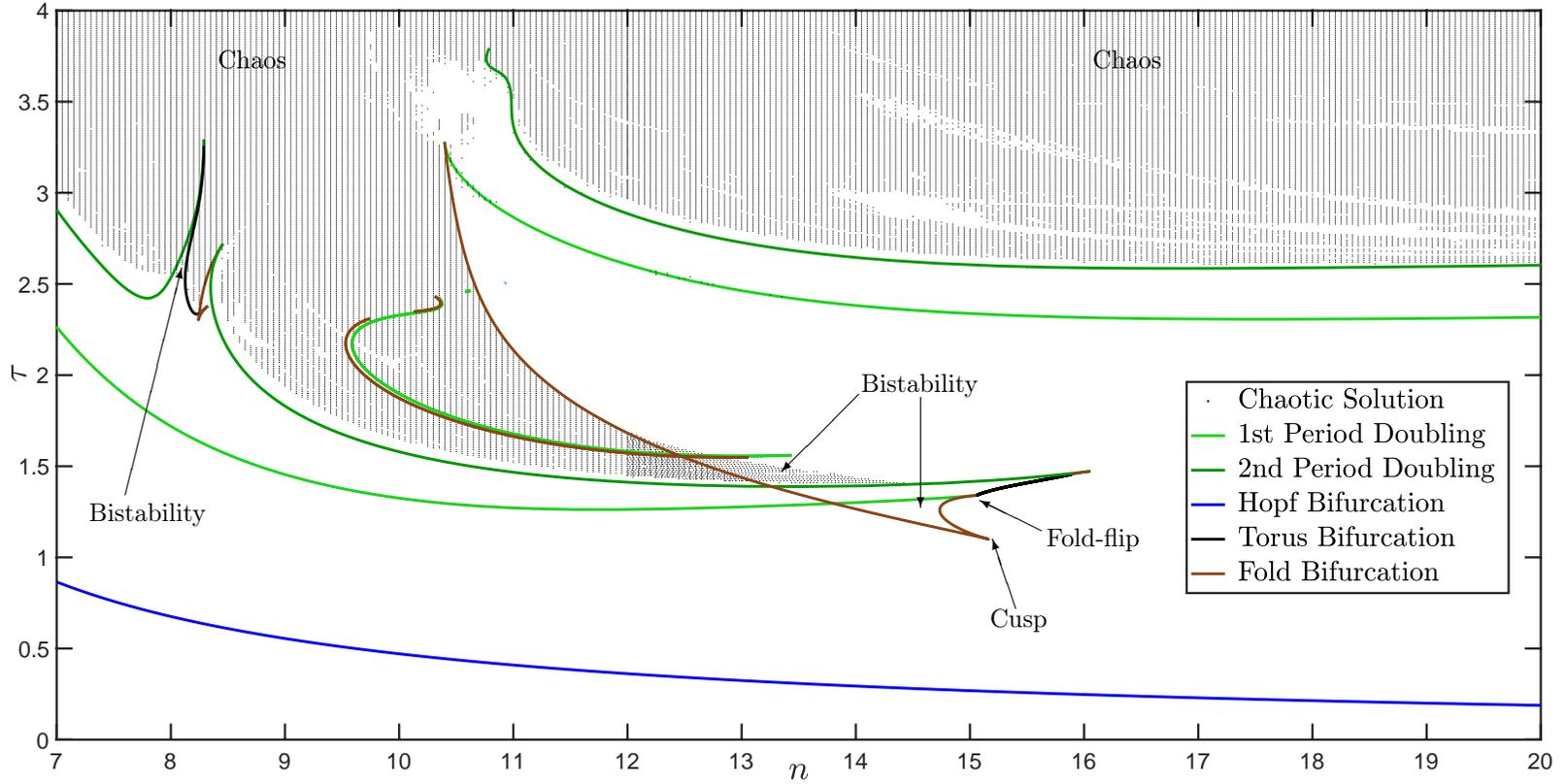
\end{landscape}
}

Our approach is not to directly determine the codimension-two bifurcations, but rather to determine the parts of the two-parameter bifurcation curves which bound stable solutions, and discard the parts that do not,
as detailed in Section~\ref{sec:ddebiftool}.
The resulting full bifurcation diagram is shown in Figure~\ref{fig:2paramcont}.
Only small parts of the first two period-doubling bifurcations from
Figure~\ref{fig:contn20r2p}(b) are retained once the solution stability is calculated. The additional bifurcation curves are computed by performing one-parameter continuation in $n$ or $\tau$ of a stable periodic orbit until a stability boundary is reached, and then performing a two-parameter continuation of the resulting bifurcation and only retaining the part on the stability boundary. The codimension-two bifurcations are then
revealed as the junctions between the different bifurcation curve segments.

In Figure~\ref{fig:2paramcont} we show the bifurcation curves where the periodic orbit originating at the Hopf bifurcation is destroyed or loses stability. Additionally if there is a bifurcation to a stable period-doubled orbit then the bifurcations where it loses stability are also included.

The regions where chaotic dynamics are detected are also shown in Figure~\ref{fig:2paramcont}.
We test for chaotic dynamics as described in Section~\ref{sec:LyapDim}. A black dot marks each parameter pair at which chaos was detected; but because the steps in $n$ are much larger the chaotic region appears as a striped region in the upper part of the diagram.
In our original computations, using the constant initial function \eqref{eq:if2},
the chaotic region stopped at the fold bifurcation near $(n,\tau)=(13,1.5)$, but later computations (to be discussed in Section~\ref{sec:cusp}) revealed bistability between a chaotic attractor and a stable periodic orbit to the right of the fold near those parameters. In this region the dynamics were recomputed
using the solution at the adjacent parameter set as the initial function $u_0$, to reveal the chaotic dynamics in the bistable region which were not revealed by the original computation.
The use of a finer parameter mesh in
this part of the chaotic region makes it appear as the more densely shaded region near $(n,\tau)=(13,1.5)$ in Figure~\ref{fig:2paramcont}.

Parts of the bifurcation diagram in Figure~\ref{fig:2paramcont} are as we might expect with a Hopf bifurcation leading to a stable periodic orbit which then period-doubles multiple times leading to chaotic dynamics.
Within the chaotic regions there appear to be strips of non-chaotic behaviour. These are the natural extension to two-parameter continuation of the periodic windows seen for one-parameter continuation in
Figures~\ref{fig:ODtau2} and~\ref{fig:ODtau2details}. Indeed, for the periodic window shown in
Figure~\ref{fig:ODtau2details}(b) we used DDE-BIFTOOL to find a fold bifurcation of periodic orbits at the left side of the window, and also the first period-doubling bifurcation that occurs in the window. These bifurcations are then continued in two parameters to reveal a \emph{valley} of periodic behaviour. We refer to these features as valleys, since the Lyapunov dimension is $1$ on the periodic orbits in the valley, and larger than $2$ in the chaotic region outside the valley, and on a surface plot of the Lyapunov dimension they appear as valleys (see Figure~\ref{fig:dimensionB} in Section~\ref{sec:dimension} for an example).

The bifurcation diagram in Figure~\ref{fig:2paramcont} does contain a number of surprises, beginning with a cusp bifurcation of periodic orbits near to $(n,\tau)=(15.16,1.1)$. This leads to a shark tooth shaped region between two fold bifurcations of periodic orbits. Within this region there is bistability of periodic orbits. If $\tau$ is increased one of these orbits undergoes a period-doubling cascade leading to bistability between the other stable periodic orbit and a chaotic attractor. For $n$ between $10.5$ and $12$ a fold bifurcation of periodic orbits appears to delineate the edge of the chaotic region.
There is also a fold-flip bifurcation of periodic orbits near $(n,\tau)=(15.05,1.34)$. Finally near $(n,\tau)=(8.1,2.6)$ chaotic behaviour is detected \emph{before} the second period doubling. The bifurcation
structures at the edge of the region of stable period-doubled orbits near to $(n,\tau)=(11,3.5)$ are complicated and delicate, and as yet incomplete, and so omitted from the figure.
We will investigate the other phenomena mentioned above in the following sections.

\subsection{Bistability in the Mackey-Glass Equation}
\label{sec:bistab}

\subsubsection{Cusp Bifurcation and Bistability of Attractors}
\label{sec:cusp}

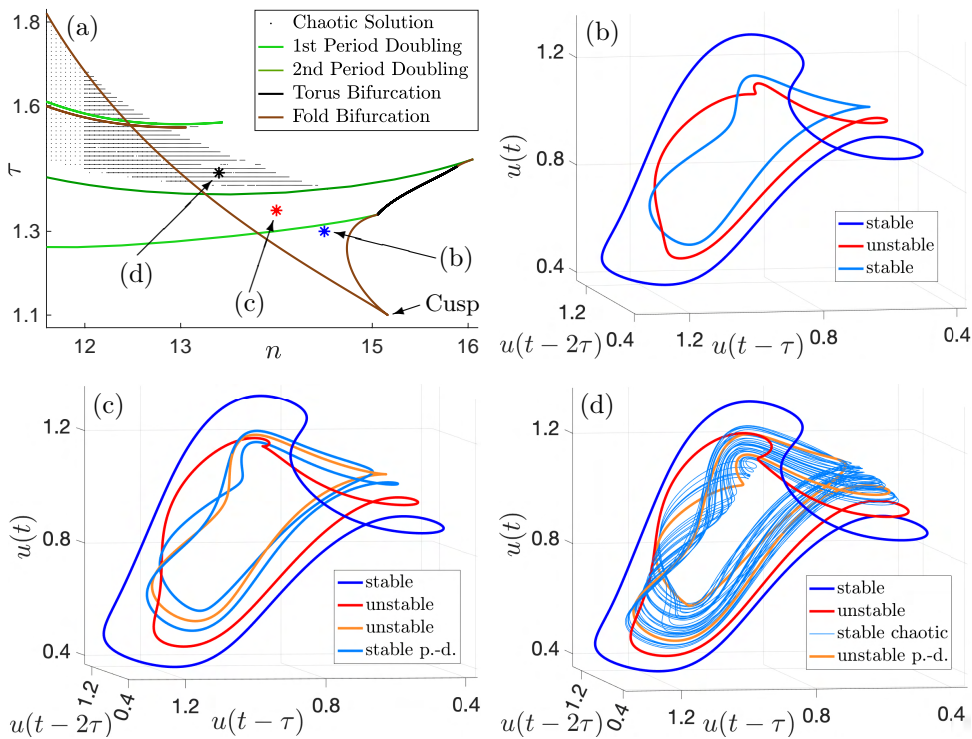
\begin{figure}[ht!]	\includegraphics[width=0.49\textwidth]{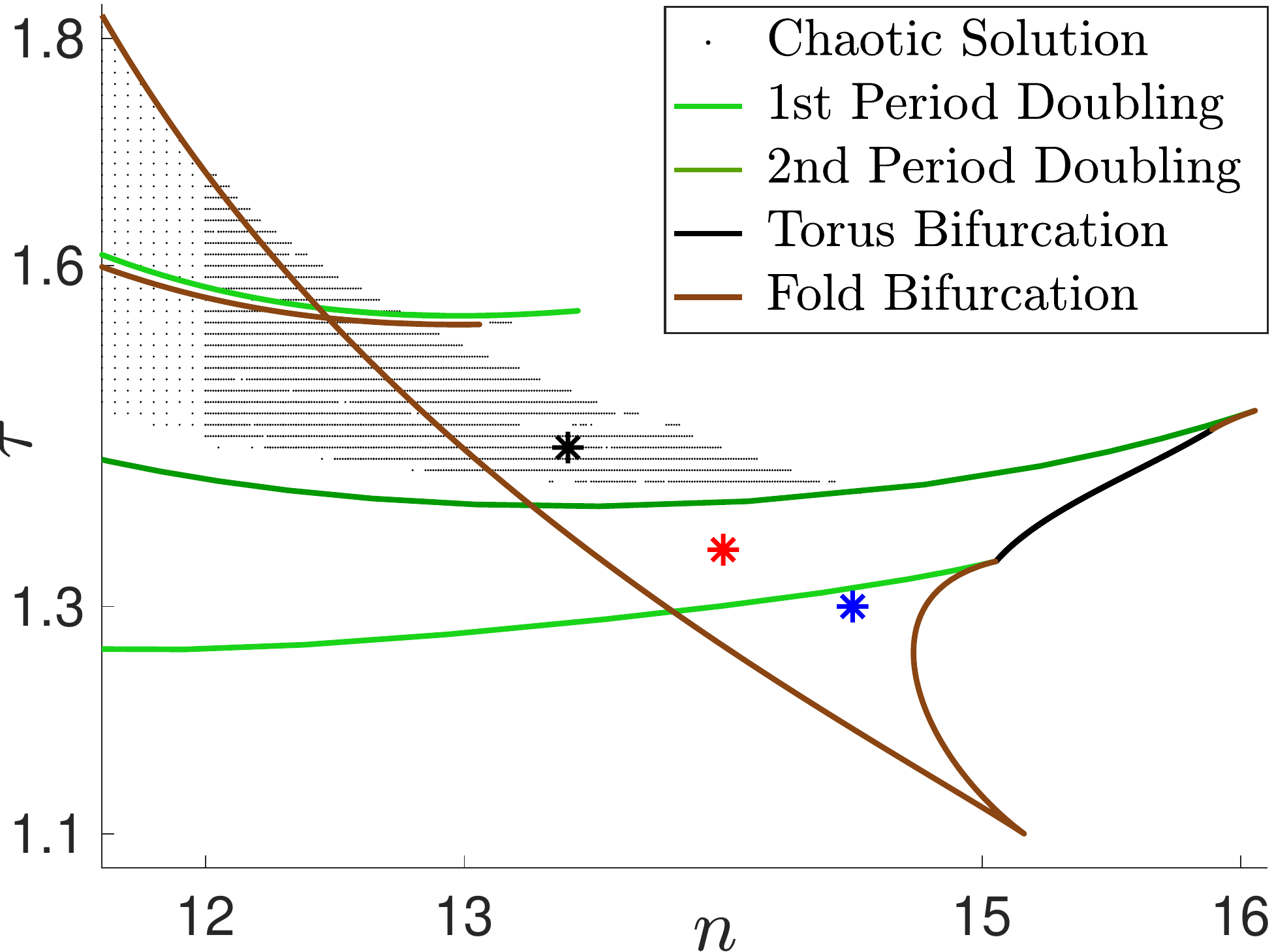}\hspace*{0.02\textwidth}\includegraphics[width=0.49\textwidth]{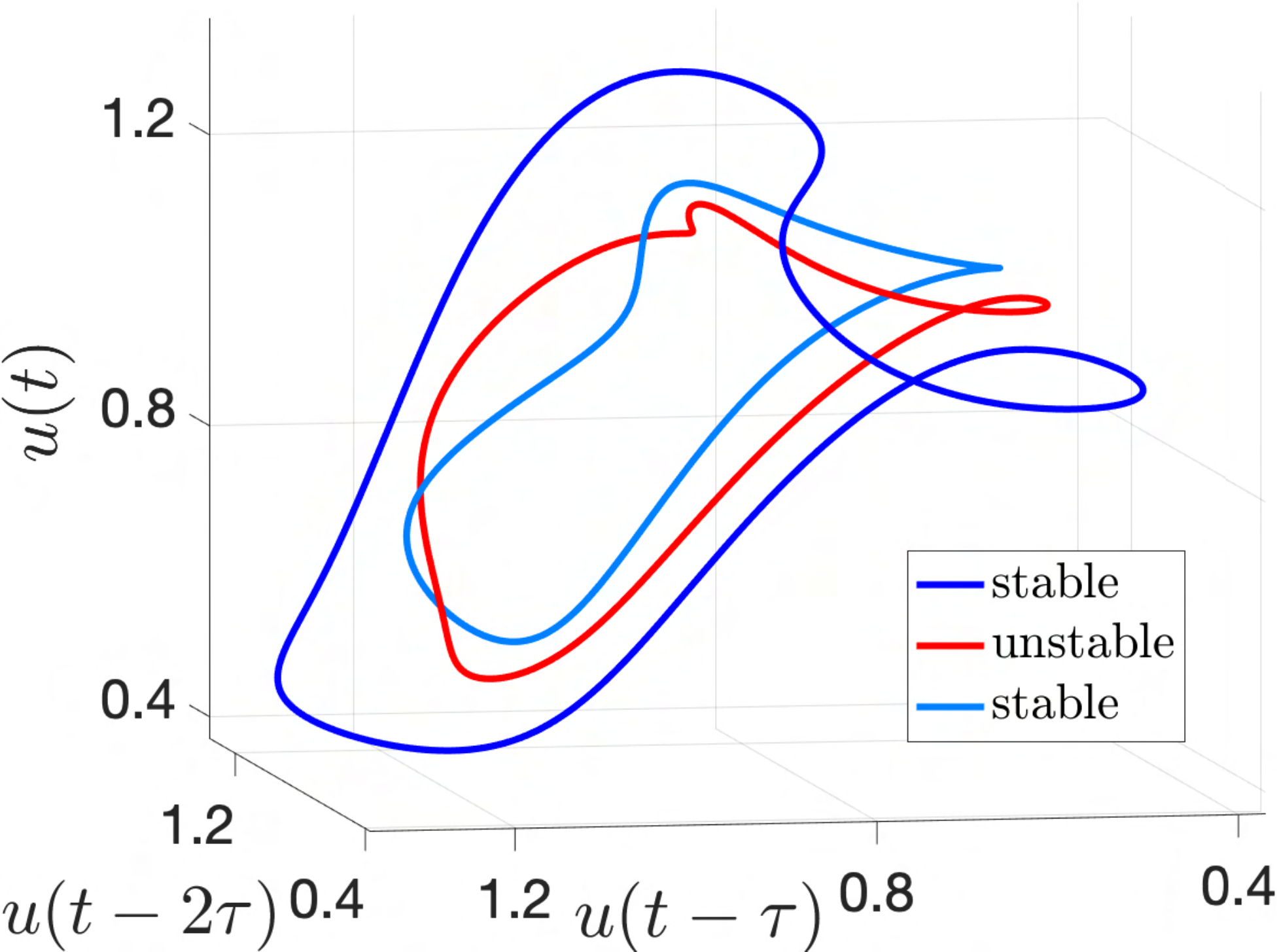}
\put(-340,122){(a)}
\put(-145,120){(b)}
\put(-200,35){(b)}
\put(-205,18){\small Cusp}
\put(-207,20){\vector(-3,-1){10}}
\put(-277,18){(c)}
\put(-320,29){(d)}
\put(-201,38){\vector(-4,1){38}}
\put(-271,25){\vector(1,3){9}}
\put(-313,39){\vector(1,1){28}}

\vspace*{2ex}

\includegraphics[width=0.49\textwidth]{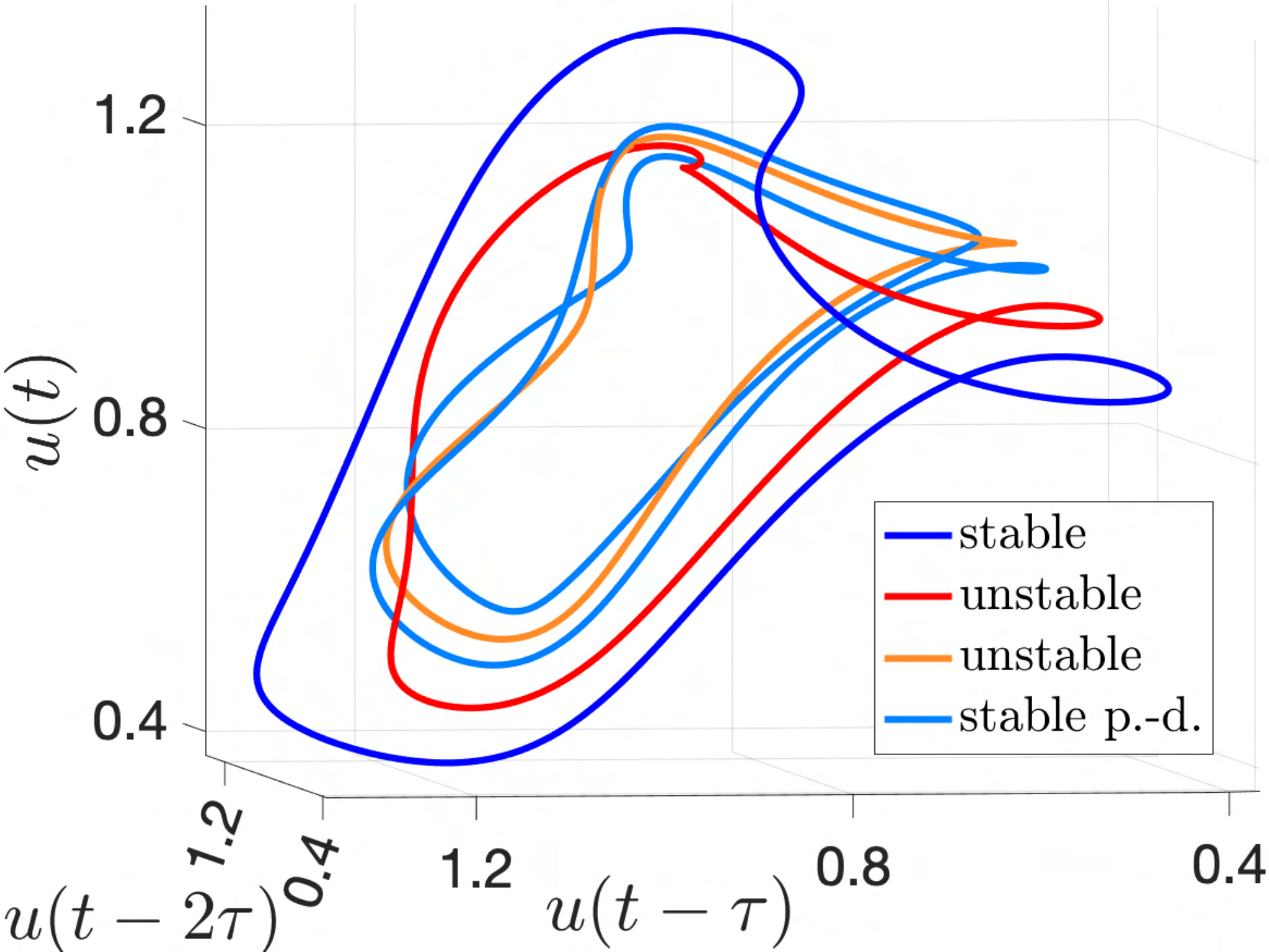}\hspace*{0.02\textwidth}\includegraphics[width=0.49\textwidth]{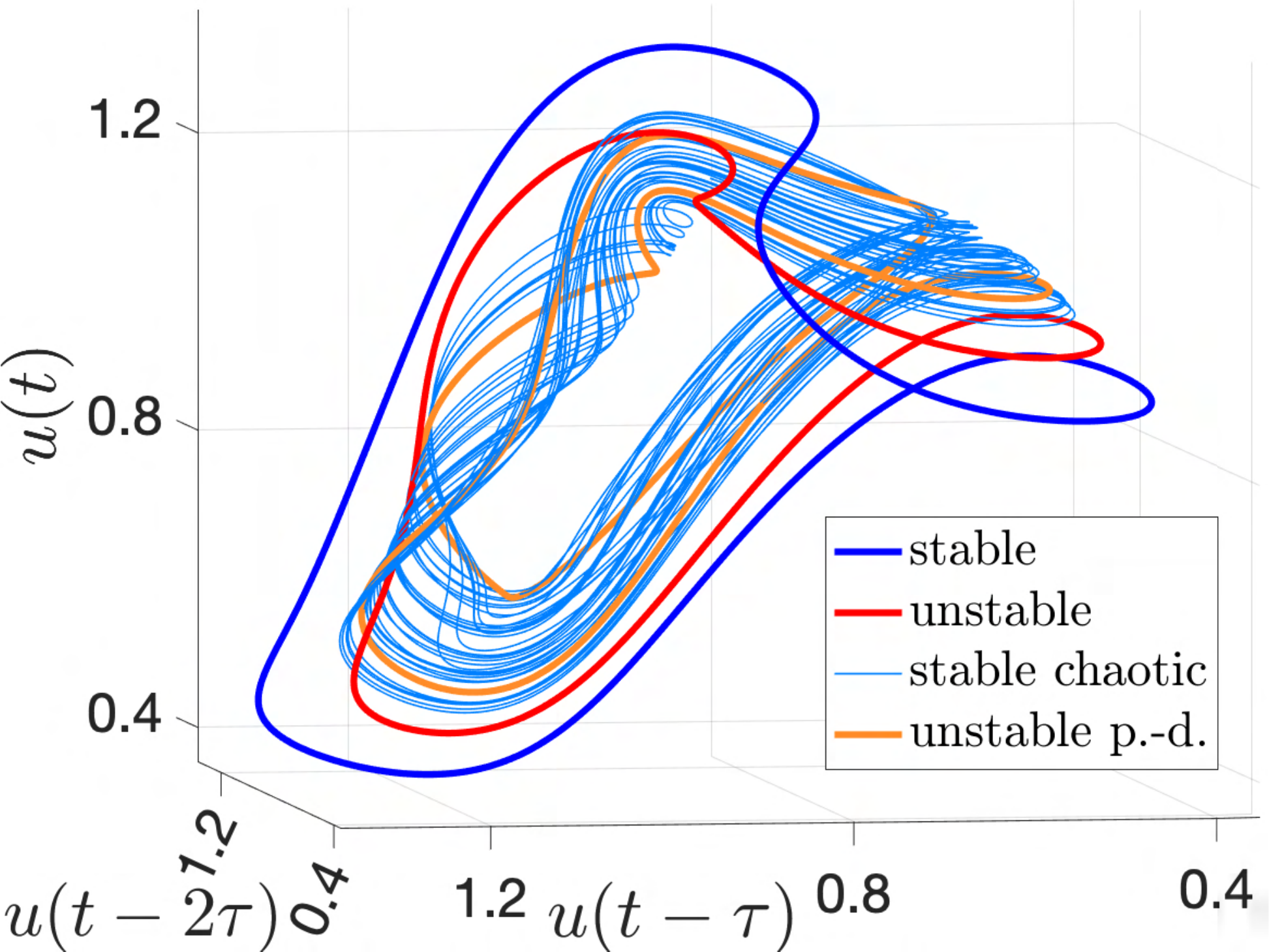}
\put(-330,122){(c)}
\put(-147,122){(d)}

\caption{(a) Enlarged detail of Bifurcation Diagram from Figure~\ref{fig:2paramcont} near the cusp point
at $(n,\tau)=(15.16,1.1)$. Phase portraits for (b) $(n,\tau)=(14.5,1.3)$ showing two stable and one unstable periodic orbits, (c)  $(n,\tau)=(14,1.35)$ showing two stable (one period-doubled) and two unstable periodic orbits, and, (d) $(n,\tau)=(13.4,1.44)$
showing one stable and two unstable periodic orbits coexisting with a stable chaotic attractor.}
\label{fig:cusp}
\end{figure}

Figure~\ref{fig:cusp}(a) shows an enlarged view of the Mackey-Glass bifurcation diagram around the cusp point at
$(n,\tau)=(15.16,1.1)$. At this point two curves of fold bifurcations of periodic orbits are born. Between these curves in the shark tooth shaped region there is bistability between two stable periodic orbits.
Figure~\ref{fig:cusp}(b) illustrates these orbits and the associated unstable periodic orbit between them at
$(n,\tau)=(14.5,1.3)$ within this region.

\sloppy{
To clearly show the orbits in Figures~\ref{fig:cusp}(b)-(d) we display them as three-dimensional plots.
Since periodic orbits of constant delay DDEs are analytic \cite{Nussbaum73}, a natural projection of the infinite-dimensional phase space into $\R^3$ would be to project $u_t\in C_+$ onto its first three Taylor coefficients.} However, these often do not scale well and lead to unsatisfactory images. Instead, we note that differentiating
\eqref{eq:MG} and then substituting \eqref{eq:MG} into the resulting expression we obtain
\be
\label{eq:MGdiff}
\uddot(t) = \gamma^2 u(t)-\beta\gamma[f(u(t-\tau))+f'(u(t-\tau))u(t-\tau))
+\beta^2 f'(u(t-\tau))f(u(t-2\tau)).
\ee
Then \eqref{eq:MG} together with \eqref{eq:MGdiff} provide a surjective mapping from
$(u(t),u(t-\tau),u(t-2\tau))$ onto $(u(t),\dot{u}(t),\ddot{u}(t))=(u_t(0),\dot{u}_t(0),\ddot{u}_t(0))$. Thus
plotting solutions as parameterised curves $(u(t),u(t-\tau),u(t-2\tau))\in\R^3$ is equivalent
to projecting the solution onto its first three Taylor coefficients, but in our experience leads to
clearer figures.


The dark blue periodic orbit in Figure~\ref{fig:cusp}(b) exists to the right of the left-hand branch of fold bifurcations emanating from the cusp point in Figure~\ref{fig:cusp}(a), and will be referred to below as the \emph{right-hand periodic orbit}. It remains stable if the delay $\tau$ is increased, until it eventually loses stability in a period-doubling bifurcation with $\tau\approx2.5$ as seen in Figure~\ref{fig:2paramcont}.
The light blue periodic orbit seen in Figure~\ref{fig:cusp}(b) exists to the left of the right-hand branch of fold bifurcations emanating from the cusp point, and hence we refer to it as the \emph{left-hand periodic orbit}. As $\tau$ is increased this orbit quickly undergoes a period-doubling cascade to chaos. Figure~\ref{fig:cusp}(c) shows a phase portrait after the first period doubling where the left-hand periodic orbit has become unstable (shown in orange) and a new stable (light blue) period-doubled orbit wraps around it.
The red unstable periodic orbit separates these orbits from the right-hand periodic orbit which remains stable.

Figure~\ref{fig:cusp}(d) shows a phase portrait in the chaotic region.
The (dark blue) right-hand periodic orbit is still stable and is separated by the (red) unstable periodic orbit from the chaotic attractor. The left-hand periodic orbit and its period doublings are all unstable, and are embedded in the chaotic attractor. The chaotic attractor is found by simulating using \verb+dde23+ using the unstable period-doubled orbit as the initial function, and discarding the initial transient dynamics.

The chaotic attractor coexists with the stable right-hand periodic orbit in the shaded region to the right of the left-hand branch of fold bifurcations emanating from the cusp point in Figure~\ref{fig:cusp}(a). If the parameters are changed to cross to the left of this curve of fold bifurcations then the stable chaotic attractor continues to exist, but the stable and unstable periodic orbits are destroyed at the fold bifurcation. On the other hand, if the parameters are altered  towards the top of the chaotic parameter region then the
chaotic attractor approaches the (red) unstable periodic orbit, until it seems that the attractor is destroyed at the edge of this parameter region in a boundary crisis where the attractor collides with the boundary of its own basin of attraction \cite{GOY82}.

Thus we have shown that the cusp point gives rise to both bistability of periodic orbits and also bistability between a stable periodic orbit and a stable chaotic attractor. The combined use of numerical bifurcation analysis with numerical simulation of the DDE was crucial to finding this behaviour. It would have been very
difficult to find this bistability by simulation alone.

\subsubsection{Chaotic Attractor Interior Crisis}
\label{sec:snoca}

\begin{figure}[ht!]	\includegraphics[width=0.49\textwidth]{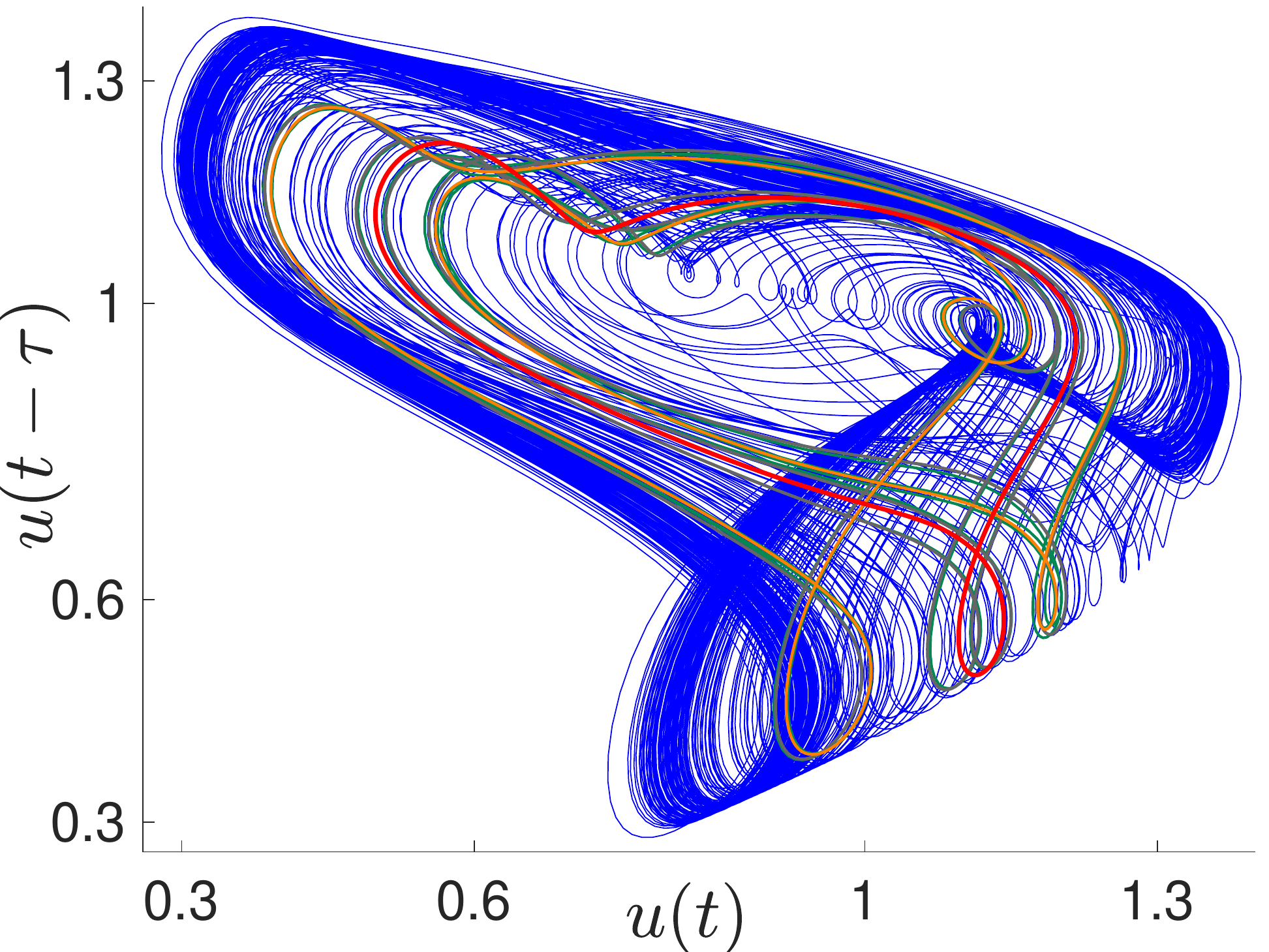}\hspace*{0.02\textwidth}\includegraphics[width=0.49\textwidth]{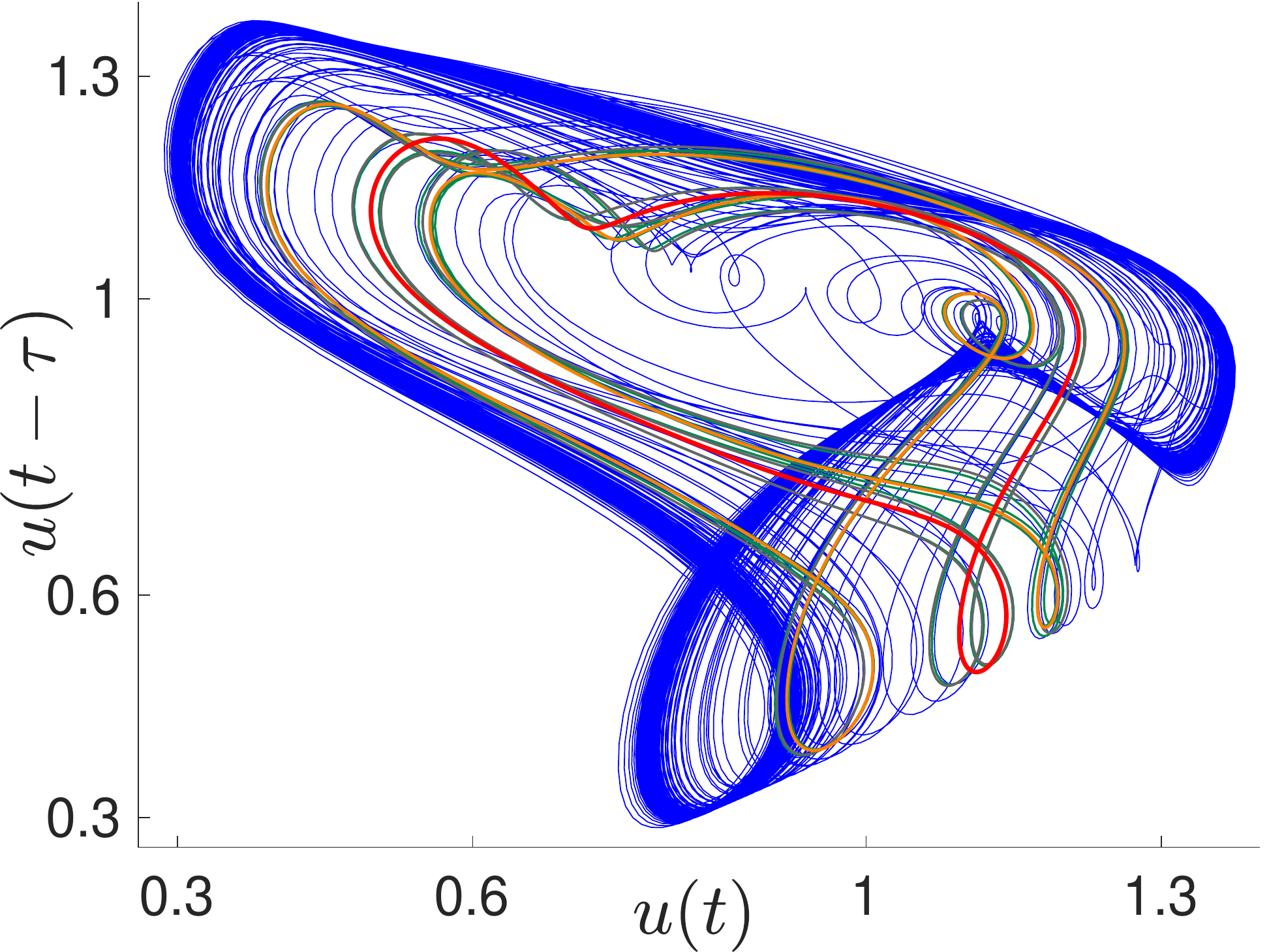}
\put(-337,25){(a)}
\put(-152,25){(b)}	

\vspace*{2ex}

\includegraphics[width=0.49\textwidth]{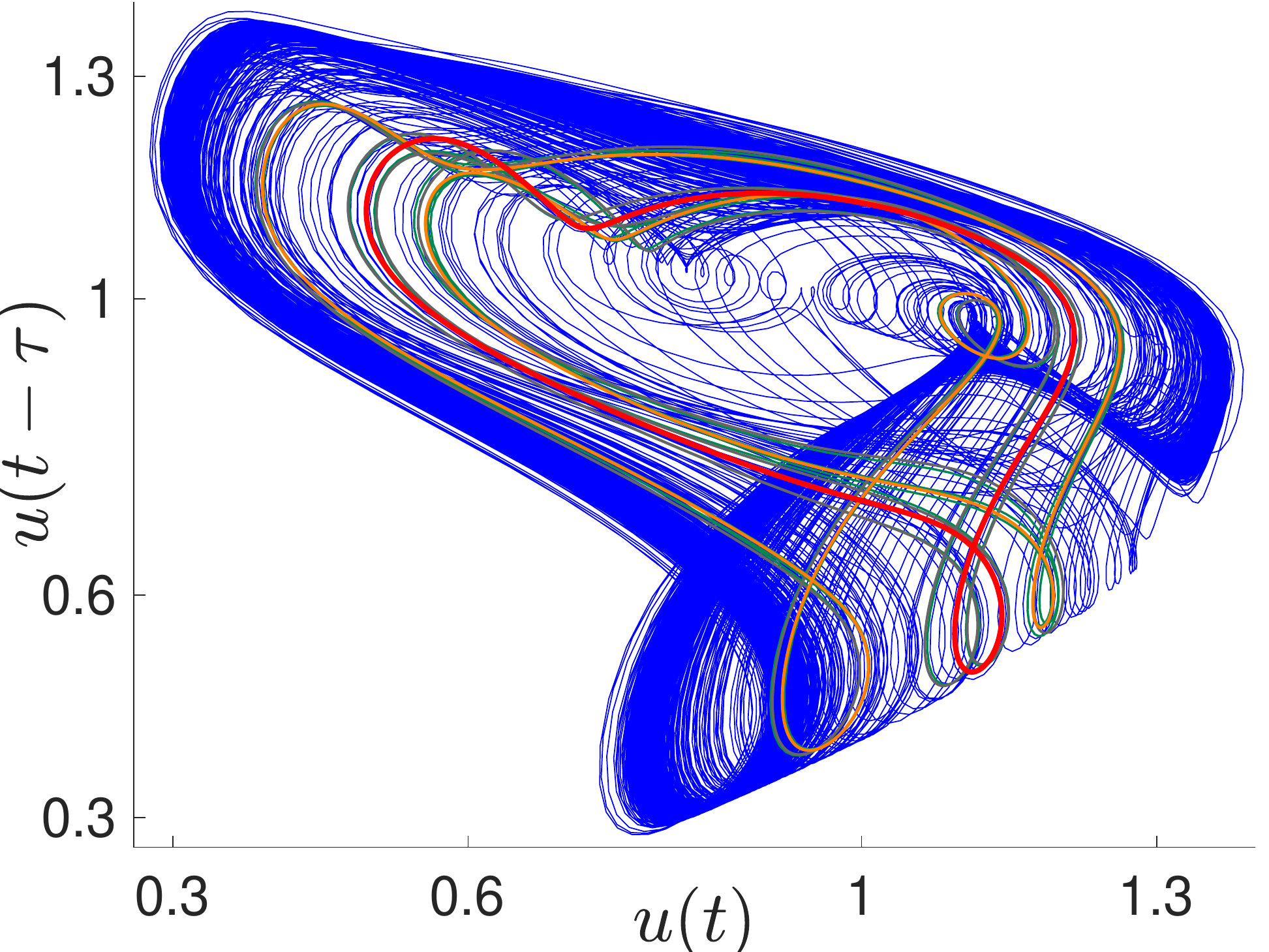}\hspace*{0.02\textwidth}\includegraphics[width=0.49\textwidth]{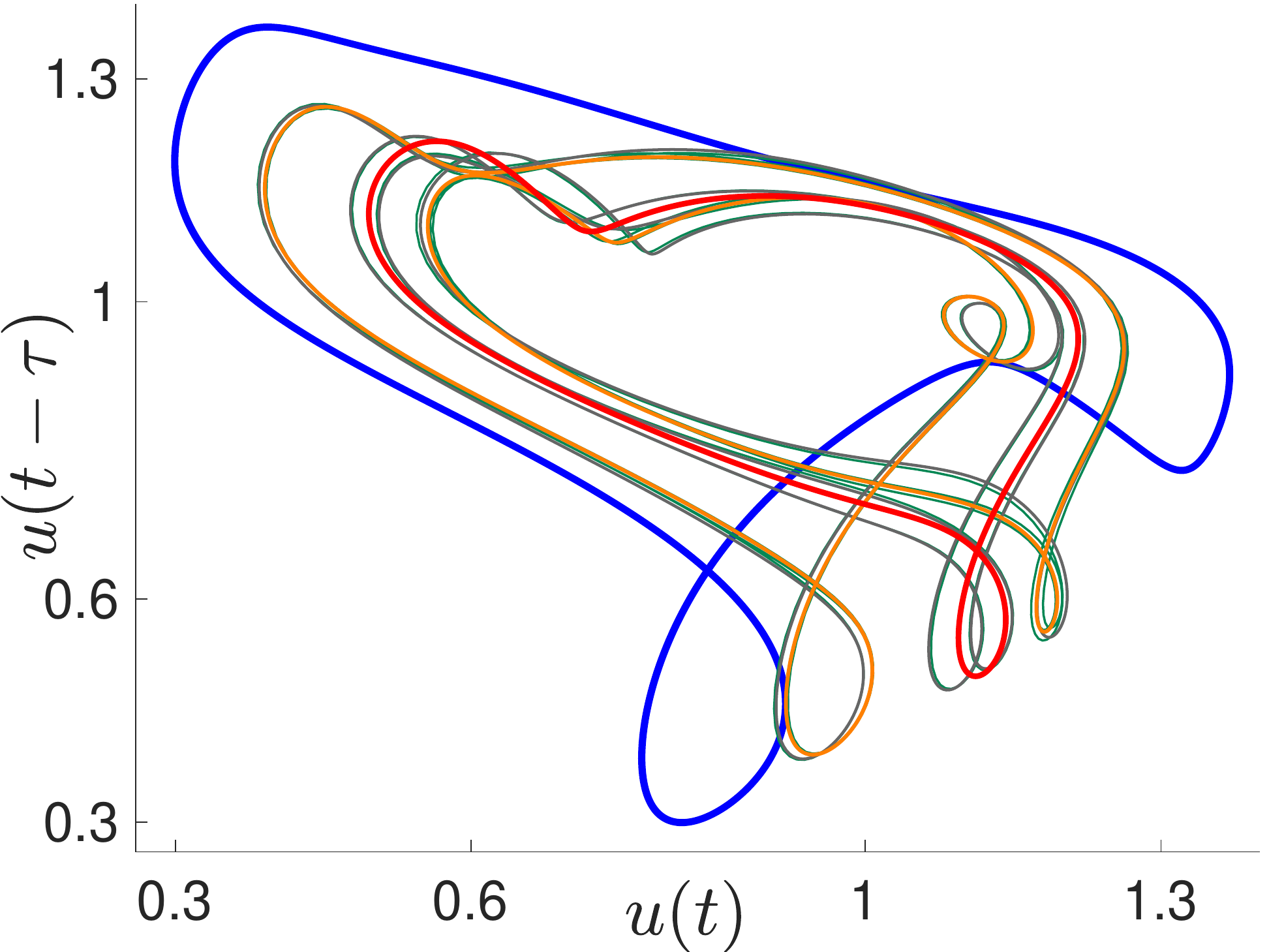}
\put(-335,25){(c)}
\put(-152,25){(d)}	

\caption{Phase portraits for the Mackey-Glass equation with $\tau=2$ showing the destruction of the chaotic attractor as the fold bifurcation is crossed at $n=11.21485$.
The stable dynamics (in blue) is found by simulation.
The unstable periodic orbits (all other colours) computed by DDE-BIFTOOL correspond to the left-hand periodic orbit and its period doublings.
(a) $n=11.1$ (b and c) $n=11.2$, with integration time tripled in
(c). (d) $n=11.215$.}
	\label{fig:bistabpoca}
\end{figure}

In Figures~\ref{fig:2paramcont} and~\ref{fig:cusp}(a) the chaotic region is seen to extend to the right of the curve of fold bifurcations of periodic orbits for $\tau<1.7$ but not for larger values of $\tau$. This suggests that the bifurcation structures are different in these two cases. We already investigated the behaviour for $\tau<1.7$ in Section~\ref{sec:cusp}. Now we investigate the behaviour for larger $\tau$ by returning to the one-parameter continuation in $n$ for $\tau=2$, that we first considered in Section~\ref{sec:contn}.

Note first that the dynamics shown in Figure~\ref{fig:contn20r2p}(a) is now easily explained. The stable periodic orbit in Figure~\ref{fig:contn20r2p}(a) (first found by Glass and Mackey \cite{Glass_Mackey_1979}) is the orbit referred to as the right-hand periodic orbit in Section~\ref{sec:cusp}. The unstable orbits in
Figure~\ref{fig:contn20r2p}(a) correspond to the left-hand periodic orbit and its period doublings.

For $\tau=2$ the right-hand periodic orbit (along with its unstable counterpart) exists and is stable to the right of the fold bifurcation which occurs at $n=11.21485$ for this value of $\tau$. We can find no chaotic dynamics to the right of the fold bifurcation, and the chaotic attractor is apparently destroyed at the fold bifurcation. To investigate the mechanism by which this occurs, in Figure~\ref{fig:bistabpoca} we show phase portraits for $n\in[11.1,11.215]$ as the fold bifurcation is approached and crossed.
The left-hand periodic orbit and its period doublings are all unstable and  do not change appreciably over this parameter range, as seen in Figure~\ref{fig:bistabpoca}.
The right-hand periodic orbit is stable but only exists for $n>11.21485$ and is seen only in Figure~\ref{fig:bistabpoca}(d).

Despite the stable right-hand periodic orbit only existing for $n>11.21485$, `ghosts' of this periodic orbit  can be seen in intermittent periodic dynamics for  $n<11.21485$.
Already for $n=11.1$ in Figure~\ref{fig:bistabpoca}(a)
the attractor looks very different than it did for $n=9.65$ in Figure~\ref{fig:contnps}(f).
For $n=11.1$ the trajectory on the attractor spends much more time on the part of the attractor near where the periodic orbits will be created in the fold bifurcation than it does on the rest of the attractor.
This is even more apparent just to the left of the fold bifurcation for $n=11.2$ in Figure~\ref{fig:bistabpoca}(b), where the shape of the periodic orbit from Figure~\ref{fig:bistabpoca}(d)
is clearly distinguishable within the chaotic attractor in Figure~\ref{fig:bistabpoca}(b).

We emphasise that the behaviour seen in Figure~\ref{fig:bistabpoca}(b) is not transient. For $\tau=2$ the chaotic attractor exists for all $n<11.21485$; that is to the left of the fold bifurcation. This is confirmed by computing the Lyapunov exponents, as seen in Figure~\ref{fig:2paramcont} where every black dot represents a distinct trajectory with a positive Lyapunov exponent. We also demonstrate that the dynamics is persistent in
Figure~\ref{fig:bistabpoca}(c). This is computed for exactly the same parameters as in
Figure~\ref{fig:bistabpoca}(b), but over a time interval that is three times longer than in
Figure~\ref{fig:bistabpoca}(b). For this longer simulation it is apparent that the orbit still visits all parts of the attractor, even though it spends more time in the vicinity of the ghost of the periodic orbit.

From our simulations it is apparent that as $n$ is increased towards the fold bifurcation at
$(n,\tau)=(11.21485,2)$ the attractor remains chaotic, but trajectories spend
longer and longer
on the weakly expansive part of the attractor near the ghost of stable periodic orbit.

At the fold bifurcation a stable periodic orbit is born (along with its unstable counterpart), and these are
embedded within the chaotic attractor destroying it in the process. This
saddle node of periodic orbits on a chaotic attractor bifurcation is an example of an interior crisis \cite{GOY82}.
While this phenomenon is mainly studied in maps \cite{RO17}, it is also found in ODEs \cite{RAOY2000,WKL01}, and so it should not be a surprise to find interior crises in DDEs.
This bifurcation is also very reminiscent of a SNIC or saddle node on an invariant circle bifurcation in which
an invariant circle is destroyed when two fixed points are created on the circle in a saddle-node bifurcation. The SNIC bifurcation is covered briefly by many authors as one of the simplest examples of a global bifurcation, and consequently it goes by many names (including saddle-node homoclinic bifurcation \cite{Kuznetsov_2004}, SNIPER or infinite-period bifurcation \cite{Strogatz}, and saddle-node on a limit cycle or SNLC \cite{HopIz97}).


In contrast for smaller values of $\tau<1.7$ we saw in Section~\ref{sec:cusp} that the periodic orbits created in the fold bifurcation can coexist with the chaotic attractor. In that case the fold bifurcation creates periodic orbits outside the chaotic attractor, which is only destroyed if the parameters are varied (for example by increasing $\tau$) so that the chaotic attractor grows and then undergoes a boundary crisis
when it collides with the unstable periodic orbit.

\subsection{Subcritical Period Doubling}
\label{sec:subcritpd}

\begin{figure}[ht!]	\includegraphics[width=0.49\textwidth]{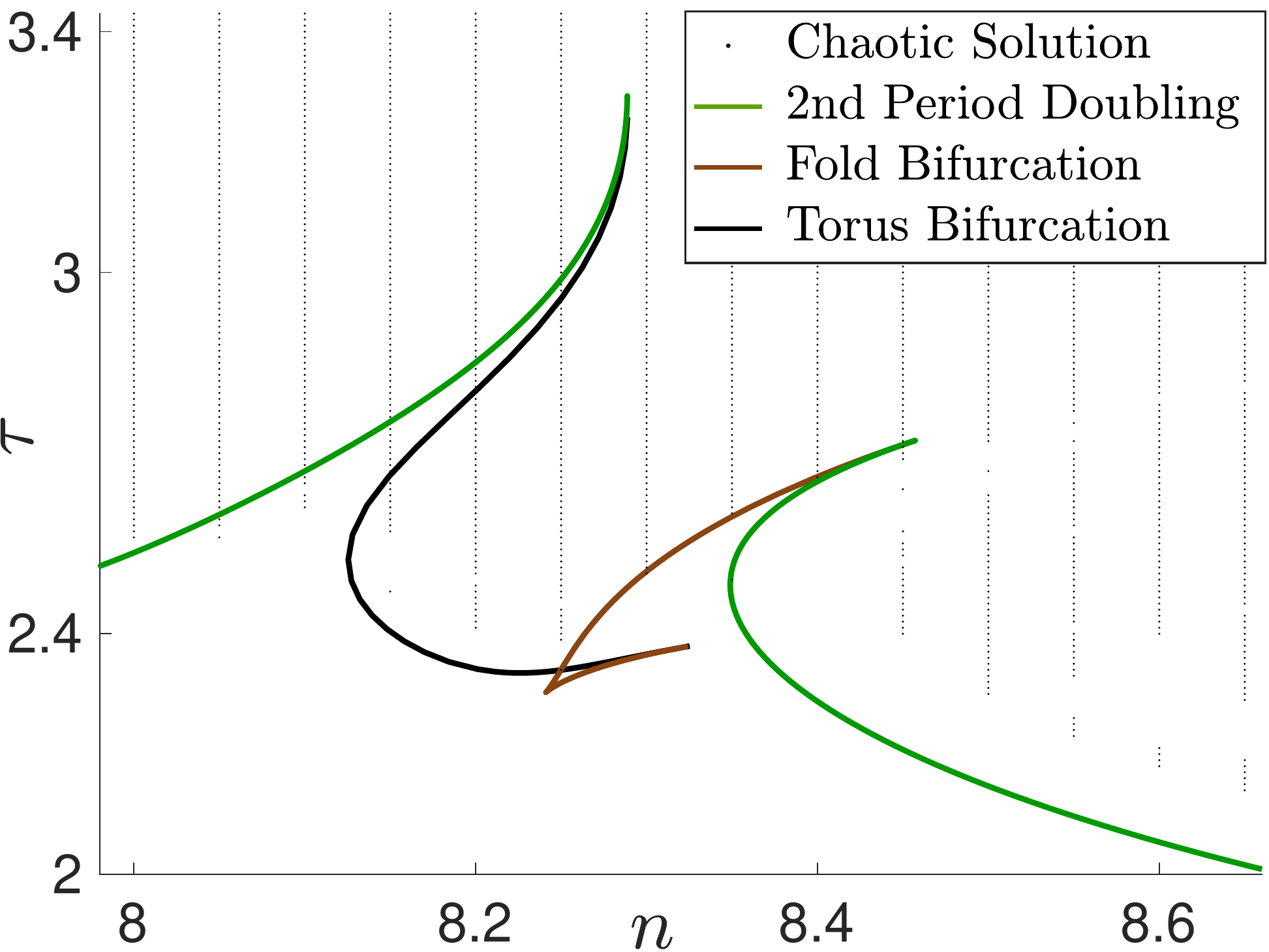}\hspace*{0.02\textwidth}\includegraphics[width=0.49\textwidth]{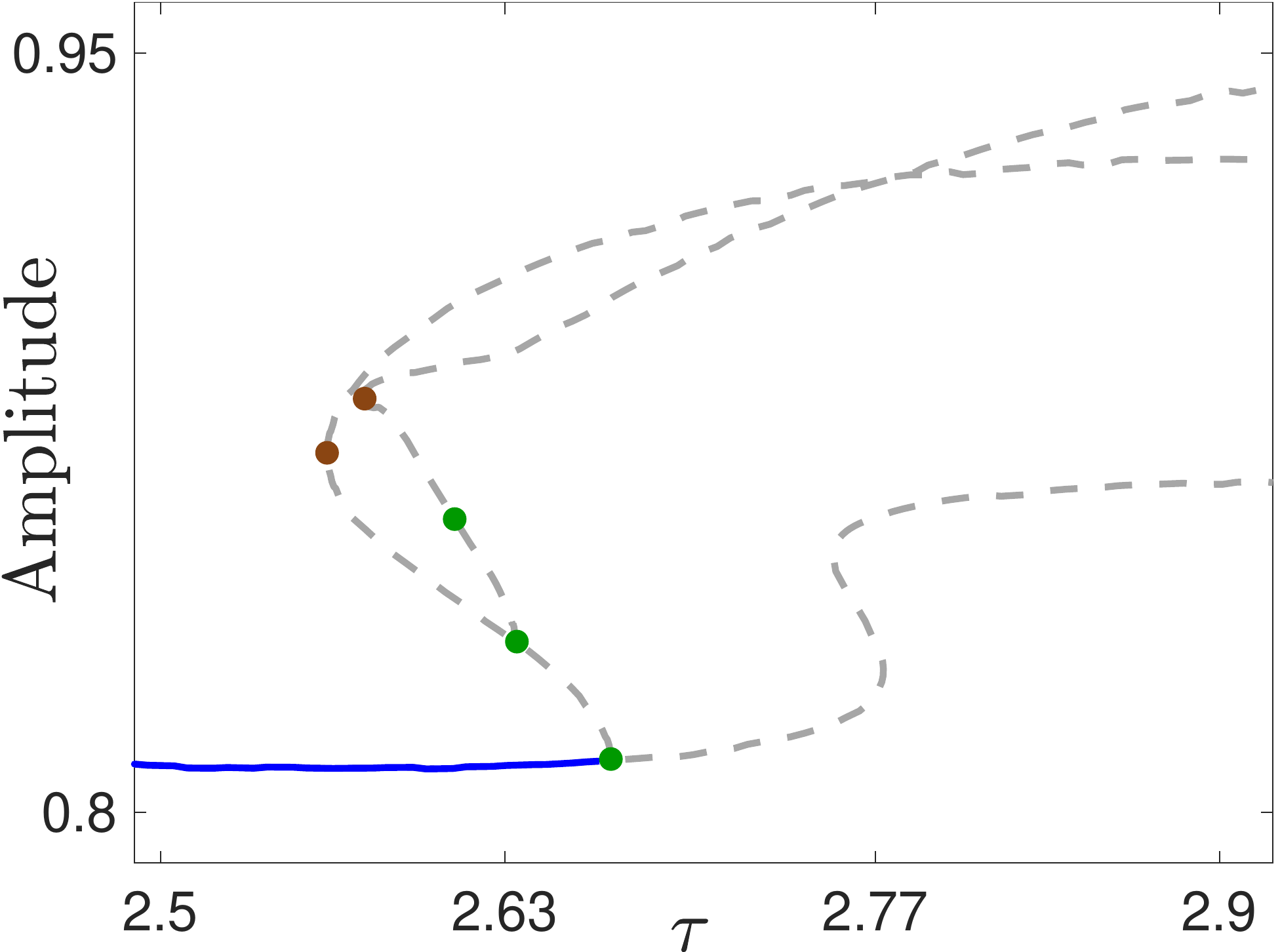}
\put(-342,17){(a)}
\put(-22,17){(b)}	
\put(-152,46){\scriptsize PD2}	
\put(-85,48){\scriptsize PD3}	
\put(-90,69){\scriptsize PD4}	
\put(-155,73){\scriptsize F1}	
\put(-145,93){\scriptsize F2}	
\put(-145,75){\vector(2,-1){10}}
\put(-140,90){\vector(1,-1){11}}
\put(-135,49){\vector(2,-1){40}}
\put(-86,50){\vector(-3,-1){16}}
\put(-91,71){\vector(-2,-1){20}}

\vspace*{2ex}

\includegraphics[width=0.49\textwidth]{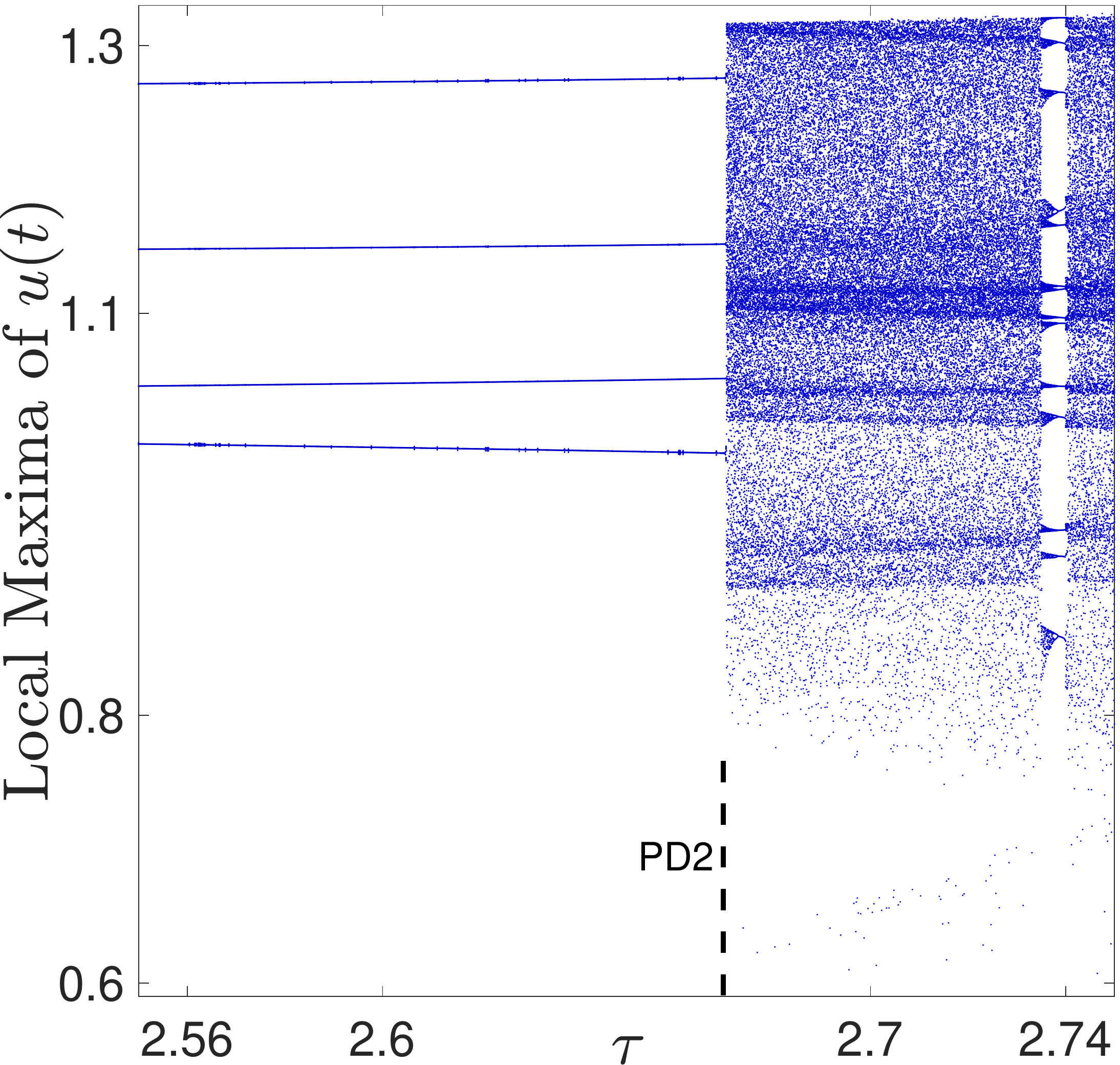}\hspace*{0.02\textwidth}\includegraphics[width=0.49\textwidth]{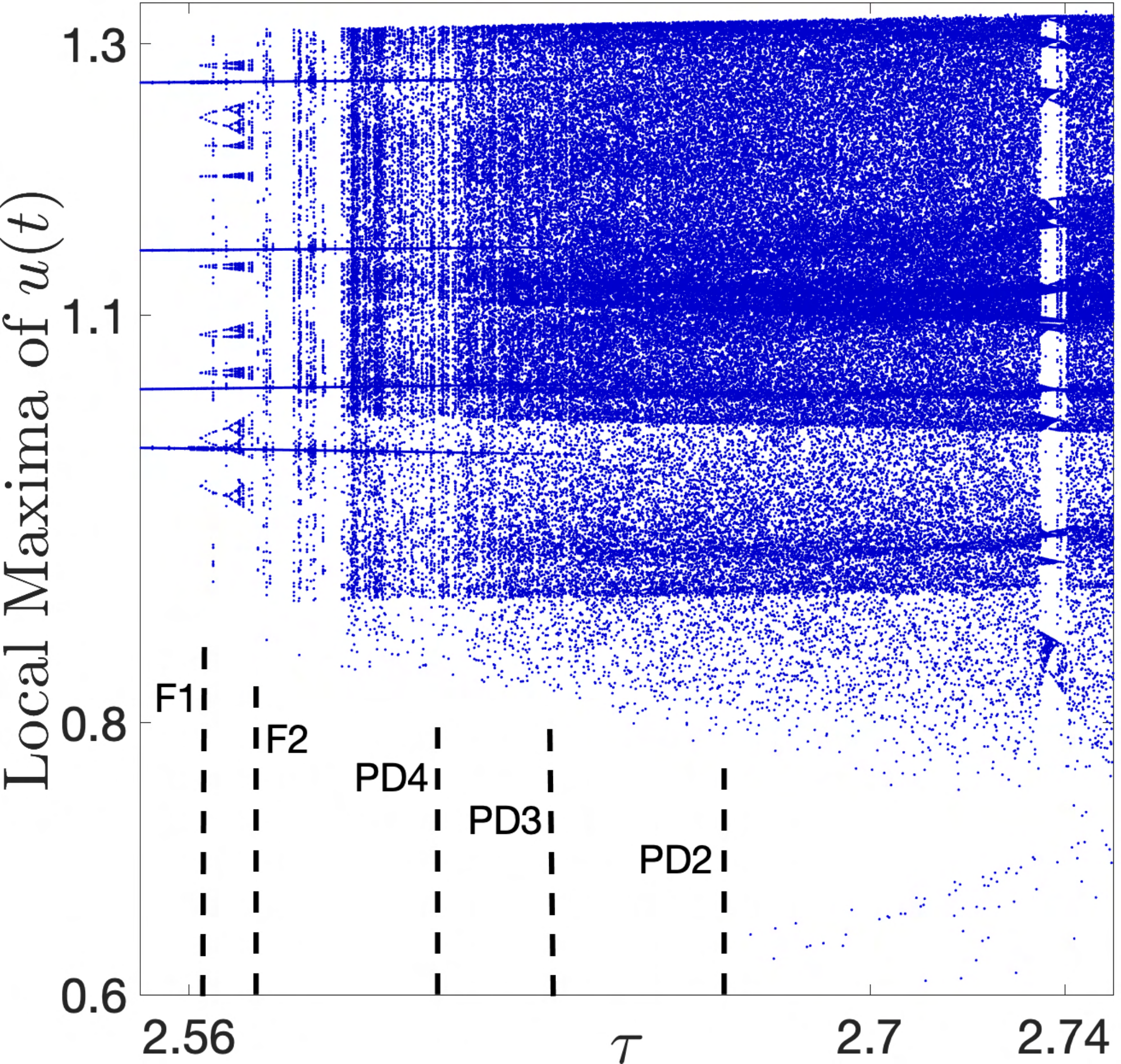}
\put(-337,17){(c)}
\put(-20,17){(d)}	

\caption{(a) Enlarged detail of part of the Bifurcation Diagram from Figure~\ref{fig:2paramcont}.
(b) DDE-BIFTOOL bifurcation diagram as $\tau$ is varied with $n=8.1$ fixed, showing period-doubling bifurcations from the first stable period-doubled orbit, and from the resulting period-doubled orbits, and fold bifurcations on these orbits.
(c) Orbit Diagram for the Mackey-Glass Equation \eqref{eq:MG12} computed with \texttt{dde23} with $n=8.1$ for increasing $\tau$ in increments of $10^{-4}$, with initial function given by solution at the previous $\tau$ value.
(d) As (c) except using the constant initial function \eqref{eq:if2} for every simulation. In (c) and (d) parameter values for the bifurcations in (b) are also marked for comparison.}
	\label{fig:subcritpd}
\end{figure}
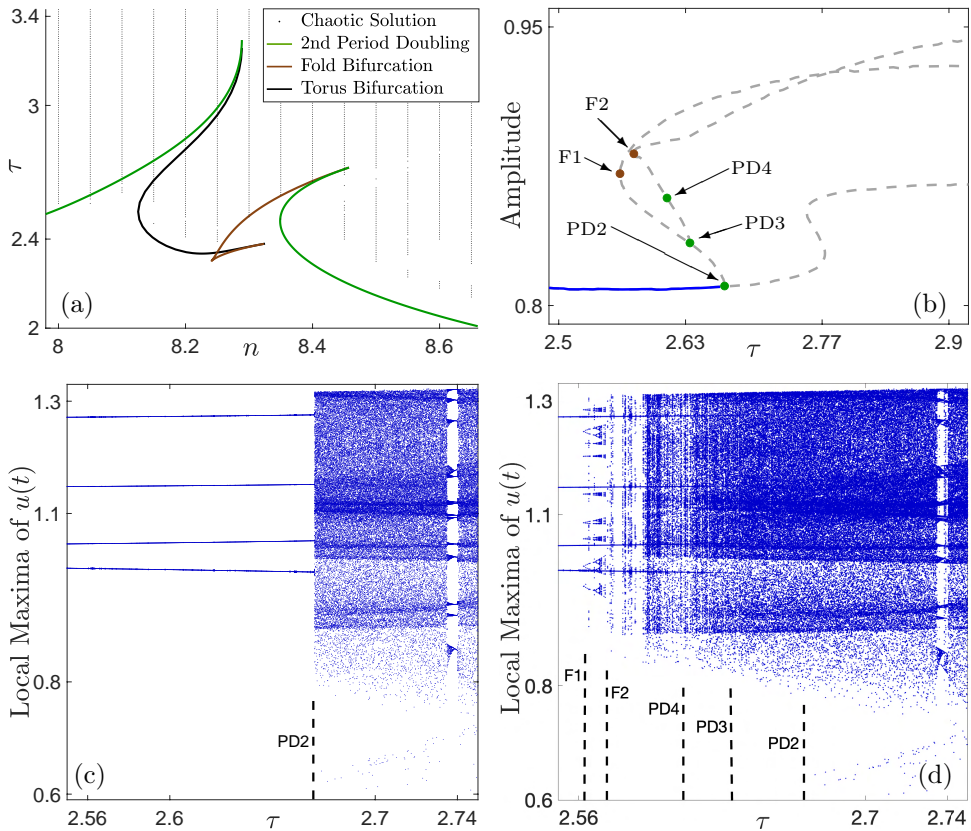

Another interesting part of the bifurcation diagram in Figure~\ref{fig:2paramcont} is near $(n,\tau)=(8.3,2.5)$, where the behaviour does not correspond to a straightforward period-doubling cascade to chaos. Figure~\ref{fig:subcritpd}(a) shows an enlarged view of this part of the bifurcation diagram. This reveals a gap in the branch of period-doubling bifurcations, where the first period-doubled orbit loses stability in either a fold or a torus bifurcation. In the preceding sections, we studied the effect of fold bifurcations of period orbits on the dynamics, while torus bifurcations in DDEs with either constant or state-dependent delay have been studied extensively in \cite{Calleja_2017,DeSouza2019,Hum-DeM-Mag-Uph-12}.
So, we will not study those bifurcations here, and instead focus on another intriguing aspect of
Figure~\ref{fig:subcritpd}(a).

There appears to be bistability between a chaotic attractor and a stable periodic orbit in Figure~\ref{fig:subcritpd}(a) near $(n,\tau)=(8.1,2.6)$. The stable periodic orbit is the one created in the first period-doubling bifurcation. In part of the parameter region where the periodic orbit is stable, simulations starting from the fixed initial condition \eqref{eq:if2} do not converge to this periodic orbit, and computation of the Lyapunov exponents reveals a positive Lyapunov exponent and hence chaotic dynamics. Although there is a cusp bifurcation of periodic orbits nearby at $(n,\tau)=(8.24,2.3)$, the bistability observed here is not between the resulting branches of fold bifurcations, and so does not result from the same mechanism as studied in Section~\ref{sec:cusp}.

To study this occurrence of bistability between a chaotic attractor and a stable periodic orbit, we begin by performing a one-parameter continuation in DDE-BIFTOOL, varying $\tau$ with $n=8.1$ held fixed.
As $\tau$ is increased, this reveals the Hopf bifurcation at $\tau=0.661$ leading to a stable periodic orbit, which then period doubles at $\tau=1.684$, leading to a stable period-doubled orbit. The further bifurcations of this period-doubled orbit are shown in Figure~\ref{fig:subcritpd}(b). It becomes unstable in a period-doubling bifurcation at $\tau=2.67$, labelled PD2 in the figure. But this bifurcation is subcritical and the resulting period-doubled orbit exists for $\tau<2.67$ and is itself unstable. This new orbit also undergoes a further subcritical period-doubling bifurcation at PD3 with $\tau=2.634$ to create another branch of unstable periodic orbits which exist for $\tau<2.634$. A further period-doubling bifurcation is detected at PD4 with $\tau=2.611$, but we do not compute the branch of periodic orbits emanating from this point.

Thus we have found a sequence of period-doubling bifurcations for decreasing values of $\tau$ at least the first two of which are subcritical and lead to unstable period-doubled orbits which coexist with the stable periodic orbit. Many authors state conditions for a single period-doubling bifurcation to be subcritical. But sequences of period doublings are universally studied by simulation, most often using orbit diagrams, to reveal sequences of supercritical period doublings leading to stable period-doubled orbits and potentially a period-doubling cascade to chaos. We are not aware of any studies of sequences of subcritical period doublings in differential equations, with or without delay.

As seen in Figure~\ref{fig:subcritpd}(b), the first two unstable subcritical period-doubled branches
subsequently undergo fold bifurcations (at F1 with $\tau=2.563$ and at F2 with $\tau=2.574$), so for $n=8.1$ there is an interval for $\tau\in(2.563,2.67)$ for which the stable periodic orbit coexists with unstable period-doubled orbits.
It seems very likely that these subcritical period-doubled branches are in some way connected to the emergence
of chaotic dynamics and we explore this in  Figure~\ref{fig:subcritpd}(c) and (d).

Figure~\ref{fig:subcritpd}(c) shows an orbit diagram with $n=8.1$ for increasing values of $\tau$. For this plot we use the solution from the previous value of $\tau$ as the starting function for the next value of $\tau$, use \verb+dde23+ to simulate the DDE, discard the resulting transient dynamics, and plot the local maxima of the resulting solution for each value of $\tau$. We see that the simulations converge to the stable periodic orbit for $\tau<2.67$, until the subcritical period-doubling bifurcation PD2 is reached, at which point the solution jumps to the chaotic attractor.

It would be natural to try to show hysteresis in the system by repeating this computation, but for decreasing $\tau$.
We were not able to perform such a computation successfully. Probably due to errors introduced during the numerical solution of \eqref{eq:MG12} by \verb+dde23+, the numerically computed solutions would jump from the chaotic attractor to the stable periodic orbit at some point, often very close to PD2. While the computation might be possible with tolerances close to machine precision and a restriction on the maximum integration step-size, such a computation would take prohibitively long to complete.

Figure~\ref{fig:subcritpd}(d) was ultimately computed by taking the fixed initial function \eqref{eq:if2} for each value of $\tau$. With this initial function the solution converges to the stable periodic orbit
for a few values of $\tau\in(2.563,2.67)$ particularly near the left side of this interval, but for most values of $\tau$ in this range converges to the chaotic attractor, thus demonstrating
bistability between the stable periodic orbit and a chaotic attractor
for these parameter values. The interval of bistability corresponds almost exactly to the parameter interval on which the DDE-BIFTOOL computations found coexistence
between the stable periodic orbit and the unstable period-doubled orbits.
We remark that in Figure~\ref{fig:subcritpd}(d)
there appears to be a sequence of period-doublings for $\tau\approx2.57$ just before the chaotic attractor is established. We conjecture that at some point the sequence of subcritical period doublings described above becomes supercritical, thus creating a sequence of supercritical period doublings leading to stable solutions and the observed stable chaotic attractor.

\subsection{Fold-Flip Bifurcation}
\label{sec:foldflip}

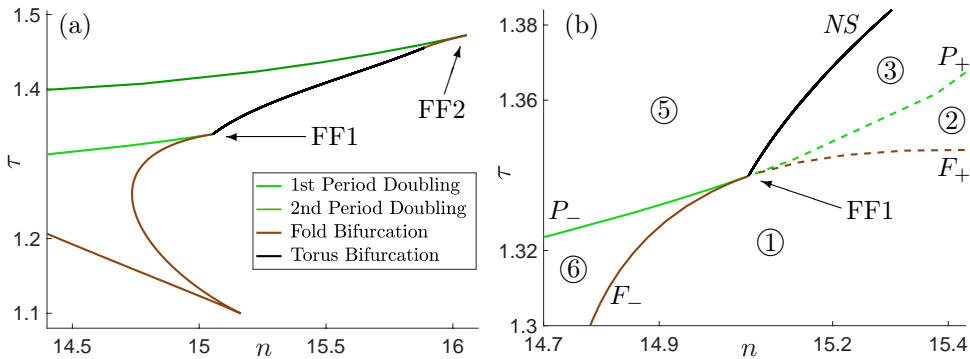
\begin{figure}[t!]	
\includegraphics[width=0.49\textwidth]{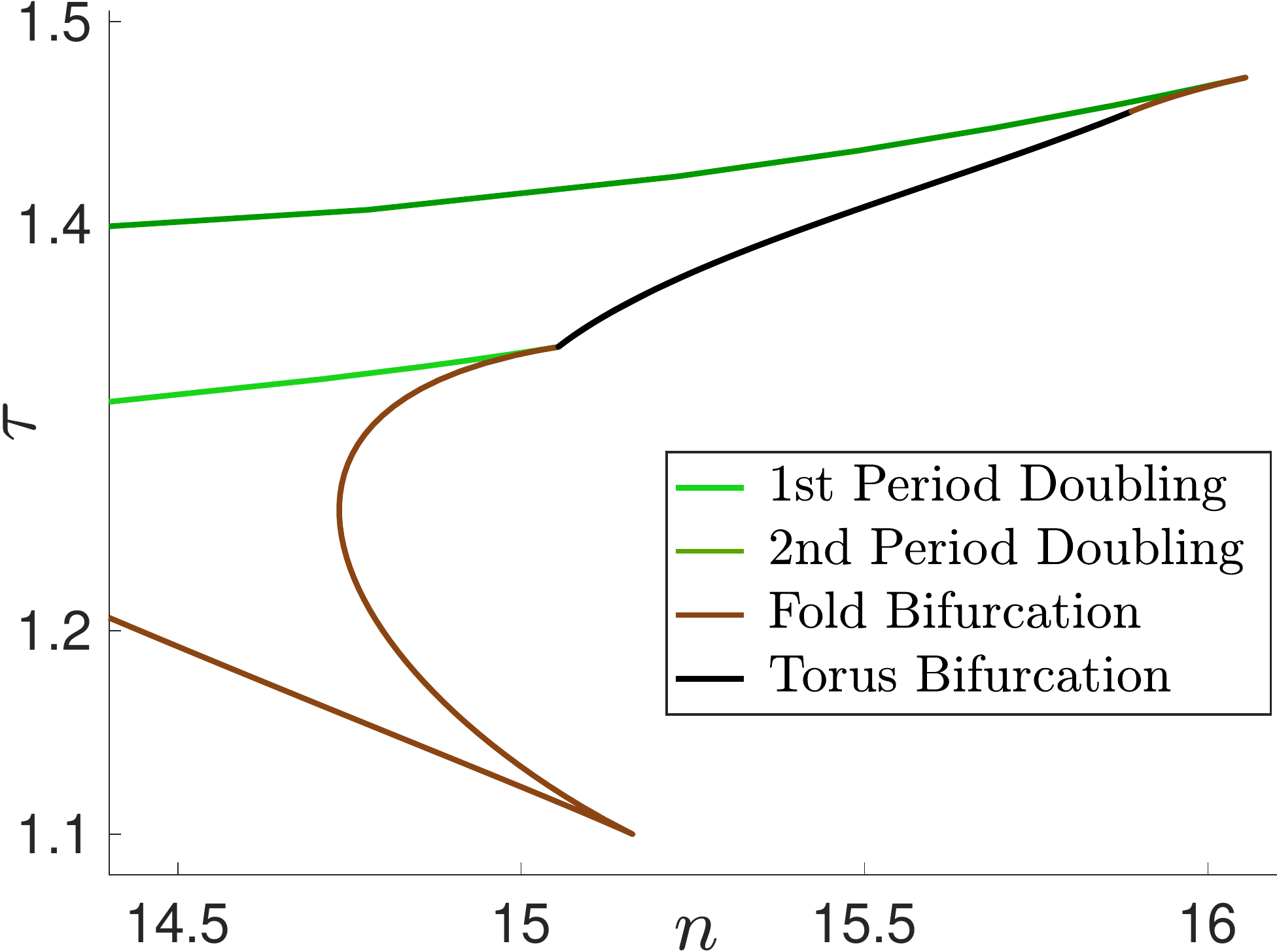}\hspace*{0.02\textwidth}\includegraphics[width=0.49\textwidth]{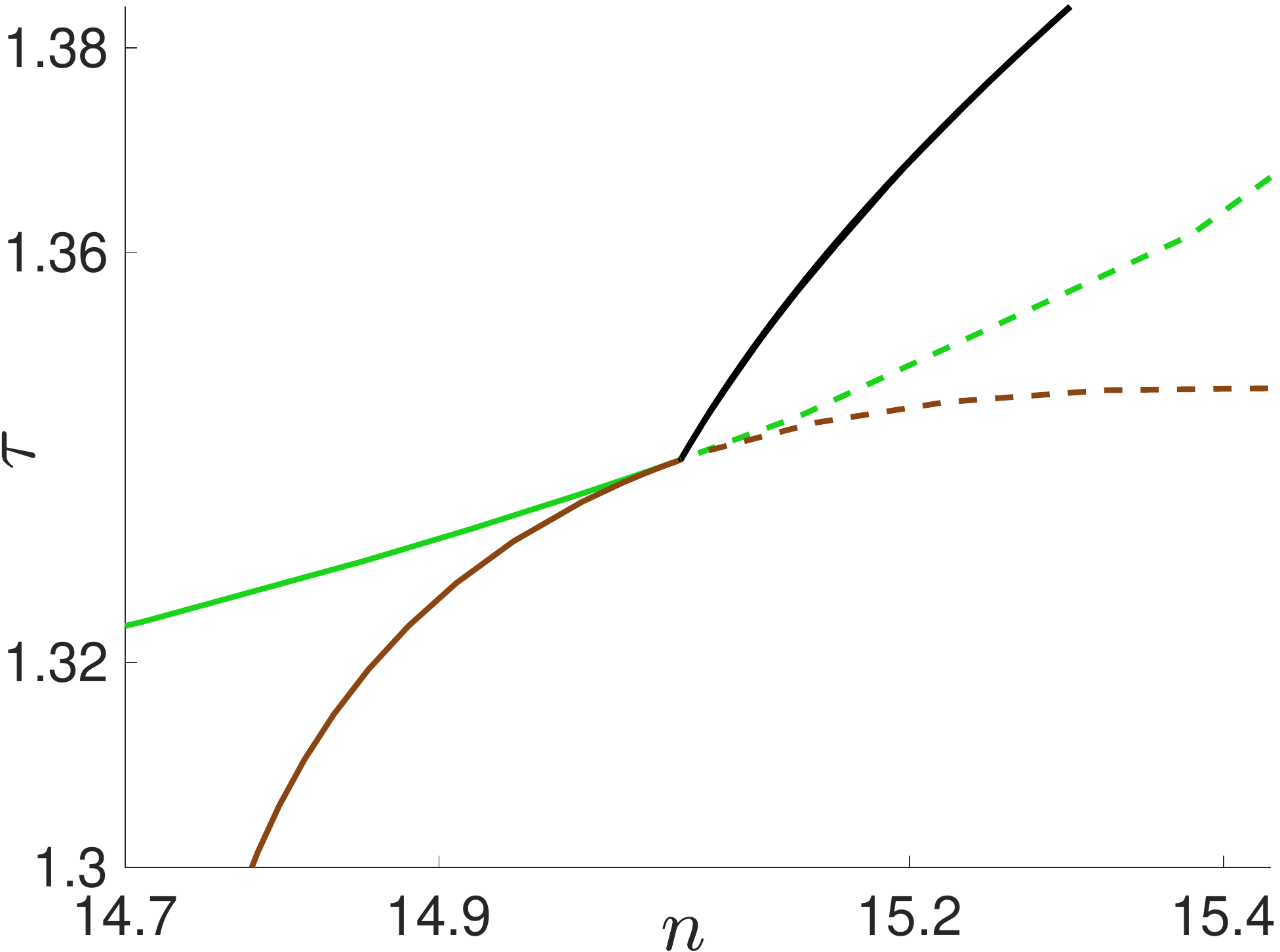}
\put(-120,90){\raisebox{.5pt}{\textcircled{\raisebox{-.9pt}{5}}}}
\put(-154,30){\raisebox{.5pt}{\textcircled{\raisebox{-.9pt}{6}}}}
\put(-80,40){\raisebox{.5pt}{\textcircled{\raisebox{-.9pt}{1}}}}
\put(-12,86){\raisebox{.5pt}{\textcircled{\raisebox{-.9pt}{2}}}}
\put(-35,105){\raisebox{.5pt}{\textcircled{\raisebox{-.9pt}{3}}}}
\put(-12,68){$F_+$}
\put(-46,52){FF1}
\put(-48,56){\vector(-3,1){30}}
\put(-12,110){$P_+$}
\put(-55,122){$\textit{NS}$}
\put(-158,52){$P_-$}
\put(-135,20){$F_-$}
\put(-342,122){(a)}
\put(-247,80){FF1}
\put(-207,90){FF2}
\put(-249,83){\vector(-1,0){30}}
\put(-197,99){\vector(1,3){6}}
\put(-152,122){(b)}	
\caption{(a) Enlarged detail of Bifurcation Diagram from Figure~\ref{fig:2paramcont} with
two fold-flip bifurcations indicated. (b) Bifurcation curves near the fold-flip bifurcation FF1, numbered and labelled as in \cite{KMV04}; see text for details.}
\label{fig:foldflip}
\end{figure}

A fold-flip bifurcation is a codimension-two bifurcation of maps where a branch of period-doubling bifurcations intersects (tangentially) with a branch of fold bifurcations. The essential details of the different cases of this bifurcation are explained in \cite{Kuznetsov_2004}, while a thorough treatment can be found in \cite{KMV04}. Although the Mackey-Glass equation \eqref{eq:MG12} is a delay differential equation rather than a map, the periodic orbits of the DDE define a map on any Poincar\'e section of the phase space $C_+$, and can thus undergo bifurcations of mappings.

Figure~\ref{fig:foldflip}(a) shows an area of the bifurcation diagram for the Mackey-Glass equation \eqref{eq:MG12} near $(n,\tau)=(15.05,1.34)$,
which reveals a tangential intersection of a curve of fold bifurcations and a curve of period-doubling bifurcations at the point labelled FF1.
To see that this is indeed a fold-flip bifurcation, we consider the bifurcation diagram in a small neighbourhood of FF1 as shown in Figure~\ref{fig:foldflip}(b).
Recall that in the main bifurcation diagram in Figure~\ref{fig:2paramcont}, we only showed the bifurcation curves involving
bifurcations from stable solutions. This resulted in the curves of fold and period-doubling bifurcations terminating at the point FF1 in Figure~\ref{fig:foldflip}(a) and Figure~\ref{fig:2paramcont}. These curves actually continue beyond the point FF1, but as bifurcations only involving unstable solutions. In Figure~\ref{fig:foldflip}(b) the unstable parts of these curves are reinstated as dashed lines labelled $P_+$ and $F_+$ respectively for the unstable period-doubling and unstable fold curves, while
their stable parts on the left side of FF1 are labelled $P_-$ and $F_-$.
The curve of torus bifurcations is labelled $\textit{NS}$ in Figure~\ref{fig:foldflip}(b); a Neimark-Sacker bifurcation being the equivalent bifurcation in maps to a torus bifurcation of a differential equation.

Considering the ordering of the curves  $F_\pm$, $P_\pm$ and $\textit{NS}$ around FF1
in Figure~\ref{fig:foldflip}(b),
and comparing with the four different cases for the normal form fold-flip bifurcations described in \cite{KMV04,Kuznetsov_2004}, we immediately see that FF1 corresponds to Kuznetsov's Case 1 of a fold-flip bifurcation
shown in Fig.~7 in \cite{KMV04} (also in Fig.~9.25 in \cite{Kuznetsov_2004}).
The curves $F_\pm$, $P_\pm$ and $\textit{NS}$ divide the neighbourhood of FF1 into five regions, which are numbered 1, 2, 3, 5 and 6 in Figure~\ref{fig:foldflip}(b), to correspond to the numbering in Fig.~7 in \cite{KMV04}.

Comparing Figure~\ref{fig:foldflip}(b) with Fig.~7 in \cite{KMV04} we see that there should be one additional bifurcation curve in Figure~\ref{fig:foldflip}.
This curve is labelled $J$ in \cite{KMV04}, and will emanate from the point FF1 adjacent to the curve of Neimack-Sacker bifurcations, thus creating another region between $J$ and $\textit{NS}$, which is numbered 4 in \cite{KMV04}.
The two periodic orbits created in the fold bifurcation on $F_\pm$ exist above the fold curve, and from the
analysis in \cite{KMV04} there is always a heteroclinic connection between them. The $J$ curve implies the existence of an additional transversal heteroclinic connection between the periodic orbits, and also nearby heteroclinic tangencies.

Case 1 has supercritical and subcritical subcases, depending on the sign of the first Lyapunov coefficient in the normal form analysis in \cite{KMV04}. In the supercritical case, the curve $J$ lies between $\textit{NS}$ and $P_+$. In this case, the stable period-doubled periodic orbit that exists in region 5 gives rise to a stable torus when crossing $\textit{NS}$ into region 4. The stable torus is then destroyed in a series of bifurcations associated with the heteroclinic structure around $J$. In the subcritical case the curve $J$ lies between $P_-$ and $\textit{NS}$. In this case, the stable
period-doubled periodic orbit that exists in region 5 gives rise to an unstable torus in region 4 between $J$ and $\textit{NS}$.

To determine which subcase FF1 corresponds to would require either a centre manifold and normal form reduction, or if in the supercritical case to numerically locate the stable torus.
Centre manifold and normal form reductions are tractable if technical and tedious for DDEs; see \cite{Calleja_2017}. While it is possible to compute stable invariant tori in DDEs \cite{Hum-DeM-Mag-Uph-12,Calleja_2017}, such a computation would be very challenging near FF1
because of the small range of parameters the torus might exist for and its proximity to other invariant objects. We did not pursue either of these approaches.

There are also other fold-flip bifurcations in the Mackey-Glass equation. While FF1 is at the right end of the first period-doubling curve in Figures~\ref{fig:2paramcont} and~\ref{fig:foldflip}, the right-hand end of the second period-doubling curve also ends at a codimension-two bifurcation where the period-doubling curve intersects a fold bifurcation. This second fold-flip point is labelled FF2 in Figure~\ref{fig:foldflip}(b). Fold-flip bifurcations can also be found in other parts of parameter space for the Mackey-Glass equation \eqref{eq:MG12}; for example there is one
visible in Figure~\ref{fig:subcritpd}(a) near $(n,\tau)=(8.46,2.72)$.

%

\subsection{Attractor Dimension}
\label{sec:dimension}

\begin{figure}[ht!]	
\includegraphics[width=\textwidth]{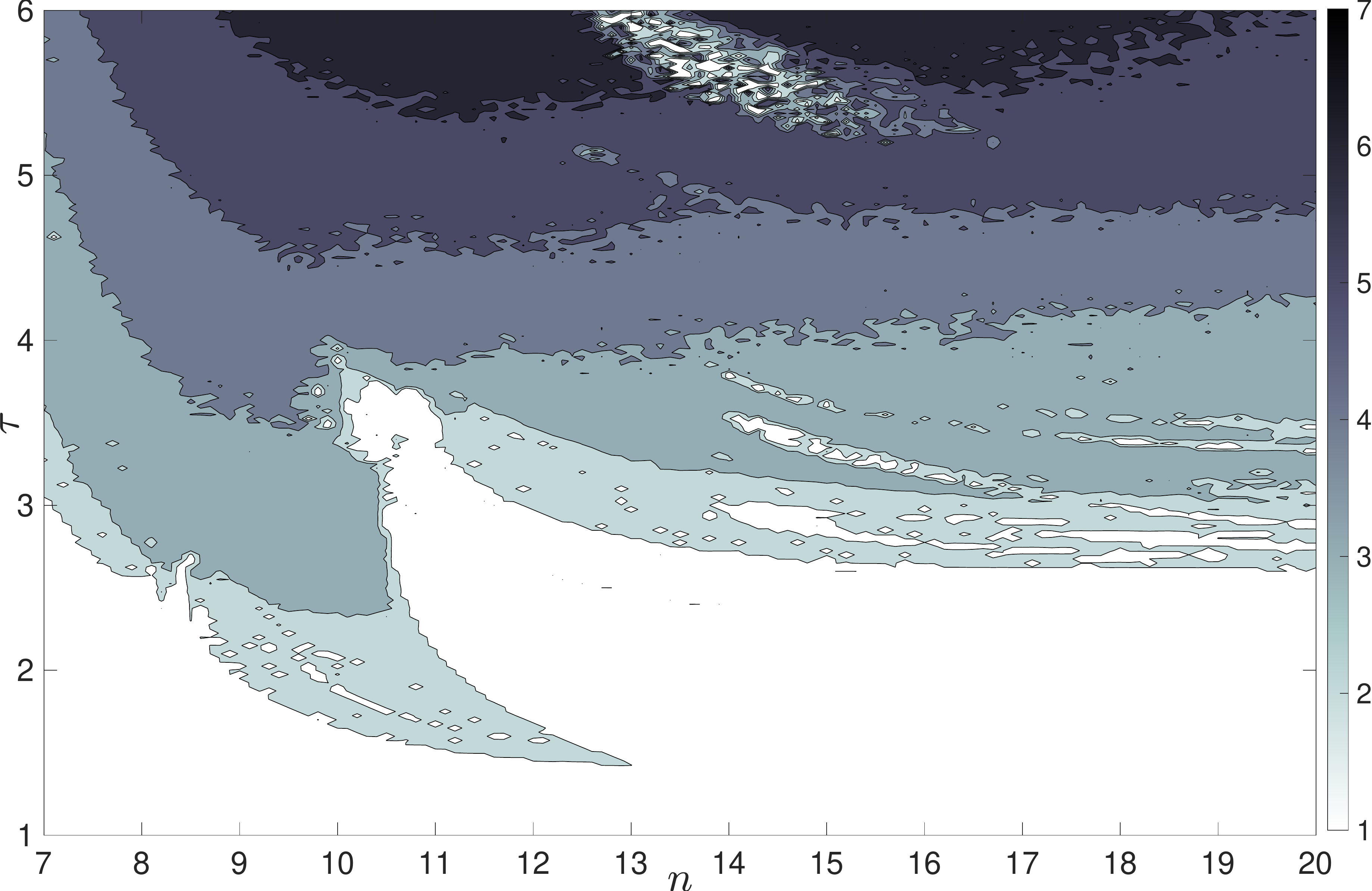}
\caption{Chaotic attractor dimension for the Mackey-Glass equation \eqref{eq:MG12} computed using \eqref{eq:lyapdim}.} \label{fig:dimensionA}
\end{figure}

\begin{figure}[p!]	
\includegraphics[width=0.37\textwidth]{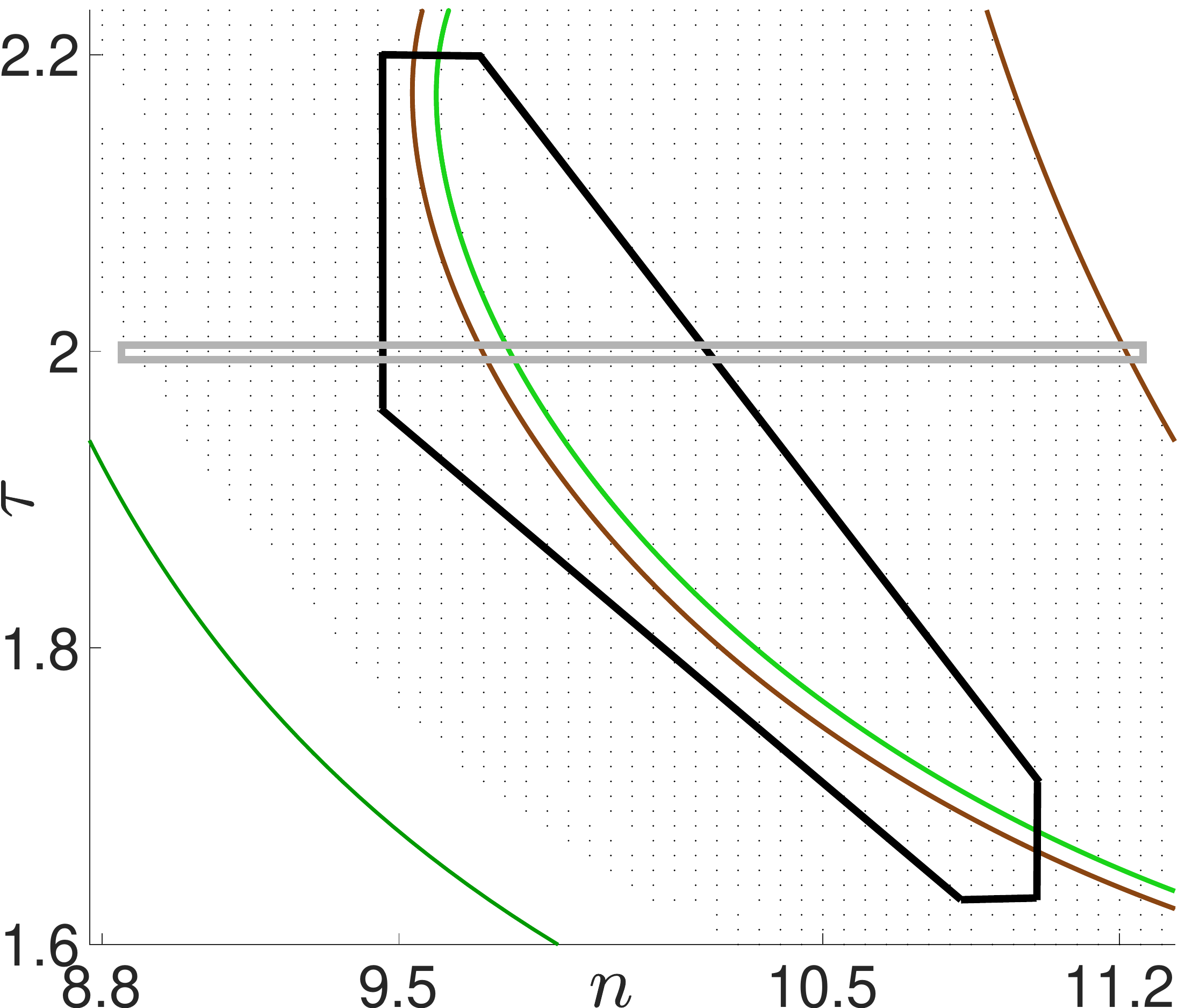}\hspace*{0.03\textwidth}\includegraphics[width=0.60\textwidth]{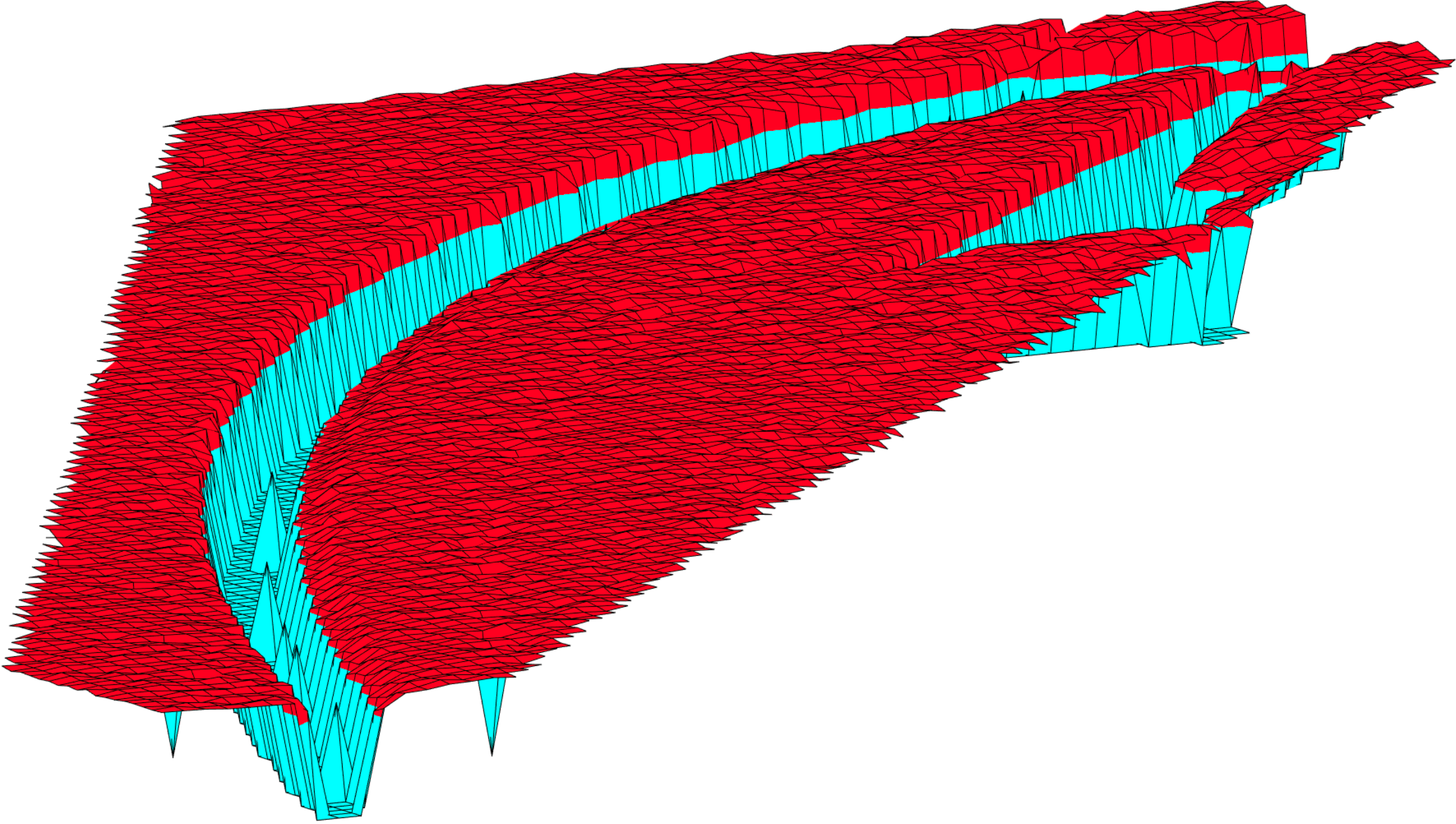}
\put(-345,14){(a)}
\put(-70,14){(b)}
\vspace*{1em}
\includegraphics[width=\textwidth]{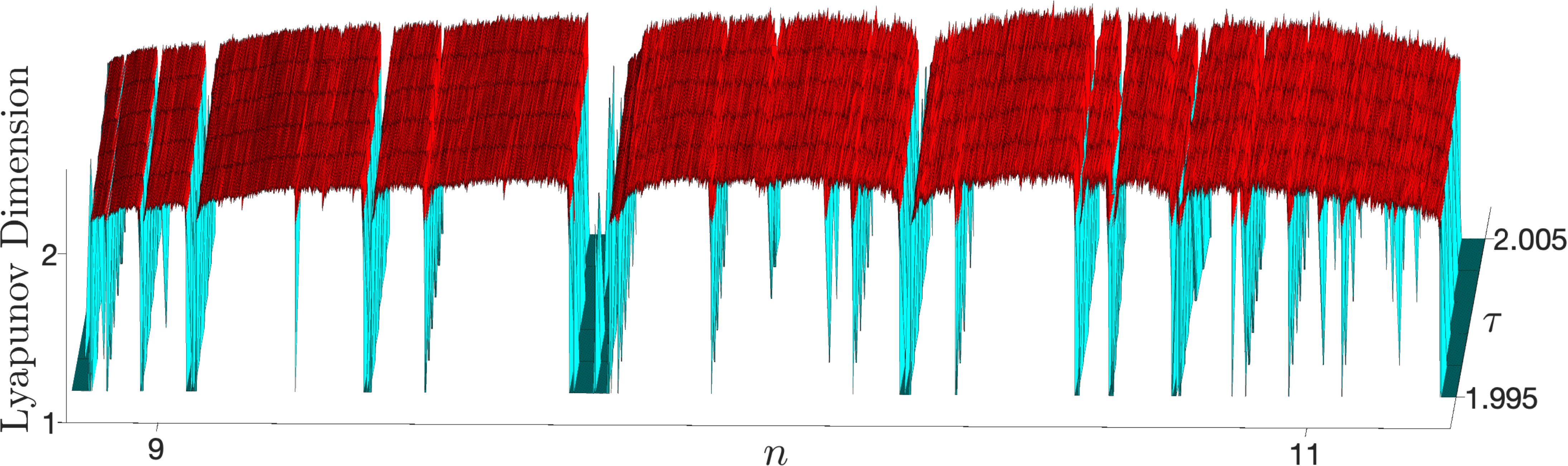}
\put(-311,18){(c)}
\caption{(a) A detail from Figure~\protect\ref{fig:2paramcont} indicating the parameter range in the $(n,\tau)$ plane considered in panels (b) and (c).
(b) Rotated surface plot of the Lyapunov dimension for the parameters within the black box in panel (a) near to the valley of periodic behaviour first seen in Figure~\protect\ref{fig:2paramcont}. The surface is coloured red where the Lyapunov dimension $d>2$.
(c) Surface plot of the Lyapunov dimension for the parameters within the grey box in panel (a) showing many valleys of periodic behaviour.} \label{fig:dimensionB}
\end{figure}

\begin{figure}[p!]	
\includegraphics[width=0.49\textwidth]{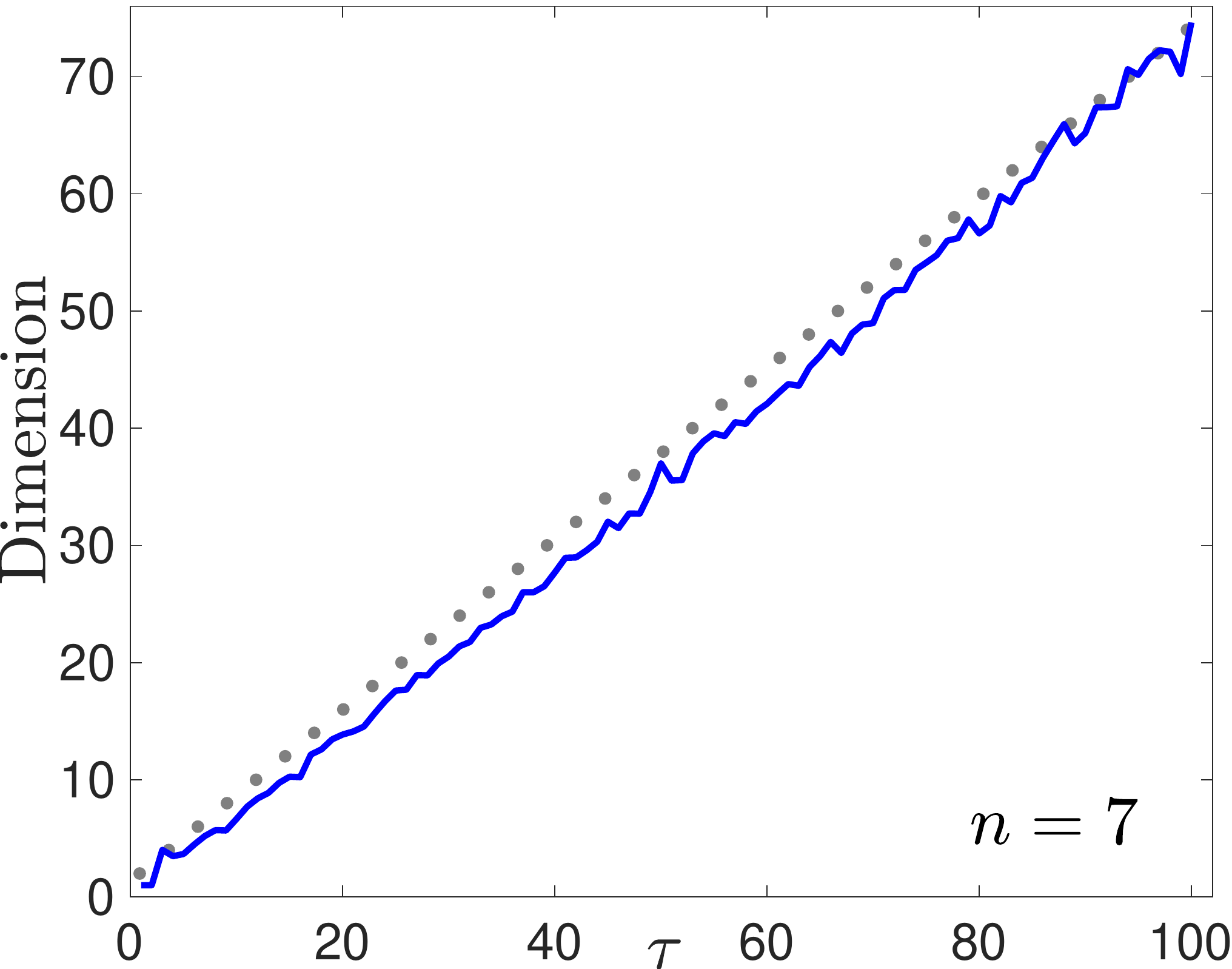}\hspace*{0.02\textwidth}\includegraphics[width=0.49\textwidth]{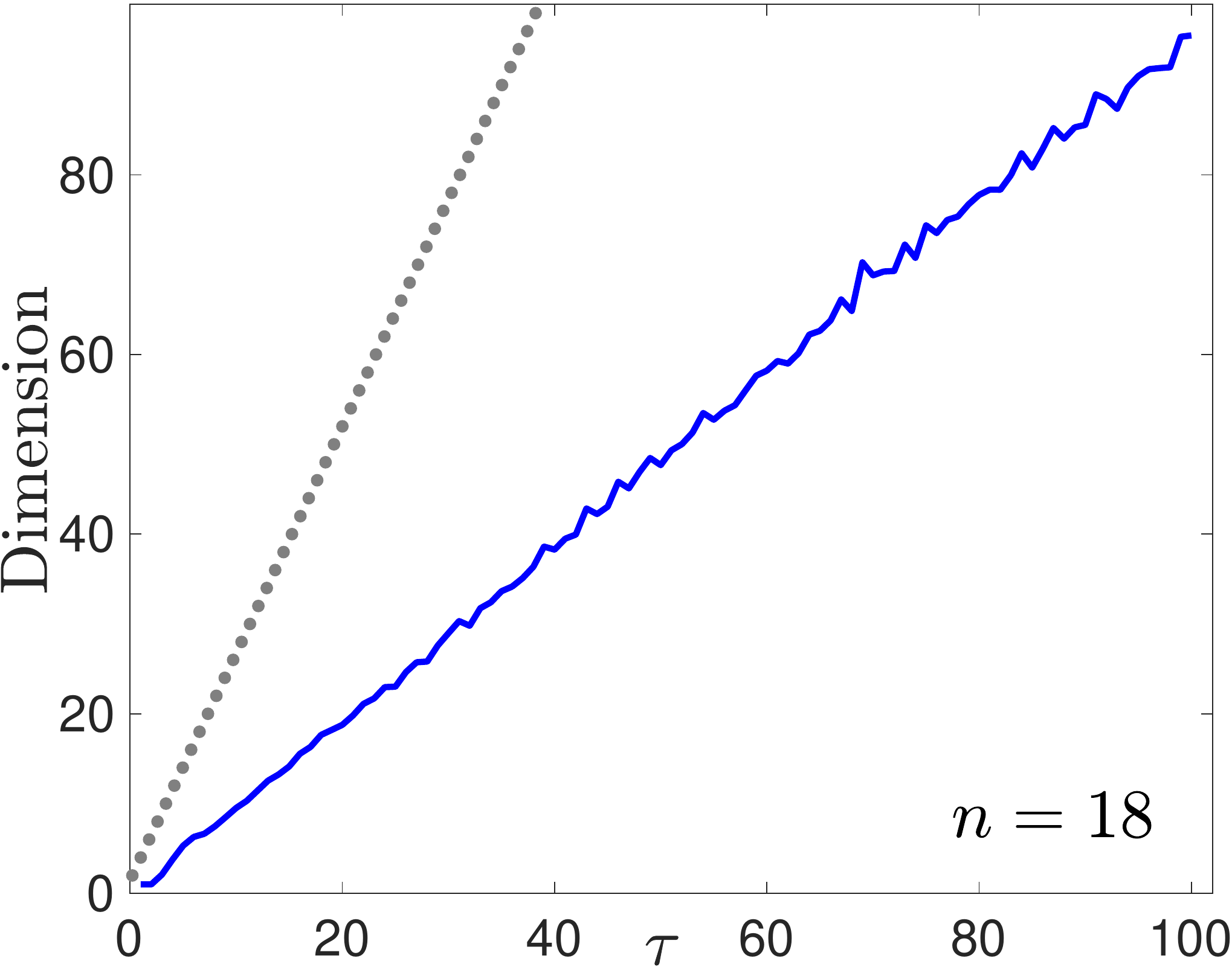}
\put(-337,124){(a)}
\put(-152,124){(b)}	
\caption{(a) For $n=7$ growth of attractor dimension with $\tau$ (blue line). Also shown is the dimension of the unstable manifold of the steady state $\xi_1$ (dotted line). (b) As (a) but for $n=18$.} \label{fig:dimensionC}
\end{figure}

In Figure~\ref{fig:2paramcont} we used the presence of a positive Lyapunov exponent to detect chaotic dynamics, but through equation \eqref{eq:lyapdim} such computations can be used to determine the Lyapunov dimension of the attractor, as described in Section~\ref{sec:LyapDim}.
High-dimensional chaos and Lyapunov dimension have already been studied in detail in many DDE systems,
including the Mackey-Glass equation \cite{Farmer82,Lepri94,Longtin1998}, so we briefly present our
results in this section.

Figure~\ref{fig:dimensionA} shows a contour plot of the Lyapunov dimension of the attractor over part of the two-dimensional parameter space $(n,\tau)$.
Interesting features are visible including the valleys of
periodic behaviour seen in Figure~\ref{fig:2paramcont} (though they mostly appear as chains of islands
because the valleys are too narrow to be resolved in the contour plot), and the Lyapunov dimension generally increasing with $\tau$.
On the whole, the attractor dimension does not seem to depend strongly on $n$, with two exceptions. Firstly, the dimension is lower for very small values of $n$, which should not be surprising as \eqref{eq:ndelind} shows that $\xi_1$ is asymptotically stable for $n<4$.
Secondly, there is the tongue of chaotic behaviour that we have already seen reaching down towards the cusp point.

Figure~\ref{fig:dimensionB}(b) shows the Lyapunov dimension as a surface plot on part of the tongue of chaotic behaviour above the cusp point. The parameter values were chosen to encompass part of the valley of periodic behaviour that was first seen in Figure~\ref{fig:2paramcont}.
Unlike Figure~\ref{fig:dimensionA} where the valley was not well resolved, the valley is clearly visible on the finer computational mesh used in Figure~\ref{fig:dimensionB}(b). Because there is always a zero Lyapunov exponent, the formula \eqref{eq:lyapdim} for the Lyapunov dimension gives dimension $d=1$ if zero is the largest Lyapunov exponent and dimension $d\geq2$ if there is a positive Lyapunov exponent. Because Lyapunov dimensions between $1$ and $2$ are not possible, the valley of periodic behaviour appears as a deep canyon in Figure~\ref{fig:dimensionB}(b).

Panel (b) of Figure~\ref{fig:dimensionB} is rotated to clearly show the valley, so in panel (a) we reproduce a part of
Figure~\ref{fig:2paramcont} with a black box added to denote the parameter range used in panel (b).
Notice that the periodic window for $\tau=2$ seen in Figure~\ref{fig:ODtau2details}(b)
cuts across the valley. In panel (c) we show a cross-section across the entire chaotic tongue for increasing $n$ with $\tau\approx2$. This shows the existence of many non-chaotic valleys; some resolved better than others. These valleys correspond to the periodic windows seen in Figures~\ref{fig:ODtau2} and~\ref{fig:ODtau2details}, but now continued into two-dimensional parameter space. Each valley is bounded on one side by a fold bifurcation of periodic orbits (an interior crisis of the chaotic attractor) and on the other by a period-doubling cascade. Rankin and Osinga \cite{RO17} (who preferred the term `channel' to valley) found similar behaviour in the Ikeda map. For that map they also studied how the valleys create gaps in the locus of a boundary crisis. Similar behaviour likely occurs in the Mackey-Glass equation, for which the wide periodic valley illustrated in Figure~\ref{fig:dimensionB} is seen in Figure~\ref{fig:cusp}(a) to cross the locus of the boundary crisis curve at the edge of the chaotic region.

In Figure~\ref{fig:dimensionC} we explore the growth of the Lyapunov dimension as $\tau$ increases for two different fixed values of $n$. Also shown on these graphs is the dimension of the unstable manifold of the steady state $\xi_1$. Since any chaotic attractor must be embedded in the global attractor of the dynamical system, the dimension of the global attractor provides an upper bound for the dimension of the attractor.
The global attractor is the  maximal compact invariant set, and must contain the unstable manifold of $\xi_1$. Thus the dimension of this unstable manifold provides a lower bound on the dimension of the global attractor, and hence a very simple estimate for its dimension as well as that of the chaotic attractor.
From Figure~\ref{fig:dimensionC}(a) it is clear that for $n=7$ this estimate is amazingly good, even up to very large values of $\tau$. However, for $n=18$, Figure~\ref{fig:dimensionC}(b) suggests that the chaotic attractor has much smaller dimension than the global attractor. Farmer~\cite{Farmer82} already suggested this estimate and reported some results for a time-rescaled version of \eqref{eq:MG12}, but only found behaviour similar to Figure~\ref{fig:dimensionC}(b).

The proportionality constant between the Lyapunov dimension and the delay has been shown to be inversely proportional to the autocorrelation time of the feedback function \cite{Longtin1998}. Another
plausible explanation for the difference between Figures~\ref{fig:dimensionC}(a) and (b) involves the
cusp bifurcation studied in Section~\ref{sec:cusp}.
For $n=7$ the left-hand periodic orbit undergoes a sequence of period-doubling bifurcations leading to chaos, while the steady state undergoes a sequence of Hopf bifurcations. Since the periodic orbits are in the unstable manifold of the steady state, it is not surprising that these bifurcations are linked (as argued in \cite{Farmer82,Longtin1998}), leading to the close agreement between the Lyapunov dimension of the chaotic attractor and the dimension of the unstable manifold of $\xi_1$.
On the other hand,
as seen in Section~\ref{sec:cusp},
to the right of the left-hand curve of folds above the cusp bifurcation, the left-hand periodic orbit and the steady state undergo the same bifurcations, while the right-hand periodic orbit remains stable. But for these larger values of $n$ it is ultimately the right-hand periodic orbit that undergoes a period-doubling cascade leading to the chaotic dynamics seen for larger $\tau$ values.
So, the cusp bifurcation in partitioning the parameter space, may also break or weaken the connection between the dimension of the chaotic attractor and the unstable manifold of $\xi_1$.

\section{Conclusions}
\label{sec:conclusions}

The Mackey-Glass equation has long been known to exhibit very complicated dynamics. In this study
we have found several phenomena which we believe have not been previously observed  for this DDE. In particular, in Section~\ref{sec:cusp} we found a cusp bifurcation of periodic orbits leading immediately to bistability of periodic orbits, and ultimately through a period-doubling cascade to bistability between a stable periodic orbit and a stable chaotic attractor.
In Sections~\ref{sec:cusp} and~\ref{sec:snoca} we found evidence of two different global bifurcations through which the chaotic attractor is destroyed in a bifurcation with a periodic orbit. In the example in Section~\ref{sec:cusp} the chaotic attractor and periodic orbit can coexist, and the chaotic attractor is destroyed in a boundary crisis on the edge of the parameter region of coexistence when the two invariant objects collide. In the example in Section~\ref{sec:snoca} the chaotic attractor and the periodic orbit cannot coexist,
and the attractor is destroyed at an interior crisis when a stable and unstable pair of periodic orbits are created in a fold bifurcation of periodic orbits on the attractor.

In Section~\ref{sec:subcritpd}
we again found
bistability between a stable periodic orbit and a stable chaotic attractor, but through a different mechanism beginning with a sequence of subcritical period-doubling bifurcations.
In Section~\ref{sec:foldflip} we also found examples of fold-flip bifurcations.

Whereas many previous studies of the Mackey-Glass equation have considered chaotic or periodic behaviour for $\tau\gg0$ or $n\gg0$, in the current study we considered small and moderate values of these bifurcation
parameters near to the onset of chaos. All of the bifurcations discussed above were found to occur with $\tau<4$ and $n<17$.

We also investigated the Lyapunov dimension of the chaotic attractor for both small and large delays in Section~\ref{sec:dimension}. There we showed how to visualise the chaotic valleys, and also found a difference in the relationship between the global attractor dimension and chaotic attractor dimension, depending on the parameter choice relative to the location of the cusp bifurcation.

In finding these bifurcation structures we demonstrated how
numerical continuation and bifurcation tools can be invaluable for studying dynamical systems defined by differential equations.
There are a number of such tools available for ODEs,
but since the Mackey-Glass equation \eqref{eq:MG} has delays we used
DDE-BIFTOOL \cite{DDEBiftool15}. This allowed us to perform one and two-parameter continuation
of steady states and periodic orbits, and determine their stability.
In particular, branches of unstable periodic orbits are found just as easily as stable orbits.
DDE-BIFTOOL also enables us to find both codimension-one and codimension-two bifurcations.
The ability to compute unstable solutions allows us to find bifurcations that
simulation alone could never reveal, such as the subcritical period-doubling bifurcations found in Section~\ref{sec:subcritpd}.

We also showed that numerical bifurcation analysis is very powerful when applied in consort with simpler numerical simulation of the differential equation. This is particularly evident when exploring multi-dimensional parameter spaces. Even in the modestly sized two-dimensional window of parameter space in which we considered the Mackey-Glass equation \eqref{eq:MG12},
it was prohibitively expensive to simulate the solution for every parameter combination, and we were only able to do this using a supercomputer.
In contrast, the ability of DDE-BIFTOOL to perform two-parameter continuation of bifurcations allowed us to explore this parameter space on a laptop computer. In this way we were able to
identify codimension-two bifurcations and parameter regimes where interesting dynamics may occur. To the best of our knowledge these bifurcations had not previously been found in the Mackey-Glass equation; and without the use of two-parameter continuation it is unlikely that they ever would have been found.
In some cases (especially Figure~\ref{fig:subcritpd}(d)) the numerical simulations were quite problematical, and the insight into the dynamics provided by DDE-BIFTOOL was essential to get to the final result.

The Mackey-Glass equation \eqref{eq:MG} has now been studied for over 40 years, but remains a fertile topic for research. Progress continues to be made, as both our work and the recent results in \cite{Krisztin_StableOrbitsMG_JDE2021} demonstrate. However,
the bifurcations that solutions can undergo, as well as the different possible routes to chaos that the equation can present, remain only partially explored. A rigorous proof of chaotic dynamics in \eqref{eq:MG}  also remains elusive \cite{walther2020impact}.

\section*{Acknowledgments}

ARH is grateful to the New Zealand Mathematics Research Institute (NZMRI) for their invitation and funding
to lecture at the NZMRI Summer School on Continuation Methods in Dynamical Systems at Raglan, New Zealand
in 2016. The current work started with those lectures and
was greatly extended during the Undergraduate Research Traineeship (URT) of VD. A version was presented at the Leon Glass and Michael C. Mackey Diamond Symposium at McGill in Montreal in 2018, and we thank Leon Glass and Mike Mackey for their feedback.
We are also grateful to the Natural Science and Engineering Research Council, Canada for funding; ARH through Discovery Grants RGPIN-2013-261389 and RGPIN-2018-05062, and VD through the use of those funds to finance his URT. VD is also supported in part through NSF grant CCF-2112665.
We thank Dimitri Breda for sharing his \textsc{Matlab}~code for the computation of Lyapunov exponents in DDEs. Some computations were made on the supercomputer Guillimin, managed by Calcul Qu\'ebec and Compute Canada. The operation of this supercomputer is funded by the Canada Foundation for Innovation, Minist\`ere de l’\'Economie et de l’Innovation du Qu\'ebec and le Fonds de recherche du Qu\'ebec.

\appendix

\section{Supplemental Material}

\textsc{Matlab} drivers to reproduce some of the results in this article are supplied as ancillary files
at \url{https://doi.org/10.48550/arXiv.2203.00181}. This includes a one-parameter continuation driver
which uses both DDE-BIFTOOL and \verb+dde23+ and can be used to reproduce Figures~\ref{fig:contnamp}, \ref{fig:contnps}, \ref{fig:contn20r2p}(a), \ref{fig:cusp}(b)-(d), and \ref{fig:bistabpoca}. A \verb+dde23+
driver produces orbit diagrams such as those in Figures~\ref{fig:ODtau2} and \ref{fig:ODtau2details}.
Also shared is a two-parameter DDE-BIFTOOL driver which can be used to find segments of two-parameter bifurcation curves that bound stable solutions; as in Figure~\ref{fig:2paramcont}.
The code that we used to compute the Lyapunov exponents is not shared because
copyright of major components of that code belongs to the authors of \cite{Breda_2014}.


%

\providecommand{\href}[2]{#2}
\providecommand{\arxiv}[1]{\href{http://arxiv.org/abs/#1}{arXiv:#1}}
\providecommand{\url}[1]{\texttt{#1}}
\providecommand{\urlprefix}{URL }


\end{document}